\documentclass[aps,prd,twocolumn,superscriptaddress,nofootinbib]{revtex4-1}

\usepackage{latexsym}
\usepackage{amsmath}
\usepackage{amssymb}
\usepackage{amsfonts}
\usepackage{bm}

\usepackage{color}
\definecolor{purple}{rgb}{0.5,0,0.5}
\definecolor{blue}{rgb}{0.0,0,0.9}
\definecolor{prdblue}{rgb}{0.133,0.118,0.498}
\usepackage[colorlinks=true, pdfstartview=FitV, linkcolor=prdblue, citecolor= prdblue, urlcolor=prdblue]{hyperref}

\usepackage{supertabular} 
\usepackage{placeins}
\usepackage{epsfig}
\usepackage{graphicx}

\begin{document}


\title{Triply charm and bottom tetraquarks in a constituent quark model}


\author{Gang Yang}
\email[]{yanggang@zjnu.edu.cn}
\affiliation{Department of Physics, Zhejiang Normal University, Jinhua 321004, China}

\author{Jialun Ping}
\email[]{jlping@njnu.edu.cn}
\affiliation{Department of Physics and Jiangsu Key Laboratory for Numerical Simulation of Large Scale Complex Systems, Nanjing Normal University, Nanjing 210023, P. R. China}

\author{Jorge Segovia}
\email[]{jsegovia@upo.es}
\affiliation{Departamento de Sistemas F\'isicos, Qu\'imicos y Naturales, Universidad Pablo de Olavide, E-41013 Sevilla, Spain}



\begin{abstract}
Singly, doubly and fully charmed tetraquark candidates, \emph{e.g.}, $T_{c\bar{s}}(2900)$, $T^+_{cc}(3875)$ and $X(6900)$ have been recently reported by the LHCb collaboration. Therefore, it is timely to implement a theoretical investigation on triply heavy tetraquark systems; herein, the S-wave triply charm and bottom tetraquarks, $\bar{Q}Q\bar{q}Q$ $(q=u,\,d,\,s;\,Q=c,\,b)$, with spin-parity $J^P=0^+$, $1^+$ and $2^+$, isospin $I=0$ and $\frac{1}{2}$, are systematically studied in a constituent quark model. Besides, all tetraquark configurations, \emph{i.e.} meson-meson, diquark-antidiquark and K-type arrangements, along with any allowed color structure, are comprehensively considered. The Gaussian expansion method (GEM), in combination with the complex-scaling method (CSM), which is quite ingenious in dealing with either bound or resonances, is the approach adopted in solving the complex scaled Schr\"odinger equation. This theoretical framework has already been applied in various tetra- and penta-quark systems. In a fully coupled-channel calculation within the GEM$+$CSM, narrow resonances are found in each $I(J^P)$ channel of the charm and bottom sector. In particular, triply charm and bottom tetraquark resonances are obtained in $5.6-5.9$ GeV and $15.3-15.7$ GeV, respectively. We provide also some insights of the compositeness of these exotic states, such as the inner quark distance, magnetic moment and dominant wave function component. All this may help to distinguish them in future high energy nuclear and particle experiments.
\end{abstract}


\maketitle


\section{Introduction}
In the past two decades, a large amount of crucial findings on exotic hadrons in heavy flavor sectors have been reported experimentally. In particular, tetraquark candidates in the charm-strange sector, $X_{0,(1)}(2900)$ and $T^{0(++)}_{c\bar{s}}(2900)$ were reported by the LHCb collaboration in studying $B$ meson three-body strong decays~\cite{LHCb:2020pxc, LHCb:2020bls, LHCb:2022sfr, LHCb:2022lzp}. In 2021, a double charm tetraquark candidate, $T^+_{cc}(3875)$, which is quite close to the $D^{*+}D^0$ threshold, was announced by the same collaboration~\cite{LHCb:2021vvq, LHCb:2021auc}. Moreover, in the fully charmed sector, four exotic states, which are denoted as $X(6400)$, $X(6600)$, $X(6900)$ and $X(7200)$, were reported by the LHCb, CMS and ATLAS collaborations~\cite{LHCb:2020bwg, CMS:2023owd, ATLAS:2023bft}. At the same time, an enormous theoretical effort with a wide variety of approaches has been carried out in order to reveal the nature of exotic hadrons. Generally, there are extensive reviews~\cite{Dong:2020hxe, Chen:2016qju, Chen:2016spr, Guo:2017jvc, Liu:2019zoy, Yang:2020atz, Dong:2021bvy, Chen:2021erj, Cao:2023rhu, Mai:2022eur, Meng:2022ozq, Chen:2022asf, Guo:2022kdi, Ortega:2020tng, Huang:2023jec, Lebed:2023vnd, Zou:2021sha, Du:2021fmf, Liu:2024uxn, Johnson:2024omq}, which explain in detail a particular theoretical method and thus capture a certain interpretation of exotic states.

One can realize from above that singly, doubly and fully heavy tetraquarks have been observed experimentally and thus asking oneself if there are tetraquarks with three constituent heavy quarks. Until now, there is no experimental data on triply heavy tetraquarks. Besides, no bound states of $QQ\bar{Q}\bar{q}$ tetraquark system are obtained in several studies using the MIT bag model~\cite{Zhu:2023lbx}, an effective field theory approach~\cite{Liu:2019mxw}, the extended chromomagnetic model~\cite{Weng:2021ngd}, the relativized quark model~\cite{Lu:2021kut}, and QCD sum rules~\cite{Jiang:2017tdc}. In contrast, several bound states are obtained in the $cc\bar{c}\bar{q}$ tetraquark system using a chiral quark model~\cite{Liu:2022jdl}. Furthermore, it is found that some $bb\bar{b}\bar{q}$ tetraquark states may lie below the bottomonia plus $B^{(*)}$ thresholds~\cite{Jiang:2017tdc}. Moreover, stable excited states of triply heavy tetraquarks are also found~\cite{Zhu:2023lbx, Jiang:2017tdc, Mutuk:2023yev, Xing:2019wil, Chen:2016ont}, calculating, besides their masses, magnetic moments, charge radii and decay properties.

We perform a systematic investigation on triply heavy tetraquarks, $\bar{Q}Q\bar{q}Q$ $(q=u,\,d,\,s;\,Q=c,\,b)$, in a constituent quark model. This approach has been already applied with reasonable success in predictions and descriptions of various tetra- and penta-quark systems, \emph{e.g.}, hidden-, single- double- and fully-heavy tetraquarks~\cite{gy:2020dht, gy:2020dhts, Yang:2021hrb, Yang:2023mov, Yang:2021zhe, Yang:2022cut, Yang:2021izl, Yang:2023mov, Yang:2023wgu}, and pentaquarks~\cite{Yang:2015bmv, Yang:2018oqd, gy:2020dcp, Yang:2022bfu, Yang:2023dzb}. The wave functions of S-wave $\bar{Q}Q\bar{q}Q$ tetraquarks, with spin-parity $J^P=0^+$, $1^+$ and $2^+$, isospin $I=0$ and $\frac{1}{2}$, are comprehensively constructed by including meson-meson, diquark-antidiquark and K-type arrangements of quarks, along with all allowed color structures. A high accuracy and efficient numerical approach, the Gaussian expansion method (GEM)~\cite{Hiyama:2003cu}, along with a powerful complex scaling method (CSM), which is quite ingenious in dealing with bound and resonant states simultaneously, is employed to solve the resulting Schr\"odinger equation for the four-body system.

This article is arranged as follows. In Sec.~\ref{sec:model} the theoretical framework is presented, it includes a detailed explanation of the constituent quark model and wave functions of the $\bar{Q}Q\bar{q}Q$ tetraquarks. Section~\ref{sec:results} is devoted to the analysis of the calculated results. Finally, a summary can be found in Sec.~\ref{sec:summary}.


\section{Theoretical framework}
\label{sec:model}

The theoretical formalism employed herein has been previously published in Ref.~\cite{Yang:2020atz}, and we shall then focus on the most relevant features of the model and the numerical approach concerning the $\bar{Q}Q\bar{q}Q$ tetraquarks.

\subsection{The Hamiltonian}

We solve the complex-scaled Schr\"odinger equation to study the 4-body bound and resonant states, the equation is 
\begin{equation}\label{CSMSE}
\left[ H(\theta)-E(\theta) \right] \Psi_{JM}(\theta)=0 \,,
\end{equation}
with $E(\theta)$ and $\Psi(\theta)$ the eigenenergies and eigenfunctions, respectively, and the 4-body Hamiltonian for a QCD-inspired constituent quark model reads as
\begin{equation}
H(\theta) = \sum_{i=1}^{4}\left( m_i+\frac{(\vec{p\,}_i e^{-i\theta})^2}{2m_i}\right) - T_{\text{CM}} + \sum_{j>i=1}^{4} V(\vec{r}_{ij} e^{i\theta}) \,,
\label{eq:Hamiltonian}
\end{equation}
where $m_{i}$ is the constituent quark mass, $\vec{p}_i$ is the momentum of a quark, $T_{\text{CM}}$ is the center-of-mass kinetic energy and the last term is the two-body potential.

By introducing an artificial parameter, \emph{i.e.} the rotated angle $\theta$, three kinds of complex eigenvalues: bound, resonance and scattering states, can be simultaneously studied. Particularly, bound and resonance states are independent of the rotated angle $\theta$, with the first ones always placed on the real-axis of the complex energy plane, and the second ones located above the threshold line with a total decay width $\Gamma=-2\,\text{Im}(E)$. Meanwhile, the scattering state are unstable with respect the rotated angle $\theta$ and align along the corresponding threshold line.

The dynamics of triply charm and bottom tetraquark systems are driven by two-body complex-scaled potentials,
\begin{equation}
\label{CQMV}
V(\vec{r}_{ij} e^{i\theta}) = V_{\text{CON}}(\vec{r}_{ij} e^{i\theta}) + V_{\text{OGE}}(\vec{r}_{ij} e^{i\theta})  \,.
\end{equation}
In particular, color-confinement and perturbative one-gluon exchange interactions are the most relevant features of QCD in its low energy regime since the presence of heavy quarks make chiral symmetry explicitly broken. Since the lowest-lying $S$-wave positive parity triply heavy tetraquarks shall be investigated, only central and spin-spin terms of the interactions are considered.

Color confinement should be encoded in the non-Abelian character of QCD. On one hand, lattice-regularized QCD has demonstrated that multi-gluon exchanges produce an attractive linearly rising potential proportional to the distance between infinite-heavy quarks~\cite{Bali:2005fu}. On the other hand, the spontaneous creation of light-quark pairs from the QCD vacuum may give rise at the same scale to a breakup of the created color flux-tube~\cite{Bali:2005fu}. We can phenomenologically describe the above two observations by implementing the following confining effective potential~\cite{Vijande:2004he, Segovia:2008zz}
\begin{equation}
V_{\text{CON}}(\vec{r}_{ij} e^{i\theta})=\left[-a_{c}(1-e^{-\mu_{c}r_{ij} e^{i\theta}})+\Delta \right] 
(\lambda_{i}^{c}\cdot \lambda_{j}^{c}) \,,
\label{eq:conf}
\end{equation}
where $\lambda^c$ denote the SU(3) color Gell-Mann matrices, whereas $a_{c}$, $\mu_{c}$ and $\Delta$ are model parameters. When the rotated angle $\theta$ is $0^\circ$, one can see in Eq.~\eqref{eq:conf} that the real-range potential is linear at short inter-quark distances with an effective confinement strength $\sigma = -a_{c} \, \mu_{c} \, (\lambda^{c}_{i}\cdot \lambda^{c}_{j})$, while it becomes a constant at large distances, $V_{\text{thr.}} = (\Delta-a_c)(\lambda^{c}_{i}\cdot \lambda^{c}_{j})$.

The perturbative one-gluon exchange potential, which includes the so-called Coulomb and color-magnetic interactions, is~\cite{Vijande:2004he, Segovia:2008zz}
\begin{align}
&
V_{\text{OGE}}(\vec{r}_{ij} e^{i\theta}) = \frac{1}{4} \alpha_{s} (\lambda_{i}^{c}\cdot \lambda_{j}^{c}) \Bigg[\frac{1}{r_{ij} e^{i\theta}} \nonumber \\ 
&
\hspace*{1.60cm} - \frac{1}{6m_{i}m_{j}} (\vec{\sigma}_{i}\cdot\vec{\sigma}_{j}) 
\frac{e^{-r_{ij} e^{i\theta} /r_{0}(\mu_{ij})}}{r_{ij} e^{i\theta} r_{0}^{2}(\mu_{ij})} \Bigg] \,,
\end{align}
where $\vec{\sigma}$ denote the Pauli matrices and the regularized contact term is given by
\begin{equation}
\delta(\vec{r}_{ij} e^{i\theta}) \sim \frac{1}{4\pi r_{0}^{2}(\mu_{ij})}\frac{e^{-r_{ij} e^{i\theta} / r_{0}(\mu_{ij})}}{r_{ij} e^{i\theta} } \,.
\end{equation}
with $r_{0}(\mu_{ij})=\hat{r}_{0}/\mu_{ij}$ a regulator that depends on the reduced mass of the quark--(anti-)quark pair.

An effective scale-dependent strong coupling constant provides a consistent description of mesons and baryons from light to heavy quark sectors. Its effective parametrizacion has the expression~\cite{Vijande:2004he, Segovia:2008zz}
\begin{equation}
\alpha_{s}(\mu_{ij})=\frac{\alpha_{0}}{\ln\left(\frac{\mu_{ij}^{2}+\mu_{0}^{2}}{\Lambda_{0}^{2}} \right)} \,,
\end{equation}
where $\alpha_{0}$, $\mu_{0}$ and $\Lambda_{0}$ are model parameters.

\begin{table}[!t]
\caption{\label{tab:model} Quark model parameters.}
\begin{ruledtabular}
\begin{tabular}{llr}
Quark masses     & $m_q\,(q=u,\,d)$ (MeV) & 313 \\
                 & $m_s$ (MeV) &  555 \\
                 & $m_c$ (MeV) & 1752 \\
                 & $m_b$ (MeV) & 5100 \\[2ex]
Confinement      & $a_c$ (MeV)         & 430 \\
                 & $\mu_c$ (fm$^{-1})$ & 0.70 \\
                 & $\Delta$ (MeV)      & 181.10 \\[2ex]
OGE              & $\alpha_0$              & 2.118 \\
                 & $\Lambda_0~$(fm$^{-1}$) & 0.113 \\
                 & $\mu_0~$(MeV)           & 36.976 \\
                 & $\hat{r}_0~$(MeV~fm)    & 28.17 \\
\end{tabular}
\end{ruledtabular}
\end{table}

\begin{table*}[!t]
\caption{\label{MesonMass} Theoretical and experimental (if available) masses of $1S$ and $2S$ states of $Q\bar{Q}$ and $Q\bar{q}\,(q=u, d, s;\, Q=c,\,b)$ mesons, unit in MeV.}
\begin{ruledtabular}
\begin{tabular}{lccclccc}
Meson & $nL$ & $M_{\text{The.}}$ & $M_{\text{Exp.}}$  & Meson & $nL$ & $M_{\text{The.}}$  & $M_{\text{Exp.}}$ \\
\hline
$\eta_c$ & $1S$ &  $2989$ & $2981$       & $\eta_b$ & $1S$ &  $9454$ & $9300$ \\
                & $2S$ & $3627$ & $-$    &                  & $2S$ & $9985$ & $-$ \\[2ex]
$J/\psi$ & $1S$ &  $3097$ & $3097$       & $\Upsilon$ & $1S$ &  $9505$ & $9460$ \\
$\psi$     & $2S$ & $3685$ & $-$    &                              & $2S$ & $10013$ & $10023$ \\[2ex]
$D$ & $1S$ &  $1897$ & $1870$       & $B$ & $1S$ &  $5278$ & $5280$ \\
        & $2S$ & $2648$ & $-$    &         & $2S$ & $5984$ & $-$ \\[2ex]
$D^*$ & $1S$ &  $2017$ & $2007$  & $B^*$ & $1S$ &  $5319$ & $5325$ \\
            & $2S$ & $2704$ & $-$         &            & $2S$ & $6005$ & $-$ \\[2ex]
$D_s$ & $1S$ &  $1989$ & $1968$  & $B_s$ & $1S$ &  $5355$ & $5367$ \\
            & $2S$ & $2705$ & $-$    &                & $2S$ & $6017$ & $-$ \\[2ex]
$D^*_s$ & $1S$ &  $2115$ & $2112$  & $B^*_s$ & $1S$ &  $5400$ & $5415$ \\
            & $2S$ & $2769$ & $-$          &                 & $2S$ & $6042$ & $-$
\end{tabular}
\end{ruledtabular}
\end{table*}

All discussed model parameters are summarized in Table~\ref{tab:model}. They have been fixed attending the phenomenology of conventional heavy hadrons, studying their spectra~\cite{Segovia:2015dia, Segovia:2016xqb, Yang:2019lsg, Ortega:2020uvc}, their electromagnetic, weak and strong decays and reactions~\cite{Segovia:2011zza, Segovia:2011dg, Segovia:2014mca, Martin-Gonzalez:2022qwd}, as well as the coupling between naive meson states and meson-meson thresholds~\cite{Ortega:2016pgg, Ortega:2018cnm, Ortega:2021xst, Ortega:2021fem}; see also reviews~\cite{Segovia:2013wma, Ortega:2020tng} to infer its application to meson-meson, baryon-meson and baryon-baryon states. Furthermore, for later concern, Table~\ref{MesonMass} lists theoretical and experimental (if available) masses of ground and first excited states of $Q\bar{Q}$ and $Q\bar{q}$ $(q=u,\,d,\,s;\, Q=c,\,b)$ mesons.

\begin{figure}[ht]
\epsfxsize=3.4in \epsfbox{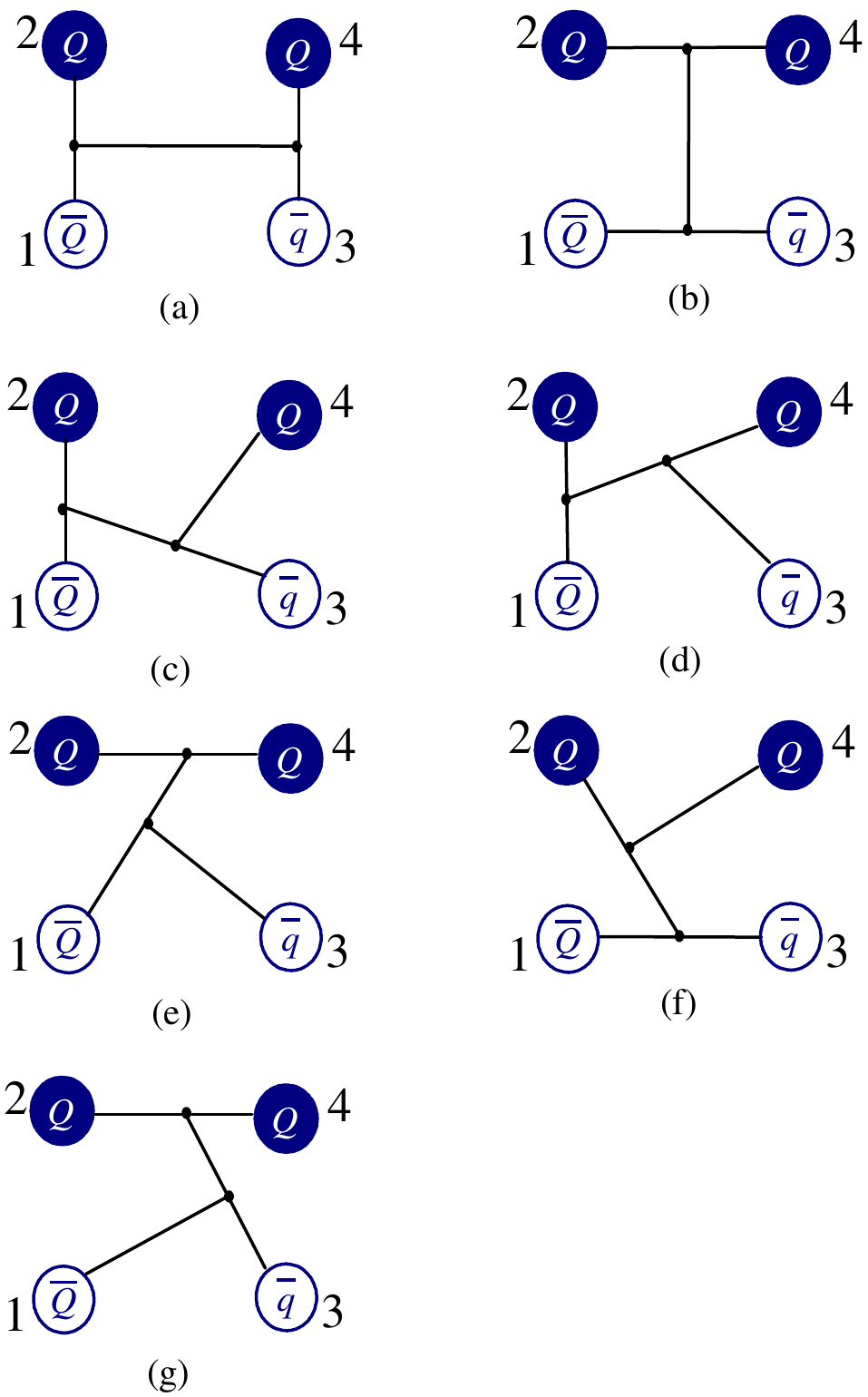}
\caption{\label{QQqq} Possible configurations of triply heavy tetraquarks $\bar{Q}Q\bar{q}Q$ $(q=u,\,d,\,s;\,Q=c,\,b)$. In particular, panel $(a)$ is meson-meson structure, panel $(b)$ is diquark-antidiquark arrangement, and the five K-type configurations are from panel $(c)$ to $(g)$.}
\end{figure}

\subsection{The wave function}

The $\bar{Q}Q\bar{q}Q$ $(q=u,\,d,\,s;\,Q=c,\,b)$ tetraquark configurations, which are under investigation in this work, are presented in Figure~\ref{QQqq}. Particularly, Fig.~\ref{QQqq}(a) is the meson-meson structure, Fig.~\ref{QQqq}(b) is the diquark-antidiquark arrangement, and panels (c) to (g) of Fig.~\ref{QQqq} are the five K-type configurations, which are rarely referred in other studies on multiquark systems.

The total wave function of a tetraquark system at the quark level is an internal product of color, spin, flavor and space wave functions. Concerning the color degree-of-freedom, the colorless wave function of a $4$-quark system in meson-meson configuration can be obtained by either two coupled color-singlet clusters, $1\otimes 1$:
\begin{align}
\label{Color1}
\chi^c_1 &= \frac{1}{3}(\bar{r}r+\bar{g}g+\bar{b}b)\times (\bar{r}r+\bar{g}g+\bar{b}b) \,,
\end{align}
or two coupled color-octet clusters, $8\otimes 8$:
\begin{align}
\label{Color2}
\chi^c_2 &= \frac{\sqrt{2}}{12}(3\bar{b}r\bar{r}b+3\bar{g}r\bar{r}g+3\bar{b}g\bar{g}b+3\bar{g}b\bar{b}g+3\bar{r}g\bar{g}r
\nonumber\\
&+3\bar{r}b\bar{b}r+2\bar{r}r\bar{r}r+2\bar{g}g\bar{g}g+2\bar{b}b\bar{b}b-\bar{r}r\bar{g}g
\nonumber\\
&-\bar{g}g\bar{r}r-\bar{b}b\bar{g}g-\bar{b}b\bar{r}r-\bar{g}g\bar{b}b-\bar{r}r\bar{b}b) \,.
\end{align}
For future reference, note herein that the first color state is the so-called color-singlet channel and the second one is the named hidden-color case.

The color wave functions associated to the diquark-antidiquark structure are the coupled color triplet-antitriplet clusters, $3\otimes \bar{3}$:
\begin{align}
\label{Color3}
\chi^c_3 &= \frac{\sqrt{3}}{6}(\bar{r}r\bar{g}g-\bar{g}r\bar{r}g+\bar{g}g\bar{r}r-\bar{r}g\bar{g}r+\bar{r}r\bar{b}b
\nonumber\\
&-\bar{b}r\bar{r}b+\bar{b}b\bar{r}r-\bar{r}b\bar{b}r+\bar{g}g\bar{b}b-\bar{b}g\bar{g}b
\nonumber\\
&+\bar{b}b\bar{g}g-\bar{g}b\bar{b}g) \,,
\end{align}
and the coupled color sextet-antisextet clusters, $6\otimes \bar{6}$:
\begin{align}
\label{Color4}
\chi^c_4 &= \frac{\sqrt{6}}{12}(2\bar{r}r\bar{r}r+2\bar{g}g\bar{g}g+2\bar{b}b\bar{b}b+\bar{r}r\bar{g}g+\bar{g}r\bar{r}g
\nonumber\\
&+\bar{g}g\bar{r}r+\bar{r}g\bar{g}r+\bar{r}r\bar{b}b+\bar{b}r\bar{r}b+\bar{b}b\bar{r}r
\nonumber\\
&+\bar{r}b\bar{b}r+\bar{g}g\bar{b}b+\bar{b}g\bar{g}b+\bar{b}b\bar{g}g+\bar{g}b\bar{b}g) \,.
\end{align}

Meanwhile, the possible color-singlet wave functions obtained from the five K-type configurations are
\begin{align}
\label{Color5}
\chi^c_5 &= \frac{1}{6\sqrt{2}}(\bar{r}r\bar{r}r+\bar{g}g\bar{g}g-2\bar{b}b\bar{b}b)+
\nonumber\\
&\frac{1}{2\sqrt{2}}(\bar{r}b\bar{b}r+\bar{r}g\bar{g}r+\bar{g}b\bar{b}g+\bar{g}r\bar{r}g+\bar{b}g\bar{g}b+\bar{b}r\bar{r}b)-
\nonumber\\
&\frac{1}{3\sqrt{2}}(\bar{g}g\bar{r}r+\bar{r}r\bar{g}g)+\frac{1}{6\sqrt{2}}(\bar{b}b\bar{r}r+\bar{b}b\bar{g}g+\bar{r}r\bar{b}b+\bar{g}g\bar{b}b) \,,
\end{align}
\begin{align}
\label{Color6}
\chi^c_6 &= \chi^c_1 \,,
\end{align}
\begin{align}
\label{Color7}
\chi^c_7 &= \chi^c_1 \,,
\end{align}
\begin{align}
\label{Color8}
\chi^c_8 &= \chi^c_2 \,,
\end{align}
\begin{align}
\label{Color9}
\chi^c_9 &= \frac{1}{2\sqrt{6}}(\bar{r}b\bar{b}r+\bar{r}r\bar{b}b+\bar{g}b\bar{b}g+\bar{g}g\bar{b}b+\bar{r}g\bar{g}r+\bar{r}r\bar{g}g+
\nonumber\\
&\bar{b}b\bar{g}g+\bar{b}g\bar{g}b+\bar{g}g\bar{r}r+\bar{g}r\bar{r}g+\bar{b}b\bar{r}r+\bar{b}r\bar{r}b)+
\nonumber\\
&\frac{1}{\sqrt{6}}(\bar{r}r\bar{r}r+\bar{g}g\bar{g}g+\bar{b}b\bar{b}b) \,,
\end{align}
\begin{align}
\label{Color10}
\chi^c_{10} &= \frac{1}{2\sqrt{3}}(\bar{r}b\bar{b}r-\bar{r}r\bar{b}b+\bar{g}b\bar{b}g-\bar{g}g\bar{b}b+\bar{r}g\bar{g}r-\bar{r}r\bar{g}g-
\nonumber\\
&\bar{b}b\bar{g}g+\bar{b}g\bar{g}b-\bar{g}g\bar{r}r+\bar{g}r\bar{r}g-\bar{b}b\bar{r}r+\bar{b}r\bar{r}b) \,,
\end{align}
\begin{align}
\label{Color11}
\chi^c_{11} &= \chi^c_9 \,,
\end{align}
\begin{align}
\label{Color12}
\chi^c_{12} &= -\chi^c_{10} \,.
\end{align}
\begin{align}
\label{Color13}
\chi^c_{13} &= \chi^c_9 \,,
\end{align}
\begin{align}
\label{Color14}
\chi^c_{14} &= \chi^c_{10} \,.
\end{align}

Let us now consider the spin wave functions, $\chi^{\sigma_i}_{S, M_S}$, for $S$-wave ground states with spin ($S$) ranging from $0$ to $2$ ($M_S$ can be set to be equal to $S$ without loss of generality). For $(S,M_S)=(0,0)$, one has
\begin{align}
\label{SWF0}
\chi_{0,0}^{\sigma_{u_1}}(4) &= \chi^\sigma_{00}\chi^\sigma_{00} \,, \\
\chi_{0,0}^{\sigma_{u_2}}(4) &= \frac{1}{\sqrt{3}}(\chi^\sigma_{11}\chi^\sigma_{1,-1}-\chi^\sigma_{10}\chi^\sigma_{10}+\chi^\sigma_{1,-1}\chi^\sigma_{11}) \,, \\
\chi_{0,0}^{\sigma_{u_3}}(4) &= \frac{1}{\sqrt{2}}\big((\sqrt{\frac{2}{3}}\chi^\sigma_{11}\chi^\sigma_{\frac{1}{2}, -\frac{1}{2}}-\sqrt{\frac{1}{3}}\chi^\sigma_{10}\chi^\sigma_{\frac{1}{2}, \frac{1}{2}})\chi^\sigma_{\frac{1}{2}, -\frac{1}{2}} \nonumber \\ 
&-(\sqrt{\frac{1}{3}}\chi^\sigma_{10}\chi^\sigma_{\frac{1}{2}, -\frac{1}{2}}-\sqrt{\frac{2}{3}}\chi^\sigma_{1, -1}\chi^\sigma_{\frac{1}{2}, \frac{1}{2}})\chi^\sigma_{\frac{1}{2}, \frac{1}{2}}\big) \,, \\
\chi_{0,0}^{\sigma_{u_4}}(4) &= \frac{1}{\sqrt{2}}(\chi^\sigma_{00}\chi^\sigma_{\frac{1}{2}, \frac{1}{2}}\chi^\sigma_{\frac{1}{2}, -\frac{1}{2}}-\chi^\sigma_{00}\chi^\sigma_{\frac{1}{2}, -\frac{1}{2}}\chi^\sigma_{\frac{1}{2}, \frac{1}{2}}) \,,
\end{align}
for $(S,M_S)=(1,1)$, the spin wave functions are
\begin{align}
\label{SWF1}
\chi_{1,1}^{\sigma_{w_1}}(4) &= \chi^\sigma_{00}\chi^\sigma_{11} \,, \\ 
\chi_{1,1}^{\sigma_{w_2}}(4) &= \chi^\sigma_{11}\chi^\sigma_{00} \,, \\
\chi_{1,1}^{\sigma_{w_3}}(4) &= \frac{1}{\sqrt{2}} (\chi^\sigma_{11} \chi^\sigma_{10}-\chi^\sigma_{10} \chi^\sigma_{11}) \,, \\
\chi_{1,1}^{\sigma_{w_4}}(4) &= \sqrt{\frac{3}{4}}\chi^\sigma_{11}\chi^\sigma_{\frac{1}{2}, \frac{1}{2}}\chi^\sigma_{\frac{1}{2}, -\frac{1}{2}}-\sqrt{\frac{1}{12}}\chi^\sigma_{11}\chi^\sigma_{\frac{1}{2}, -\frac{1}{2}}\chi^\sigma_{\frac{1}{2}, \frac{1}{2}} \nonumber \\ 
&-\sqrt{\frac{1}{6}}\chi^\sigma_{10}\chi^\sigma_{\frac{1}{2}, \frac{1}{2}}\chi^\sigma_{\frac{1}{2}, \frac{1}{2}} \,, \\
\chi_{1,1}^{\sigma_{w_5}}(4) &= (\sqrt{\frac{2}{3}}\chi^\sigma_{11}\chi^\sigma_{\frac{1}{2}, -\frac{1}{2}}-\sqrt{\frac{1}{3}}\chi^\sigma_{10}\chi^\sigma_{\frac{1}{2}, \frac{1}{2}})\chi^\sigma_{\frac{1}{2}, \frac{1}{2}} \,, \\
\chi_{1,1}^{\sigma_{w_6}}(4) &= \chi^\sigma_{00}\chi^\sigma_{\frac{1}{2}, \frac{1}{2}}\chi^\sigma_{\frac{1}{2}, \frac{1}{2}} \,,
\end{align}
and, for $(S,M_S)=(2,2)$, one has
\begin{align}
\label{SWF2}
\chi_{2,2}^{\sigma_{1}}(4) &= \chi^\sigma_{11}\chi^\sigma_{11} \,.
\end{align}
Besides, the superscripts $u_1,\ldots,u_4$ and $w_1,\ldots,w_6$ determine the spin wave function for each configuration of tetraquark system, their values are listed in Table~\ref{SpinIndex}. Furthermore, the expressions above are obtained by considering the coupling between two sub-clusters whose spin wave functions are given by trivial SU(2) algebra, whose basis reads as
\begin{align}
\label{Spin}
&\chi^\sigma_{00} = \frac{1}{\sqrt{2}}(\chi^\sigma_{\frac{1}{2}, \frac{1}{2}} \chi^\sigma_{\frac{1}{2}, -\frac{1}{2}}-\chi^\sigma_{\frac{1}{2}, -\frac{1}{2}} \chi^\sigma_{\frac{1}{2}, \frac{1}{2}}) \,, \\
&\chi^\sigma_{11} = \chi^\sigma_{\frac{1}{2}, \frac{1}{2}} \chi^\sigma_{\frac{1}{2}, \frac{1}{2}} \,, \\
&\chi^\sigma_{1,-1} = \chi^\sigma_{\frac{1}{2}, -\frac{1}{2}} \chi^\sigma_{\frac{1}{2}, -\frac{1}{2}} \,, \\
&\chi^\sigma_{10} = \frac{1}{\sqrt{2}}(\chi^\sigma_{\frac{1}{2}, \frac{1}{2}} \chi^\sigma_{\frac{1}{2}, -\frac{1}{2}}+\chi^\sigma_{\frac{1}{2}, -\frac{1}{2}} \chi^\sigma_{\frac{1}{2}, \frac{1}{2}}) \,.
\end{align}

\begin{table}[!t]
\caption{\label{SpinIndex} The values of the superscripts $u_1,\ldots,u_4$ and $w_1,\ldots,w_6$ that determine the spin wave function for each configuration of the triply heavy tetraquark systems.}
\begin{ruledtabular}
\begin{tabular}{lccccccc}
& Di-meson & Diquark-antidiquark & $K_1$ & $K_2$ & $K_3$ & $K_4$ & $K_5$ \\
\hline
$u_1$ & 1 & 3 & & & & & \\
$u_2$ & 2 & 4 & & & & & \\
$u_3$ &   &   & 5 & 7 &  9 & 11 & 13 \\
$u_4$ &   &   & 6 & 8 & 10 & 12 & 14 \\[2ex]
$w_1$ & 1 & 4 & & & & & \\
$w_2$ & 2 & 5 & & & & & \\
$w_3$ & 3 & 6 & & & & & \\
$w_4$ &   &   & 7 & 10 & 13 & 16 & 19 \\
$w_5$ &   &   & 8 & 11 & 14 & 17 & 20\\
$w_6$ &   &   & 9 & 12 & 15 & 18 & 21
\end{tabular}
\end{ruledtabular}
\end{table}

The flavor wave functions, $\chi^{f}_{I}$, of triply charm and bottom tetraquarks are simply denoted as
\begin{align}
\chi^{f}_{I} = \bar{Q}Q\bar{q}Q,\,(q=u,\,d,\,s;\, Q=c,\,b) \,,
\end{align}
where the isospin, $I$, and its third component, $M_I$, are both equal to zero for the $\bar{Q}Q\bar{s}Q$ system, while they are $\frac{1}{2}$ for the $\bar{Q}Q\bar{d}Q$ one.

We use the Rayleigh-Ritz variational principle to solve the Schr\"odinger-like four-body equation. Within a complex-scaled theoretical framework, the spatial wave function reads as below:
\begin{equation}
\label{eq:WFexp}
\psi_{LM_L}= \left[ \left[ \phi_{n_1l_1}(\vec{\rho}e^{i\theta}\,) \phi_{n_2l_2}(\vec{\lambda}e^{i\theta}\,)\right]_{l} \phi_{n_3l_3}(\vec{R}e^{i\theta}\,) \right]_{L M_L} \,,
\end{equation}
where the internal Jacobi coordinates are defined as
\begin{align}
\vec{\rho} &= \vec{x}_1-\vec{x}_{2(4)} \,, \\
\vec{\lambda} &= \vec{x}_3 - \vec{x}_{4(2)} \,, \\
\vec{R} &= \frac{m_1 \vec{x}_1 + m_{2(4)} \vec{x}_{2(4)}}{m_1+m_{2(4)}}- \frac{m_3 \vec{x}_3 + m_{4(2)} \vec{x}_{4(2)}}{m_3+m_{4(2)}} \,,
\end{align}
for the meson-meson configuration of Fig.~\ref{QQqq}(a), and as
\begin{align}
\vec{\rho} &= \vec{x}_1-\vec{x}_3 \,, \\
\vec{\lambda} &= \vec{x}_2 - \vec{x}_4 \,, \\
\vec{R} &= \frac{m_1 \vec{x}_1 + m_3 \vec{x}_3}{m_1+m_3}- \frac{m_2 \vec{x}_2 + m_4 \vec{x}_4}{m_2+m_4} \,,
\end{align}
for the diquark-antidiquark arrangement of Fig.~\ref{QQqq}(b). The remaining five K-type configurations shown in panels $(c)$ to $(g)$ of Fig.~\ref{QQqq} are ($i, j, k, l$ take values according to the panels $(c)$ to $(g)$ of Fig.~\ref{QQqq}):
\begin{align}
\vec{\rho} &= \vec{x}_i-\vec{x}_j \,, \\
\vec{\lambda} &= \vec{x}_k- \frac{m_i \vec{x}_i + m_j \vec{x}_j}{m_i+m_j} \,, \\
\vec{R} &= \vec{x}_l- \frac{m_i \vec{x}_i + m_j \vec{x}_j+m_k \vec{x}_k}{m_i+m_j+m_k} \,.
\end{align}
Obviously, the center-of-mass kinetic term, $T_\text{CM}$, can be completely eliminated for a non-relativistic system defined in any of the above sets of relative motion coordinates.

The basis expansion of the space wave function of Eq.~(\ref{eq:WFexp}) is performed by employing the Gaussian expansion method (GEM)~\cite{Hiyama:2003cu}, which has been proven to be quite efficient on solving bound- and scattering-state problems of a few- and many-body systems. This consists basically on using Gaussian basis functions, whose sizes are taken in geometric progression, for each relative motion of the tetraquark system (see Ref.~\cite{Yang:2015bmv} for details). Hence, the form of the orbital wave function $\phi_{nlm}$ in Eq.~\eqref{eq:WFexp} is simply
\begin{align}
&
\phi_{nlm}(\vec{r}e^{i\theta}\,) = \sqrt{1/4\pi} \, N_{nl} \, (re^{i\theta})^{l} \, e^{-\nu_{n} (re^{i\theta})^2} \,.
\end{align}

Finally, the complete wave function, which fulfills the Pauli principle, is written as
\begin{align}
\label{TPs}
 \Psi_{J M_J, I} &= \sum_{i, j} c_{ij} \Psi_{J M_J, I, i, j} \nonumber \\
 &=\sum_{i, j} c_{ij} {\cal A} \left[ \left[ \psi_{L M_L} \chi^{\sigma_i}_{S M_S}(4) \right]_{J M_J} \chi^{f}_I \chi^{c}_j \right] \,,
\end{align}
where $\cal{A}$ is the antisymmetry operator of $\bar{Q}Q\bar{q}Q$ tetraquark systems, which take into account the fact of involving two identical heavy quarks. Its definition, according to Fig.~\ref{QQqq}, is 
\begin{equation}
\label{Antisym}
{\cal{A}} = 1-(24) \,.
\end{equation}
This is necessary in our theoretical framework, since the complete wave function of the 4-quark system is constructed from two sub-clusters: meson-meson, diquark-antidiquark and K-type configurations.

Since our intention is to shed some light about the nature of triply heavy tetraquarks, their internal structure is studied by quantitative analyses of the inter-quark distance,
\begin{equation}
\label{quarkdistance}
{r_{q\bar{q}}} = Re(\sqrt{\langle \Psi_{J M_J, I} \vert  (r_{q\bar{q}} e^{i\theta})^2 \vert \Psi_{J M_J, I} \rangle}) \,,
\end{equation}
and magnetic moment,
\begin{equation}
\label{MM}
{\mu_m} = Re(\langle \Psi_{J M_J, I} \vert  \sum_{i=1}^{4} \frac{\hat Q_i}{2m_i} \hat \sigma^z_i \vert \Psi_{J M_J, I} \rangle) \,.
\end{equation}
Moreover, a qualitative survey of the dominant configuration or component,
\begin{equation}
\label{DCC}
C_p = Re(\sum_{i,j} \langle c^l_{ij} \Psi_{J M_J, I, i, j} \vert c^r_{ij} \Psi_{J M_J, I, i, j} \rangle) \,,
\end{equation}
is also performed. Meanwhile, in Eq.~\eqref{MM}, $\hat Q_i$ is the electric charge operator of the $i$-th quark and $\sigma^z_i$ is the $z$-component of Pauli matrix. Moreover, $c^l_{ij}$ and $c^r_{ij}$ denote, respectively, the left and right generalized eigenvectors of the complete anti-symmetric complex wave function. Finally, note that these observables are complex numbers in the complex scaling method, and thus their real-parts are analyzed for that resonance whose width is small where the approach is expected to work well. 

One inevitable issue of our theoretical framework has to be pointed. The different multiquark channels are not orthogonal to each other. Accordingly, we perform two kinds of computations related with the triply heavy tetraquarks components. Consider an $n$-dimension matrix representation of the overlap in Eq.~\eqref{DCC}:
\begin{equation*}
\begin{pmatrix}
c_{11} & c_{12} & \cdots & c_{1n} \\
c_{21} & c_{22} & \cdots & c_{2n} \\
\vdots & \vdots & \ddots & \vdots \\
c_{n1} &c_{n2} & \cdots & c_{nn}
\end{pmatrix} \,.
\end{equation*}
One can firstly consider the diagonal elements only in obtaining the relative weights of each channel or, secondly, the off-diagonal elements are employed by adding all such elements in the $i$-th row to the $i$-th diagonal element. A comparison on these two ways of proceed help us significantly in identifying the dominant component of the exotic state, and to express some conclusions about the off-diagonal contributions.
 

\section{Results}
\label{sec:results}

We proceed to discuss our results on $\bar{Q}Q\bar{q}Q$ $(q=u,\,d,\,s;\,Q=c,\,b)$ tetraquarks, considering all possible meson-meson, diquark-antidiquark and K-type arrangements. Since we are interested on $S$-wave states, the total angular momentum, $J$, coincides with the total spin, $S$, and can take values of $0$, $1$ and $2$, respectively. The systems' isospin is $\frac{1}{2}$ for $\bar{Q}Q\bar{d}Q$ and $0$ for $\bar{Q}Q\bar{s}Q$. With the purpose of solving a manageable problem of eigen-values and -vectors, the artificial parameter of rotated angle is ranged form $0^\circ$ to $6^\circ$; being the rotated angle $\theta=0^{\circ}$ the real-range case and the rest used to distinguish between bound, resonance and scattering states in the complex energy plane.

Tables~\ref{GresultCC1} to~\ref{GresultCCT} list calculated results of $\bar{Q}Q\bar{q}Q$ tetraquark states. In particular, real-range computations on the lowest-lying masses are presented in Tables~\ref{GresultCC1}, ~\ref{GresultCC2}, ~\ref{GresultCC3}, ~\ref{GresultCC4}, ~\ref{GresultCC5}, ~\ref{GresultCC6}, ~\ref{GresultCC7}, ~\ref{GresultCC8}, ~\ref{GresultCC9}, ~\ref{GresultCC10}, ~\ref{GresultCC11} and ~\ref{GresultCC12}, respectively. Therein, the considered meson-meson, diquark-antidiquark and K-type configurations are listed in the first column; if possible, the experimental value of the non-interacting di-meson threshold is labeled in parentheses. In the second column, each channel is assigned with an index which indicates a particular combination of spin ($\chi_J^{\sigma_i}$) and color ($\chi_j^c$) wave functions, that are shown explicitly in the third column. The theoretical mass calculated in each channel is shown in the fourth column, and the coupled result for each kind of configuration is presented in the last one. Note also that the last row of the table indicates the lowest-lying mass obtained in a complete coupled-channel calculation of real-range. 

\begin{figure}[!t]
\includegraphics[clip, trim={3.0cm 1.9cm 3.0cm 1.0cm}, width=0.45\textwidth]{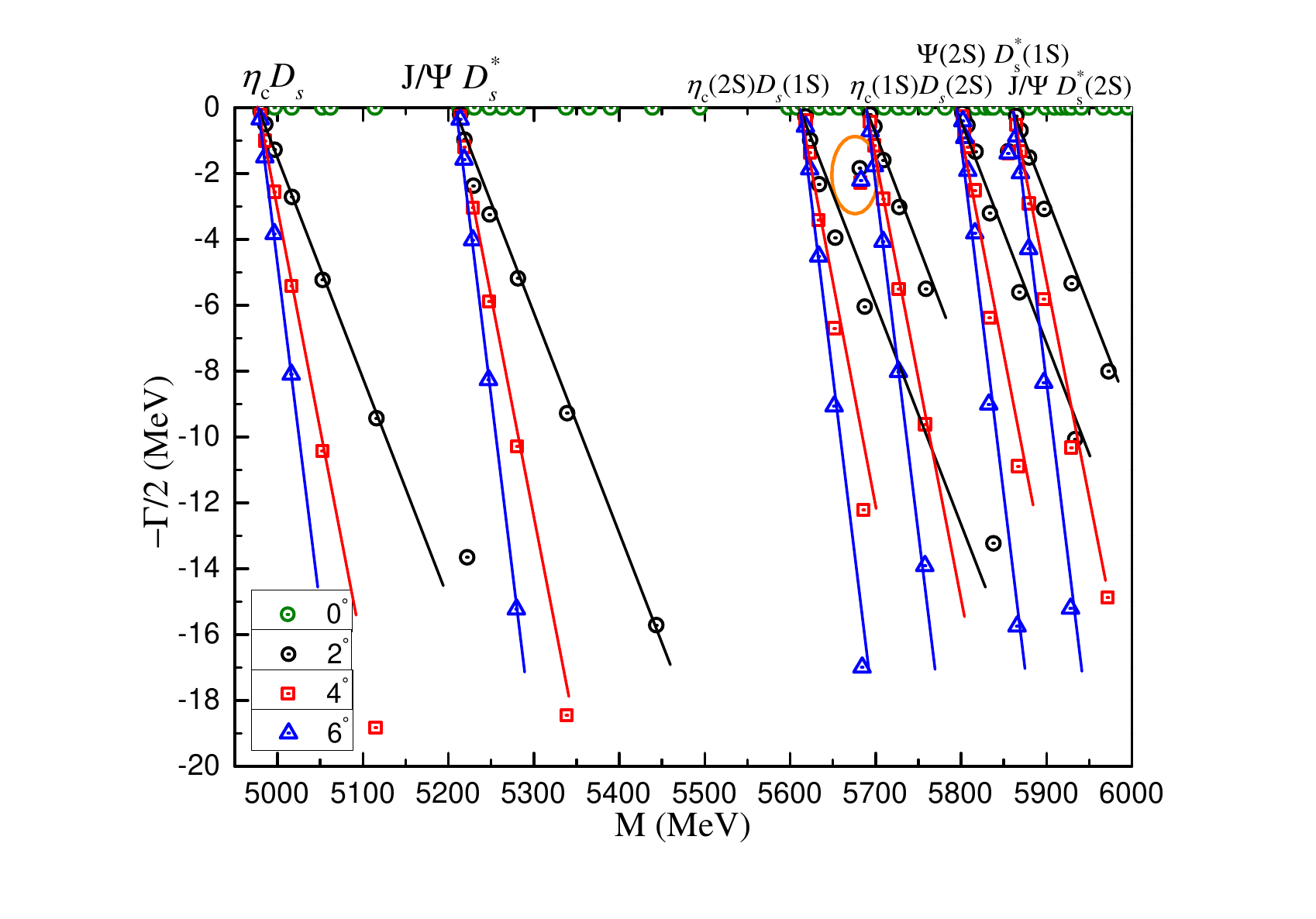} \\[1ex]
\includegraphics[clip, trim={3.0cm 1.9cm 3.0cm 1.0cm}, width=0.45\textwidth]{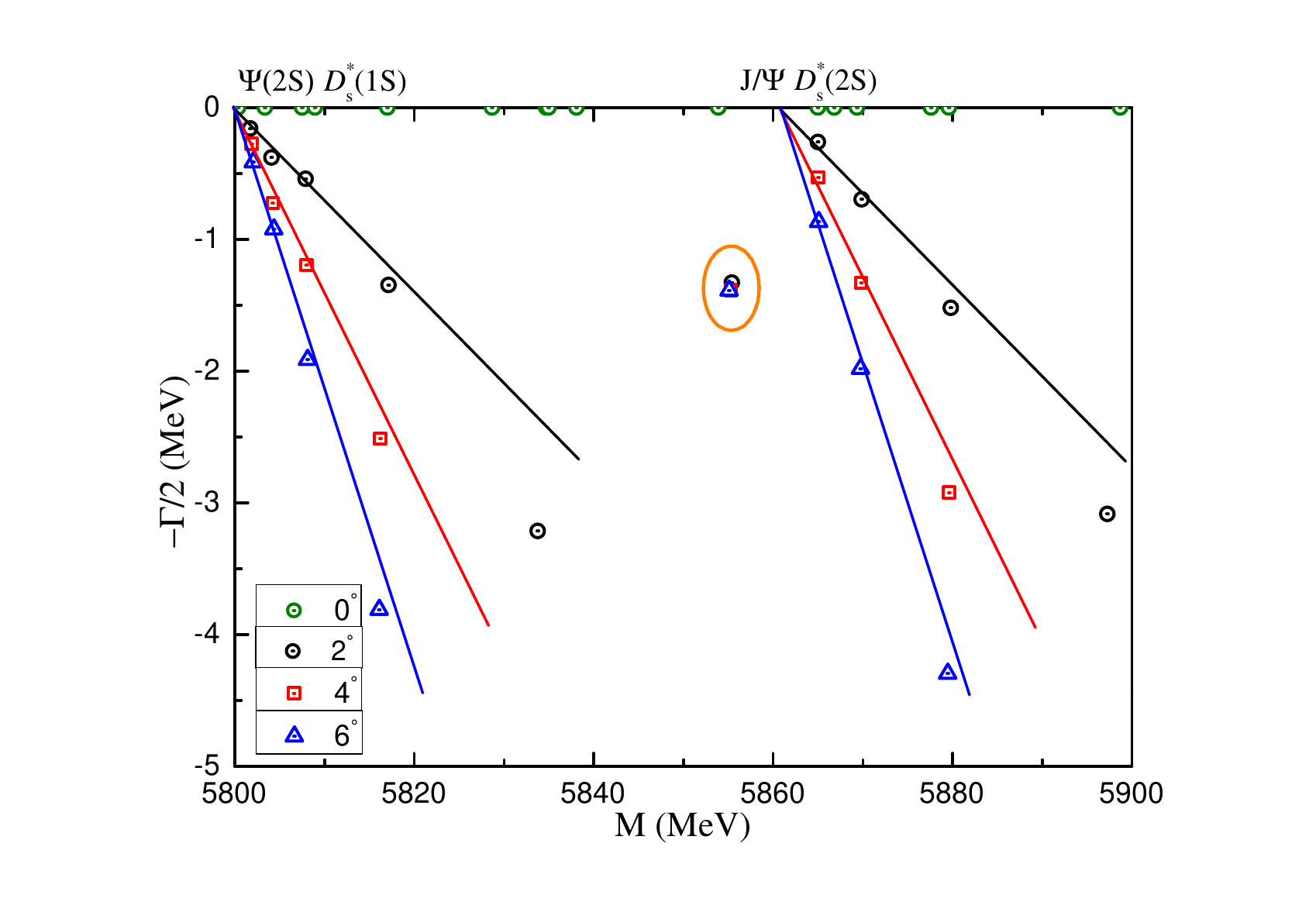}
\caption{\label{PP4} The complete coupled-channels calculation of $\bar{c}c\bar{s}c$ tetraquark system with $I(J^P)=0(0^+)$ quantum numbers. Particularly, the bottom panel is enlarged parts of dense energy region from $5.8\,\text{GeV}$ to $5.9\,\text{GeV}$.}
\end{figure}

In a further step, the CSM method is applied to the complete coupled-channel calculation. Figures~\ref{PP1} to~\ref{PP12} show the distribution of complex eigen-energies and, therein, the obtained resonance states are indicated inside circles. Several insights about the nature of found resonances are given by calculating their size, magnetic moment and dominant component, results are listed among Tables~\ref{GresultR1}, ~\ref{GresultR2}, ~\ref{GresultR3}, ~\ref{GresultR4}, ~\ref{GresultR5}, ~\ref{GresultR6}, ~\ref{GresultR7}, ~\ref{GresultR8}, ~\ref{GresultR9}, ~\ref{GresultR10}, ~\ref{GresultR11} and ~\ref{GresultR12}. Particularly, four kinds of quark distances, which are $r_{Q\bar{Q}}$, $r_{\bar{Q}\bar{q}}$, $r_{Q\bar{q}}$ and $r_{QQ}$ are computed. Besides, two sets of results on resonance's dominant components are analyzed according to the discussion in Sec.~\ref{sec:model}, \emph{viz.} only diagonal elements are considered in the calculation of set I, and the addition of off-diagonal ones are employed in set II.

Finally, a summary of our most salient results is presented in Table~\ref{GresultCCT}.


\subsection{The $\mathbf{\bar{c}c\bar{d}c}$ tetraquarks}

\begin{table}[!t]
\caption{\label{GresultCC1} Lowest-lying $\bar{c}c\bar{d}c$ tetraquark states with $I(J^P)=\frac{1}{2}(0^+)$ calculated within the real range formulation of the constituent quark model.
The allowed meson-meson, diquark-antidiquark and K-type configurations are listed in the first column; when possible, the experimental value of the non-interacting meson-meson threshold is labeled in parentheses. Each channel is assigned an index in the 2nd column, it reflects a particular combination of spin ($\chi_J^{\sigma_i}$) and color ($\chi_j^c$) wave functions that are shown explicitly in the 3rd column. The theoretical mass obtained in each channel is shown in the 4th column and the coupled result for each kind of configuration is presented in the 5th column.
When a complete coupled-channels calculation is performed, last row of the table indicates the calculated lowest-lying mass (unit: MeV).}
\begin{ruledtabular}
\begin{tabular}{lcccc}
~~Channel   & Index & $\chi_J^{\sigma_i}$;~$\chi_j^c$ & $M$ & Mixed~~ \\
        &   &$[i; ~j]$ &  \\[2ex]
$(\eta_c D)^1 (4851)$          & 1  & [1;~1]  & $4886$ & \\
$(J/\psi D^*)^1 (5104)$  & 2  & [2;~1]   & $5114$ & $4886$  \\[2ex]
$(\eta_c D)^8$          & 3  & [1;~2]  & $5367$ & \\
$(J/\psi D^*)^8$       & 4  & [2;~2]   & $5338$ & $5305$  \\[2ex]
$(cc)(\bar{c}\bar{d})$      & 5     & [3;~4]  & $5354$ & \\
$(cc)^*(\bar{c}\bar{d})^*$  & 6  & [4;~3]   & $5369$ & $5324$ \\[2ex]
$K_1$  & 7  & [5;~5]   & $5311$ & \\
            & 8  & [5;~6]   & $5363$ & \\
            & 9  & [6;~5]   & $5353$ & \\
            & 10  & [6;~6]   & $5253$ & $5157$ \\[2ex]
$K_2$  & 11  & [7;~7]   & $5305$ & \\
             & 12  & [7;~8]   & $5321$ & \\
             & 13  & [8;~7]   & $5187$ & \\
             & 14  & [8;~8]   & $5362$ & $5176$ \\[2ex]
$K_3$  & 15  & [9;~10]   & $5361$ & \\
             & 16  & [10;~9]   & $5349$ & $5313$ \\[2ex]
$K_4$  & 17  & [11;~12]   & $5370$ & \\
             & 18  & [12;~12]   & $5839$ & \\
             & 19  & [11;~11]   & $5758$ & \\
             & 20  & [12;~11]   & $5351$ & $5275$ \\[2ex]
$K_5$  & 21  & [13;~14]   & $5372$ & \\
             & 22  & [14;~13]   & $5336$ & $5307$ \\[2ex]
\multicolumn{4}{c}{Complete coupled-channels:} & $4886$
\end{tabular}
\end{ruledtabular}
\end{table}

\begin{figure}[!t]
\includegraphics[clip, trim={3.0cm 1.9cm 3.0cm 1.0cm}, width=0.45\textwidth]{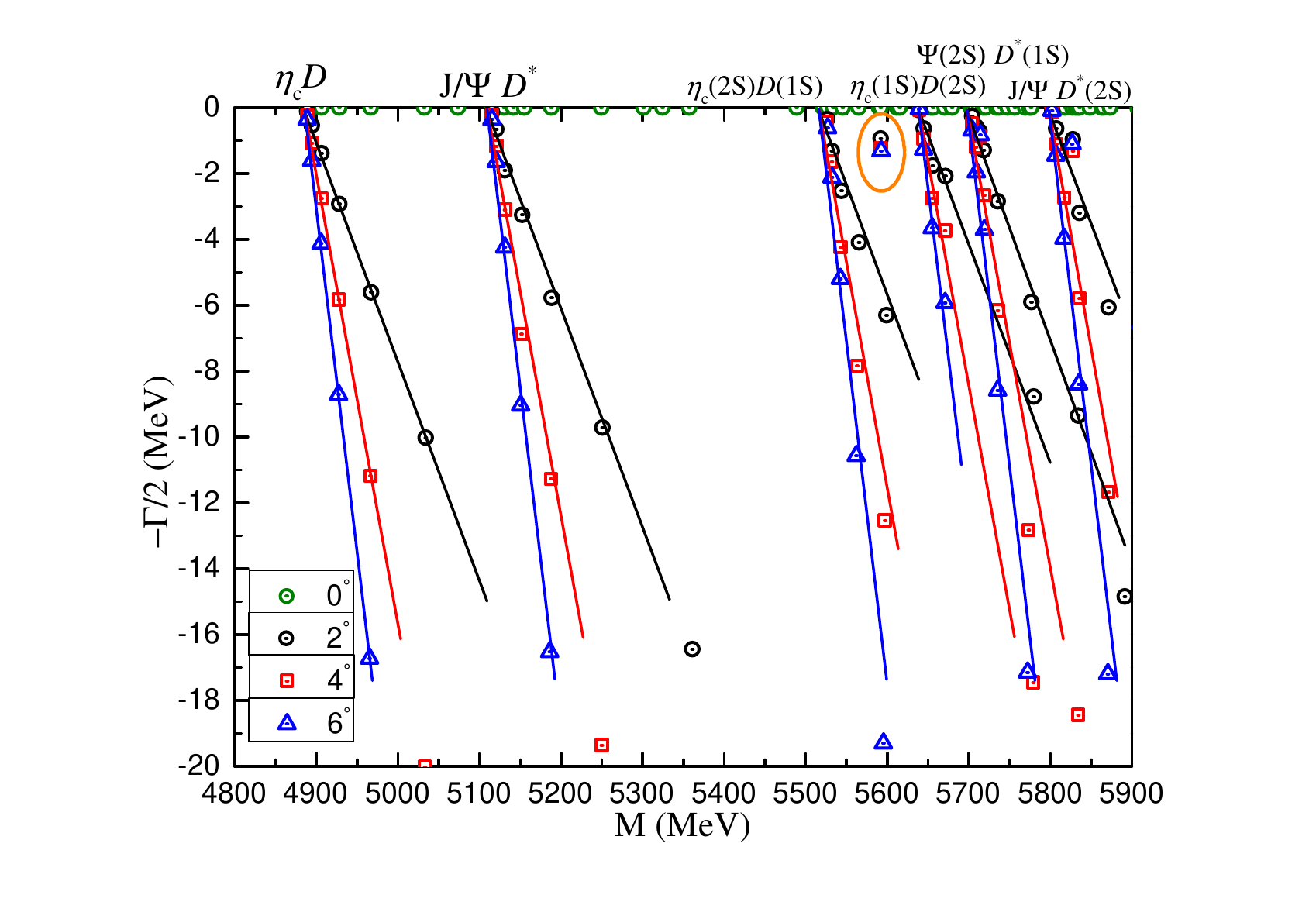} \\[1ex]
\includegraphics[clip, trim={3.0cm 1.9cm 3.0cm 1.0cm}, width=0.45\textwidth]{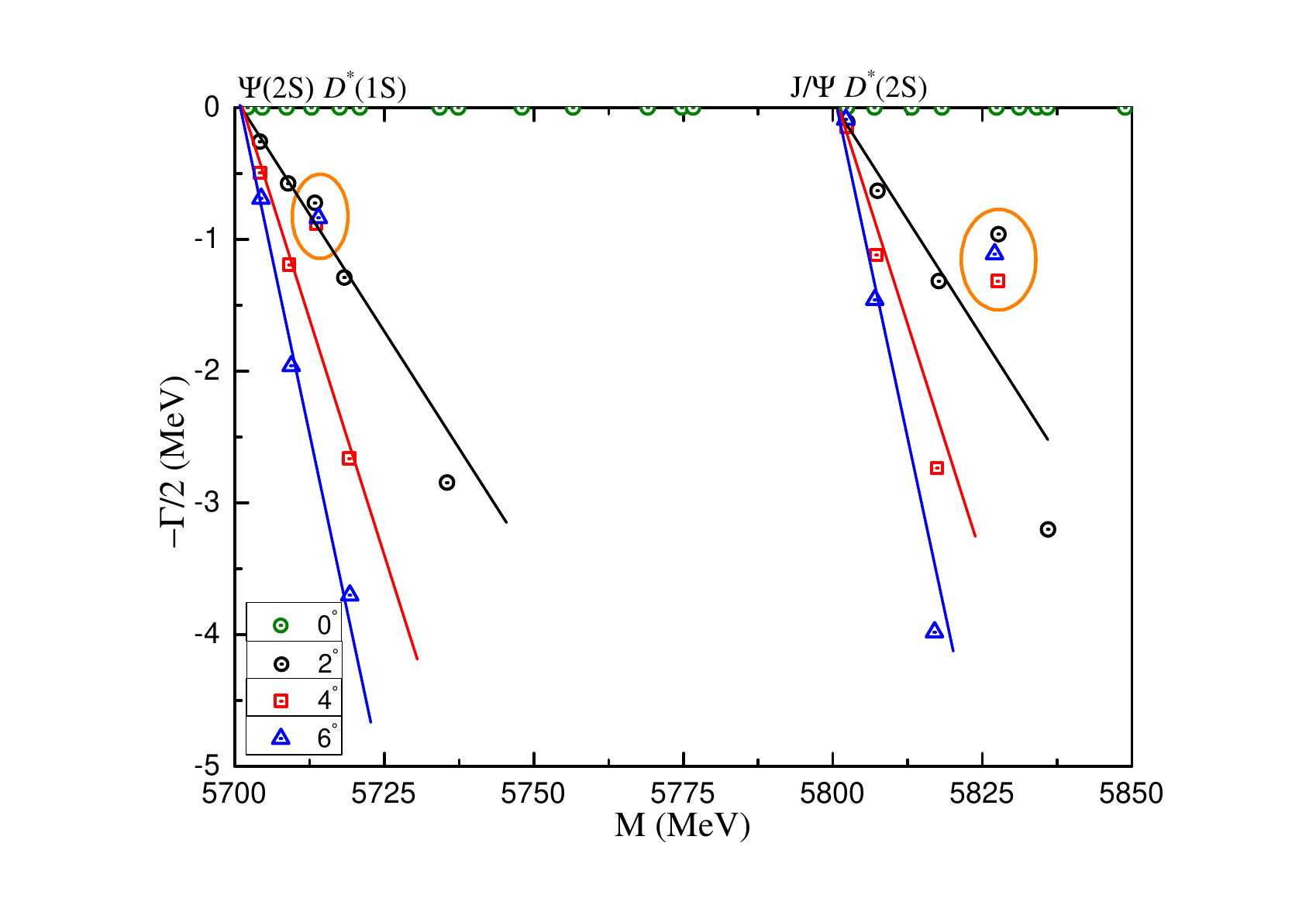}
\caption{\label{PP1} The complete coupled-channels calculation of $\bar{c}c\bar{d}c$ tetraquark system with $I(J^P)=\frac{1}{2}(0^+)$ quantum numbers. Particularly, the bottom panel is enlarged part of dense energy region from $5.70\,\text{GeV}$ to $5.85\,\text{GeV}$.}
\end{figure}

\begin{table}[!t]
\caption{\label{GresultR1} Compositeness of exotic resonances obtained in a complete coupled-channel calculation in the $\frac{1}{2}(0^+)$ state of $\bar{c}c\bar{d}c$ tetraquark. Particularly, the first column is resonance poles labeled by $M-i\Gamma$, unit in MeV; the second one is the magnetic moment of resonance, unit in $\mu_N$; the distance between any two quarks or quark-antiquark, unit in fm; and the component of resonant state ($S$: dimeson structure in color-singlet channel; $H$: dimeson structure in hidden-color channel; $Di$: diquark-antiquark configuration; $K$: K-type configuration). Herein, two sets of results on a resonance component are listed. Particularly, set I is results on components, that only diagonal elements are employed, and results, that both diagonal and off-diagonal elements are considered, are listed in set II.}
\begin{ruledtabular}
\begin{tabular}{rccc}
Resonance       & \multicolumn{3}{c}{Structure} \\[2ex]
$5592-i2.5$   & \multicolumn{3}{c}{$\mu=0$} \\
  & \multicolumn{3}{c}{$r_{c \bar{c}}:0.96$;\,\,\,\,\,$r_{\bar{c}\bar{d}}:1.42$;\,\,\,\,\,$r_{c\bar{d}}:1.28$;\,\,\,\,\,$r_{cc}:1.30$} \\
$Set$ I: & \multicolumn{3}{c}{$S$: 10.6\%;\, $H$: 1.0\%;\, $Di$: 29.7\%;\, $K$: 58.7\%}\\
$Set$ II: & \multicolumn{3}{c}{$S$: 10.0\%;\, $H$: 2.1\%;\, $Di$: 18.9\%;\, $K$: 68.9\%}\\[2ex]
$5714-i1.8$   & \multicolumn{3}{c}{$\mu=0$} \\
  & \multicolumn{3}{c}{$r_{c \bar{c}}:1.31$;\,\,\,\,\,$r_{\bar{c}\bar{d}}:1.87$;\,\,\,\,\,$r_{c\bar{d}}:1.57$;\,\,\,\,\,$r_{cc}:1.79$} \\
$Set$ I: & \multicolumn{3}{c}{$S$: 25.6\%;\, $H$: 14.2\%;\, $Di$: 6.3\%;\, $K$: 53.9\%}\\
$Set$ II: & \multicolumn{3}{c}{$S$: 29.6\%;\, $H$: 3.5\%;\, $Di$: 4.2\%;\, $K$: 62.7\%}\\[2ex]
$5828-i2.6$   & \multicolumn{3}{c}{$\mu=0$} \\
  & \multicolumn{3}{c}{$r_{c \bar{c}}:1.31$;\,\,\,\,\,$r_{\bar{c}\bar{d}}:1.82$;\,\,\,\,\,$r_{c\bar{d}}:1.57$;\,\,\,\,\,$r_{cc}:1.79$} \\
$Set$ I: & \multicolumn{3}{c}{$S$: 11.7\%;\, $H$: 6.5\%;\, $Di$: 15.9\%;\, $K$: 65.9\%}\\
$Set$ II: & \multicolumn{3}{c}{$S$: 4.3\%;\, $H$: 3.5\%;\, $Di$: 6.9\%;\, $K$: 85.3\%}\\
\end{tabular}
\end{ruledtabular}
\end{table}

{\bf The $\bm{I(J^P)=\frac{1}{2}(0^+)}$ sector:} Two meson-meson configurations, $\eta_c D$ and $J/\psi D^*$, in both singlet- and hidden-color channels, two diquark-antidiquark structures, $(cc)(\bar{c}\bar{d})$ and $(cc)^*(\bar{c}\bar{d})^*$, along with five K-type arrangements are individually computed in Table~\ref{GresultCC1}. The lowest mass corresponds to the color-singlet state $\eta_c D$, $4886$ MeV, the other di-meson channel $J/\psi D^*$ with the same color structure is located at $5114$ MeV. The scattering nature of these two meson-meson structures can be concluded when compared with threshold's mass. Furthermore, bound states are also not obtained when the single channel calculation is performed in each exotic structure. Particularly, masses of states in both hidden-color and diquark-antidiquark channels are around $5.35$ GeV whereas the K-type configurations present masses in the interval $5.19-5.84$ GeV.

Results on partially coupled-channel study of these eight configurations are then listed in the last column of Table~\ref{GresultCC1}. Firstly, the lowest mass remains at $4.89$ GeV, hence it is a scattering state of $\eta_c D$. Secondly, the coupling among different channels within the same exotic structure produce remarkable mass shifts for hidden-color, diquark-antidiquark and K-type channels. However, coupled-channel effects are still weak to produce bound states; the lowest coupled-masses of hidden-color and diquark-antidiquark channels are both $\sim5.31$ GeV and K-type configurations are within an energy region of $5.15-5.31$ GeV. In a further step, when 22 channels listed in Table~\ref{GresultCC1} are all included in a coupled-channel calculation of real-range, the scattering nature of $\eta_c D$ remains.

Three narrow resonances are obtained when a complete coupled-channel calculation is performed using the complex-range method. Therein, the six scattering states of $\eta_c D$, $J/\psi D^*$ and their radial excitations are clearly presented in Fig.~\ref{PP1} with an energy interval from $4.8$ to $5.9$ GeV. Besides, one stable pole, $5592-i2.5$ MeV, is found in the top panel of Fig.~\ref{PP1}. Meanwhile, two more resonances are obtained in the bottom panel, which is an enlarged part of a dense energy region from $5.70$ to $5.85$ GeV, whose complex eigenenergies are $5714-i1.8$ MeV and $5828-i2.6$ MeV.

The resonance electromagnetic and geometric properties, along with their dominant components, are further investigated. Results are summarized in Table~\ref{GresultR1}. In particular, the magnetic moment of these three resonances is $0$. The lowest resonance at $5.59$ GeV is a compact $\bar{c}c\bar{d}c$ tetraquark structure whose size is within $0.96-1.42$ fm. However, the other two higher resonances present larger sizes, $1.31-1.87$ fm.

After a comparison of components between Set I and II, one can find that off-diagonal elements contributions are generally small. There is only $\sim10\%$ modification for diquark-antidiquark and K-type components of the first resonance, and this result also holds for hidden-color and K-type channels of the second one. The third resonance has $\sim20\%$ modification on the K-type component when off-diagonal elements are included. Accordingly, the dominant components of the three resonances are exotic color configurations of K-type channels. Besides, there are considerable diquark-antidiquark ($\sim19\%$) and di-meson color-singlet ($\sim26\%$) components for the resonances at $5.59$ GeV and $5.71$ GeV, respectively. 


\begin{table}[!t]
\caption{\label{GresultCC2} Lowest-lying $\bar{c}c\bar{d}c$ tetraquark states with $I(J^P)=\frac{1}{2}(1^+)$ calculated within the real range formulation of the chiral quark model. Results are similarly organized as those in Table~\ref{GresultCC1} (unit: MeV).}
\begin{ruledtabular}
\begin{tabular}{lcccc}
~~Channel   & Index & $\chi_J^{\sigma_i}$;~$\chi_j^c$ & $M$ & Mixed~~ \\
        &   &$[i; ~j]$ &  \\[2ex]
$(\eta_c D^*)^1 (4988)$   & 1  & [1;~1]  & $5006$ & \\
$(J/\psi D)^1 (4967)$  & 2  & [2;~1]   & $4994$ &  \\
$(J/\psi D^*)^1 (5104)$  & 3  & [3;~1]   & $5114$ & $4994$  \\[2ex]
$(\eta_c D^*)^8$       & 4  & [1;~2]   & $5353$ &  \\
$(J/\psi D)^8$      & 5  & [2;~2]  & $5358$ & \\
$(J/\psi D^*)^8$  & 6  & [3;~2]   & $5333$ & $5316$ \\[2ex]
$(cc)^*(\bar{c}\bar{d})^*$  & 7  & [6;~3]  & $5379$ & \\
$(cc)^*(\bar{c}\bar{d})$    & 8  & [5;~4]   & $5342$ &  \\
$(cc)(\bar{c}\bar{d})^*$  & 9  & [4;~3]   & $5350$ & $5323$  \\[2ex]
$K_1$      & 10  & [7;~5]   & $5359$ &  \\
      & 11 & [8;~5]  & $5293$ & \\
      & 12  & [9;~5]   & $5334$ & \\
      & 13   & [7;~6]  & $5336$ & \\
      & 14   & [8;~6]  & $5368$ & \\
      & 15   & [9;~6]  & $5276$ & $5194$ \\[2ex]
$K_2$      & 16  & [10;~7]   & $5301$ & \\
                 & 17  & [11;~7]   & $5280$ &  \\
                 & 18  & [12;~7]   & $5219$ & \\
                 & 19  & [10;~8]   & $5319$ & \\
                 & 20  & [11;~8]   & $5353$ & \\
                 & 21  & [12;~8]   & $5345$ & $5201$ \\[2ex]
$K_3$      & 22  & [13;~10]   & $5352$ & \\
                 & 23  & [14;~10]   & $5372$ & \\
                 & 24  & [15;~9]    & $5334$ & $5313$ \\[2ex]
$K_4$      & 25  & [16;~11]   & $5385$ & \\
                 & 26  & [17;~11]   & $5391$ & \\
                 & 27  & [18;~11]   & $5805$ & \\
                 & 28  & [16;~12]   & $5382$ & \\
                 & 29  & [17;~12]   & $5381$ & \\
                 & 30  & [18;~12]   & $5350$ & $5283$ \\[2ex]
$K_5$      & 31  & [19;~14]   & $5379$ & \\
                 & 32  & [20;~14]   & $5372$ & \\
                 & 33  & [21;~13]   & $5322$ & $5306$ \\[2ex]  
\multicolumn{4}{c}{Complete coupled-channels:} & $4994$
\end{tabular}
\end{ruledtabular}
\end{table}

\begin{figure}[!t]
\includegraphics[clip, trim={3.0cm 1.9cm 3.0cm 1.0cm}, width=0.45\textwidth]{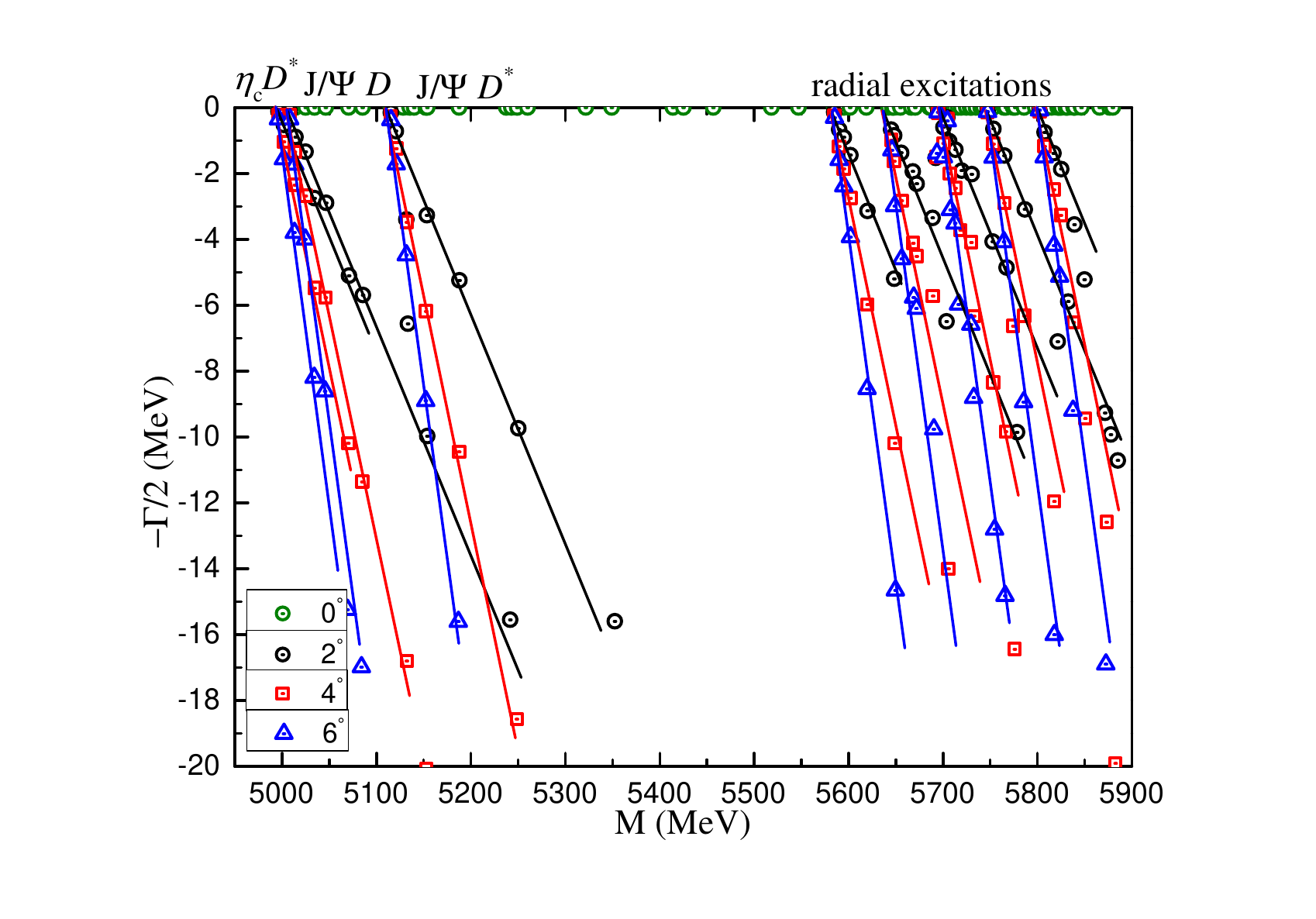} \\[1ex]
\includegraphics[clip, trim={3.0cm 1.9cm 3.0cm 1.0cm}, width=0.45\textwidth]{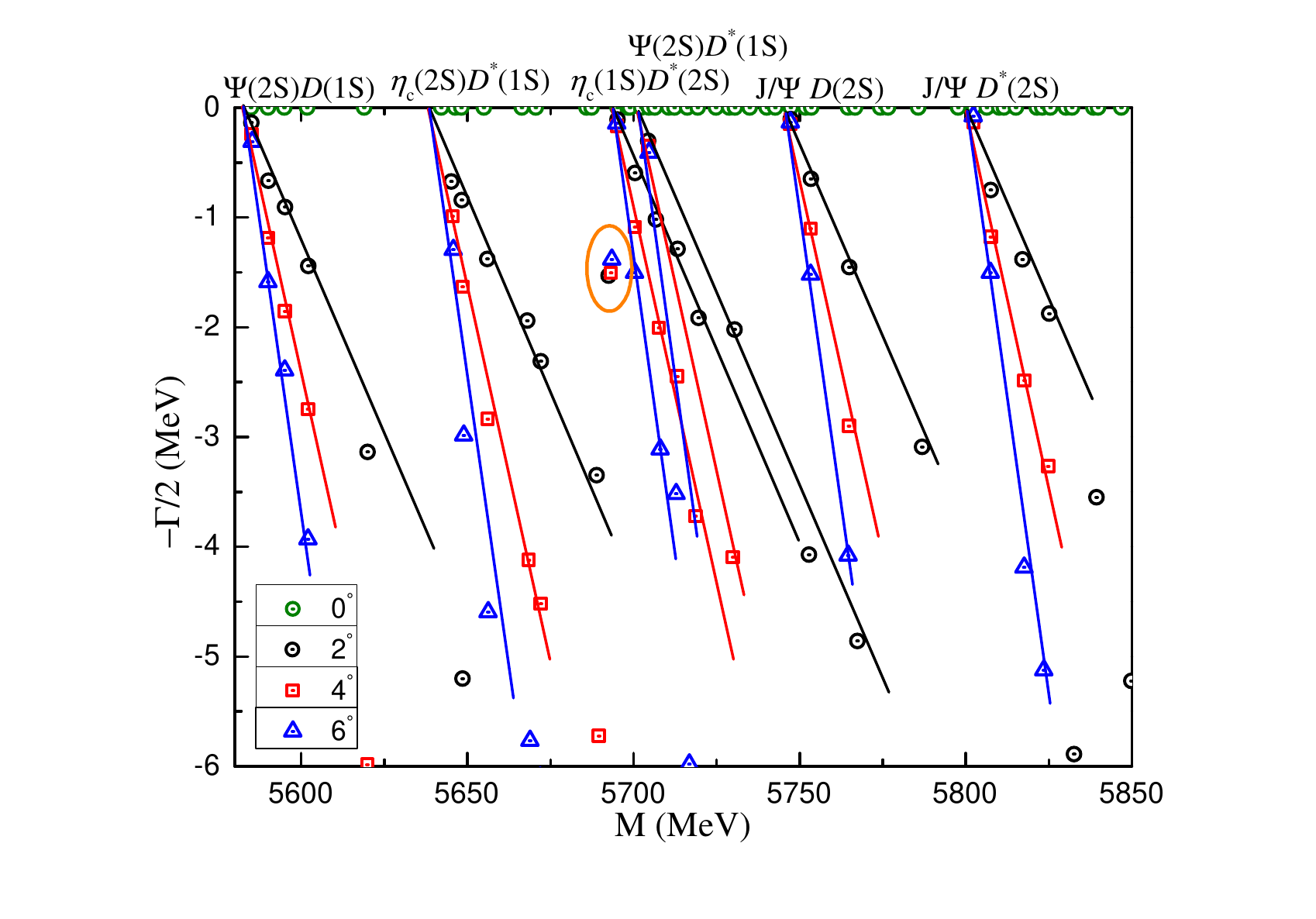}
\caption{\label{PP2} The complete coupled-channels calculation of $\bar{c}c\bar{d}c$ tetraquark system with $I(J^P)=\frac{1}{2}(1^+)$ quantum numbers. Particularly, the bottom panel is enlarged parts of dense energy region from $5.57\,\text{GeV}$ to $5.85\,\text{GeV}$.}
\end{figure}

\begin{table}[!t]
\caption{\label{GresultR2} Compositeness of exotic resonance obtained in a complete coupled-channel calculation in the $\frac{1}{2}(1^+)$ state of $\bar{c}c\bar{d}c$ tetraquark. Results are similarly organized as those in Table~\ref{GresultR1}.}
\begin{ruledtabular}
\begin{tabular}{rccc}
Resonance       & \multicolumn{3}{c}{Structure} \\[2ex]
$5693-i3.0$   & \multicolumn{3}{c}{$\mu=-1.06$} \\
  & \multicolumn{3}{c}{$r_{c \bar{c}}:1.25$;\,\,\,\,\,$r_{\bar{c}\bar{d}}:1.74$;\,\,\,\,\,$r_{c\bar{d}}:1.34$;\,\,\,\,\,$r_{cc}:1.61$} \\
$Set$ I: & \multicolumn{3}{c}{$S$: 3.2\%;\, $H$: 1.1\%;\, $Di$: 5.5\%;\, $K$: 90.2\%}\\
$Set$ II: & \multicolumn{3}{c}{$S$: 7.7\%;\, $H$: 4.0\%;\, $Di$: 9.3\%;\, $K$: 79.0\%}\\
\end{tabular}
\end{ruledtabular}
\end{table}

{\bf The $\bm{I(J^P)=\frac{1}{2}(1^+)}$ sector:} 33 channels should be considered in this case, and our results in real-range calculations are listed in Table~\ref{GresultCC2}. Among the three meson-meson configurations, the lowest channel is the color-singlet state of $J/\psi D$, whose theoretical mass, $4994$ MeV, is just equal to its non-interacting di-meson theoretical threshold value. Therefore, no bound state is found. This conclusion also holds for the other cases in both single- and coupled-channel computations. Particularly, the remaining two meson-meson channels of $\eta_c D^*$ and $J/\psi D^*$ in color-singlet states locate at $5.01$ GeV and $5.11$ GeV, respectively. Masses of the three hidden-color channels and three diquark-antidiquark ones converge at $5.35$ GeV. 24 K-type channels generally distribute within an energy region of $5.22-5.39$ GeV. The channels coupling is extremely weak in color-singlet configurations, in contrast to the fact that the mass considerably shifts in coupled-channel calculations of exotic color configurations. As one can see too, all these states are still unstable color resonances in a complete channels coupling calculation.

Figure~\ref{PP2} presents our results on a fully coupled-channel investigation by the CSM. Generally, scattering states of $\eta_c D^*$, $J/\psi D$, $J/\psi D^*$ and their radial excitations are plotted in the top panel, which are located in the energy interval $4.95-5.90$ GeV. Neither bound nor resonant states are obtained below $5.5$ GeV. However, a narrow resonance is found at $5693$ MeV, and its width is $3.0$ MeV. This stable pole can be clearly distinguished in the bottom panel of Fig.~\ref{PP2}, where six excited states of $\psi(2S)D(1S)$, $\eta_c(2S)D^*(1S)$, $\eta_c(1S)D^*(2S)$, $\psi(2S)D^*(1S)$, $J/\psi(1S)D(2S)$ and $J/\psi(1S)D^*(2S)$ are shown between $5.57$ GeV and $5.85$ GeV.

Compositeness of the found resonance is depicted in Table~\ref{GresultR2}. In particular, its magnetic moment is $-1.06\mu_N$, and its size is around $1.25-1.74$ fm, indicating an extended structure. Our theoretical uncertainty is less than $12\%$ after comparing the two sets of component results. Hence, the dominant components seems to be exotic, more than $79\%$ comes from K-type configurations.


\begin{table}[!t]
\caption{\label{GresultCC3} Lowest-lying $\bar{c}c\bar{d}c$ tetraquark states with $I(J^P)=\frac{1}{2}(2^+)$ calculated within the real range formulation of the chiral quark model. Results are similarly organized as those in Table~\ref{GresultCC1} (unit: MeV).}
\begin{ruledtabular}
\begin{tabular}{lcccc}
~~Channel   & Index & $\chi_J^{\sigma_i}$;~$\chi_j^c$ & $M$ & Mixed~~ \\
        &   &$[i; ~j]$ &  \\[2ex]
$(J/\psi D^*)^1 (5104)$  & 1  & [1;~1]   & $5114$ &  \\[2ex]
$(J/\psi D^*)^8$            & 2  & [1;~3]   & $5358$ &  \\[2ex]
$(cc)^*(\bar{c}\bar{d})^*$  & 3  & [1;~7]   & $5397$ & \\[2ex]
$K_1$  & 4  & [1;~8]   & $5348$ & \\
            & 5  & [1;~10]   & $5385$ & $5287$ \\[2ex]
$K_2$  & 6  & [1;~11]   & $5335$ & \\
             & 7  & [1;~12]   & $5357$ & $5323$ \\[2ex]
$K_3$  & 8  & [1;~14]   & $5393$ & \\[2ex]
$K_4$  & 9  & [1;~8]   & $5814$ & \\
             & 10  & [1;~10]   & $5398$ & $5370$ \\[2ex]
$K_5$  & 11  & [1;~11]   & $5404$ & \\[2ex]
\multicolumn{4}{c}{Complete coupled-channels:} & $5114$
\end{tabular}
\end{ruledtabular}
\end{table}

\begin{figure}[!t]
\includegraphics[clip, trim={3.0cm 1.9cm 3.0cm 1.0cm}, width=0.45\textwidth]{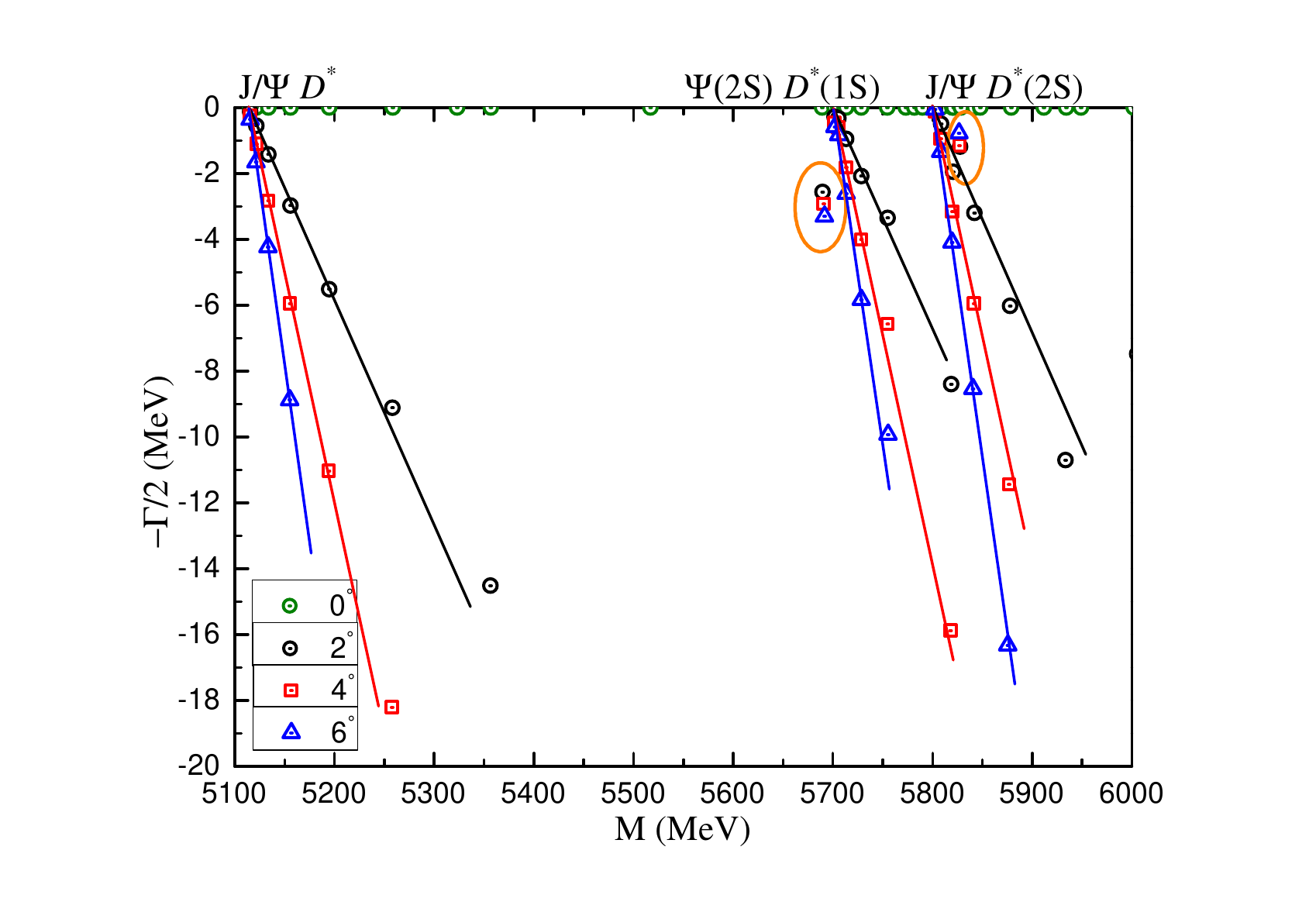}
\caption{\label{PP3} The complete coupled-channels calculation of $\bar{c}c\bar{d}c$ tetraquark system with $I(J^P)=\frac{1}{2}(2^+)$ quantum numbers.}
\end{figure}

\begin{table}[!t]
\caption{\label{GresultR3} Compositeness of exotic resonances obtained in a complete coupled-channel calculation in the $\frac{1}{2}(2^+)$ state of $\bar{c}c\bar{d}c$ tetraquark. Results are similarly organized as those in Table~\ref{GresultR1}.}
\begin{ruledtabular}
\begin{tabular}{rccc}
Resonance       & \multicolumn{3}{c}{Structure} \\[2ex]
$5691-i5.9$   & \multicolumn{3}{c}{$\mu=-1.64$} \\
  & \multicolumn{3}{c}{$r_{c \bar{c}}:0.90$;\,\,\,\,\,$r_{\bar{c}\bar{d}}:1.17$;\,\,\,\,\,$r_{c\bar{d}}:1.09$;\,\,\,\,\,$r_{cc}:0.97$} \\
$Set$ I: & \multicolumn{3}{c}{$S$: 10.4\%;\, $H$: 9.2\%;\, $Di$: 18.6\%;\, $K$: 61.8\%}\\
$Set$ II: & \multicolumn{3}{c}{$S$: 17.2\%;\, $H$: 3.8\%;\, $Di$: 20.7\%;\, $K$: 58.3\%}\\[2ex]
$5827-i2.3$   & \multicolumn{3}{c}{$\mu=-1.64$} \\
  & \multicolumn{3}{c}{$r_{c \bar{c}}:1.29$;\,\,\,\,\,$r_{\bar{c}\bar{d}}:1.77$;\,\,\,\,\,$r_{c\bar{d}}:1.55$;\,\,\,\,\,$r_{cc}:1.77$} \\
$Set$ I: & \multicolumn{3}{c}{$S$: 8.3\%;\, $H$: 11.1\%;\, $Di$: 19.0\%;\, $K$: 61.6\%}\\
$Set$ II: & \multicolumn{3}{c}{$S$: 5.3\%;\, $H$: 10.3\%;\, $Di$: 14.8\%;\, $K$: 69.6\%}\\
\end{tabular}
\end{ruledtabular}
\end{table}

{\bf The $\bm{I(J^P)=\frac{1}{2}(2^+)}$ state:} Table~\ref{GresultCC3} lists 11 channels under investigation for the highest spin state of S-wave $\bar{c}c\bar{d}c$ tetraquark. Firstly, bound states are not present in each kind of real-range calculation, which includes single channel but also partially and complete coupled-channels. Mass of $J/\psi D^*$ in color singlet configuration is $5114$ MeV, and its hidden color case is located at $5358$ MeV. Meanwhile, the other exotic color structures are located at $\sim5.40$ GeV, except for one $K_4$ channel with mass of $5.81$ GeV.

With a rotated angle varied from $0^\circ$ to $6^\circ$, Fig.~\ref{PP3} shows the distribution of the calculated complex energies. Particularly, in the $5.1-6.0$ GeV energy region, three scattering states, which are $J/\psi D^*$, $\psi(2S)D^*(1S)$ and $J/\psi D^*(2S)$, are well plotted. Apart from those unstable dots, two resonant poles are circled, their complex energies are $5691-i5.9$ and $5827-i2.3$, respectively.

The nature of these two resonances can be guessed from Table~\ref{GresultR3}. First of all, their magnetic moments are both $-1.64\mu_N$. However, the geometric structure of these states are different. The lowest resonance is a compact $\bar{c}c\bar{d}c$ tetraquark with a size around $1.1$ fm. The higher resonance at $5.8$ GeV is a loose structure within $1.29-1.77$ fm. Modifications in the composition due to off-diagonal elements are less than $8\%$ for both cases. Therefore, one may conclude that K-type configurations are dominant, with more than $60\%$, and the remaining three configurations, meson-meson in color-singlet and hidden-color plus diquark-antidiquark arrangement, occupies the rest with the same magnitude of proportion.


\begin{table}[!t]
\caption{\label{GresultCC4} Lowest-lying $\bar{c}c\bar{s}c$ tetraquark states with $I(J^P)=0(0^+)$ calculated within the real range formulation of the chiral quark model. Results are similarly organized as those in Table~\ref{GresultCC1} (unit: MeV).}
\begin{ruledtabular}
\begin{tabular}{lcccc}
~~Channel   & Index & $\chi_J^{\sigma_i}$;~$\chi_j^c$ & $M$ & Mixed~~ \\
        &   &$[i; ~j]$ &  \\[2ex]
$(\eta_c D_s)^1 (4949)$          & 1  & [1;~1]  & $4978$ & \\
$(J/\psi D^*_s)^1 (5209)$  & 2  & [2;~1]   & $5212$ & $4978$  \\[2ex]
$(\eta_c D_s)^8$          & 3  & [1;~2]  & $5462$ & \\
$(J/\psi D^*_s)^8$       & 4  & [2;~2]   & $5437$ & $5404$  \\[2ex]
$(cc)(\bar{c}\bar{s})$      & 5     & [3;~4]  & $5447$ & \\
$(cc)^*(\bar{c}\bar{s})^*$  & 6  & [4;~3]   & $5471$ & $5423$ \\[2ex]
$K_1$  & 7  & [5;~5]   & $5422$ & \\
            & 8  & [5;~6]   & $5472$ & \\
            & 9  & [6;~5]   & $5462$ & \\
            & 10  & [6;~6]   & $5363$ & $5269$ \\[2ex]
$K_2$  & 11  & [7;~7]   & $5422$ & \\
             & 12  & [7;~8]   & $5432$ & \\
             & 13  & [8;~7]   & $5311$ & \\
             & 14  & [8;~8]   & $5470$ & $5300$ \\[2ex]
$K_3$  & 15  & [9;~10]   & $5468$ & \\
             & 16  & [10;~9]   & $5458$ & $5425$ \\[2ex]
$K_4$  & 17  & [11;~12]   & $5472$ & \\
             & 18  & [12;~12]   & $5952$ & \\
             & 19  & [11;~11]   & $5881$ & \\
             & 20  & [12;~11]   & $5445$ & $5396$ \\[2ex]
$K_5$  & 21  & [13;~14]   & $5477$ & \\
             & 22  & [14;~13]   & $5446$ & $5419$ \\[2ex]
\multicolumn{4}{c}{Complete coupled-channels:} & $4978$
\end{tabular}
\end{ruledtabular}
\end{table}

\begin{table}[!t]
\caption{\label{GresultR4} Compositeness of the exotic resonances obtained in a complete coupled-channel calculation in the $0(0^+)$ state of $\bar{c}c\bar{s}c$ tetraquark. Results are similarly organized as those in Table~\ref{GresultR1}.}
\begin{ruledtabular}
\begin{tabular}{rccc}
Resonance       & \multicolumn{3}{c}{Structure} \\[2ex]
$5682-i4.6$   & \multicolumn{3}{c}{$\mu=0$} \\
  & \multicolumn{3}{c}{$r_{c \bar{c}}:1.53$;\,\,\,\,\,$r_{\bar{c}\bar{s}}:2.09$;\,\,\,\,\,$r_{c\bar{s}}:1.59$;\,\,\,\,\,$r_{cc}:2.10$} \\
$Set$ I: & \multicolumn{3}{c}{$S$: 19.7\%;\, $H$: 21.8\%;\, $Di$: 2.1\%;\, $K$: 56.4\%}\\
$Set$ II: & \multicolumn{3}{c}{$S$: 29.2\%;\, $H$: 12.2\%;\, $Di$: 3.5\%;\, $K$: 55.1\%}\\[2ex]
$5855-i2.8$   & \multicolumn{3}{c}{$\mu=0$} \\
  & \multicolumn{3}{c}{$r_{c \bar{c}}:1.07$;\,\,\,\,\,$r_{\bar{c}\bar{s}}:1.38$;\,\,\,\,\,$r_{c\bar{s}}:1.12$;\,\,\,\,\,$r_{cc}:1.36$} \\
$Set$ I: & \multicolumn{3}{c}{$S$: 17.1\%;\, $H$: 22.9\%;\, $Di$: 6.3\%;\, $K$: 53.7\%}\\
$Set$ II: & \multicolumn{3}{c}{$S$: 19.1\%;\, $H$: 10.2\%;\, $Di$: 10.0\%;\, $K$: 60.7\%}\\
\end{tabular}
\end{ruledtabular}
\end{table}

\subsection{The $\mathbf{\bar{c}c\bar{s}c}$ tetraquarks}

{\bf The $\bm{I(J^P)=0(0^+)}$ sector:} 22 channels are studied in Table~\ref{GresultCC4}. Masses of $\eta_c D_s$ and $J/\psi D^*_s$ in the singlet-color channel are $4978$ MeV and $5212$ MeV, respectively. The hidden-color and diquark-antidiquark channels are all at around $5.45$ GeV. Besides, 16 K-type configurations are locates at an energy region of $5.31-5.95$ GeV. Bound states are not obtained in studies of each single channel, and this fact remains in partially and complete coupled-channel calculations. Particularly, the lowest coupled-mass of 7 exotic configuration is $\sim5.4$ GeV, except for the $K_1$ and $K_2$ structures, whose mass is about $5.3$ GeV.

In a further step, when the CSM is employed in a fully coupled-channel investigation, two narrow resonances are obtained. In Fig.~\ref{PP4}, whose energy region is from $4.95$ GeV to $6.0$ GeV, one can find clear distributions of scattering states for $\eta_c D_s$, $J/\psi D^*_s$ and their radial excitations. Moreover, two resonances are obtained in the radial excitation region. In particular, a stable pole is obtained in the top panel, the complex energy reads $5682-i4.6$ MeV. As circled in the bottom panel of Fig.~\ref{PP4}, there is also another resonance state at $5855$ MeV, whose width is $2.8$ MeV.

Magnetic moment, quark distance and component distribution are listed in Table~\ref{GresultR4}. Firstly, the $\mu$ is 0 for these two resonances. Secondly, the first resonance is an extended object with a size of $1.6-2.1$ fm; however, the other resonance is compact with a size less than $1.4$ fm. The calculated two sets of components indicate a small contribution from off-diagonal elements, the dominant ones are color-singlet ($\sim20\%$) and K-type ($\sim55\%$) channels for both states.


\begin{table}[!t]
\caption{\label{GresultCC5} Lowest-lying $\bar{c}c\bar{s}c$ tetraquark states with $I(J^P)=0(1^+)$ calculated within the real range formulation of the chiral quark model. Results are similarly organized as those in Table~\ref{GresultCC1} (unit: MeV).}
\begin{ruledtabular}
\begin{tabular}{lcccc}
~~Channel   & Index & $\chi_J^{\sigma_i}$;~$\chi_j^c$ & $M$ & Mixed~~ \\
        &   &$[i; ~j]$ &  \\[2ex]
$(\eta_c D^*_s)^1 (5093)$   & 1  & [1;~1]  & $5104$ & \\
$(J/\psi D_s)^1 (5065)$  & 2  & [2;~1]   & $5086$ &  \\
$(J/\psi D^*_s)^1 (5209)$  & 3  & [3;~1]   & $5212$ & $5086$  \\[2ex]
$(\eta_c D^*_s)^8$       & 4  & [1;~2]   & $5450$ &  \\
$(J/\psi D_s)^8$      & 5  & [2;~2]  & $5453$ & \\
$(J/\psi D^*_s)^8$  & 6  & [3;~2]   & $5433$ & $5415$ \\[2ex]
$(cc)^*(\bar{c}\bar{d})^*$  & 7  & [6;~3]  & $5480$ & \\
$(cc)^*(\bar{c}\bar{d})$    & 8  & [5;~4]   & $5438$ &  \\
$(cc)(\bar{c}\bar{d})^*$  & 9  & [4;~3]   & $5454$ & $5424$  \\[2ex]
$K_1$      & 10  & [7;~5]   & $5464$ &  \\
      & 11 & [8;~5]  & $5410$ & \\
      & 12  & [9;~5]   & $5445$ & \\
      & 13   & [7;~6]  & $5449$ & \\
      & 14   & [8;~6]  & $5473$ & \\
      & 15   & [9;~6]  & $5383$ & $5303$ \\[2ex]
$K_2$      & 16  & [10;~7]   & $5420$ & \\
                 & 17  & [11;~7]   & $5400$ &  \\
                 & 18  & [12;~7]   & $5336$ & \\
                 & 19  & [10;~8]   & $5434$ & \\
                 & 20  & [11;~8]   & $5459$ & \\
                 & 21  & [12;~8]   & $5455$ & $5320$ \\[2ex]
$K_3$      & 22  & [13;~10]   & $5462$ & \\
                 & 23  & [14;~10]   & $5477$ & \\
                 & 24  & [15;~9]    & $5445$ & $5427$ \\[2ex]
$K_4$      & 25  & [16;~11]   & $5480$ & \\
                 & 26  & [17;~11]   & $5484$ & \\
                 & 27  & [18;~11]   & $5921$ & \\
                 & 28  & [16;~12]   & $5483$ & \\
                 & 29  & [17;~12]   & $5482$ & \\
                 & 30  & [18;~12]   & $5454$ & $5401$ \\[2ex]
$K_5$      & 31  & [19;~14]   & $5482$ & \\
                 & 32  & [20;~14]   & $5479$ & \\
                 & 33  & [21;~13]   & $5435$ & $5421$ \\[2ex]  
\multicolumn{4}{c}{Complete coupled-channels:} & $5086$
\end{tabular}
\end{ruledtabular}
\end{table}

\begin{figure}[!t]
\includegraphics[clip, trim={3.0cm 1.9cm 3.0cm 1.0cm}, width=0.45\textwidth]{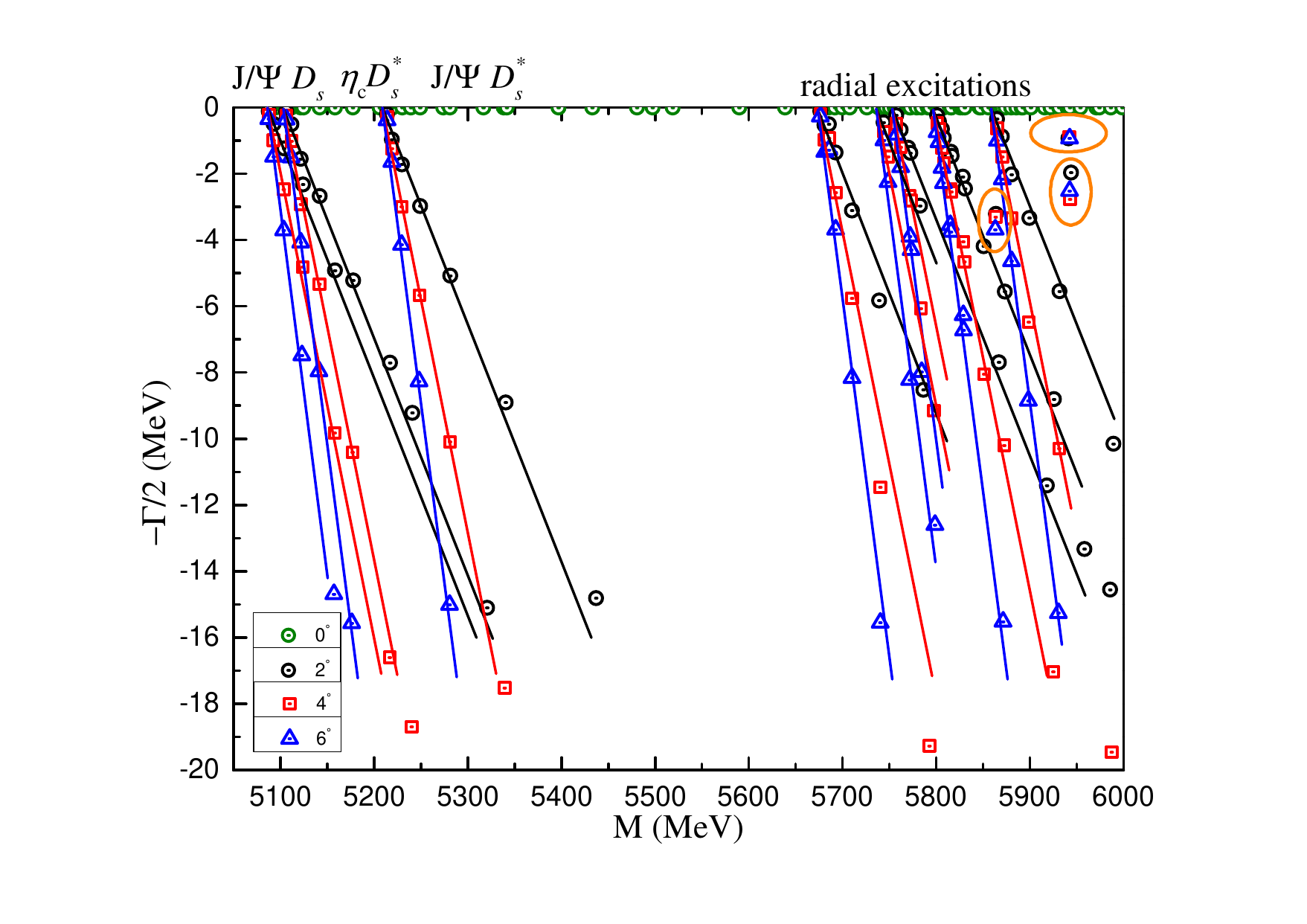} \\[1ex]
\includegraphics[clip, trim={3.0cm 1.9cm 3.0cm 1.0cm}, width=0.45\textwidth]{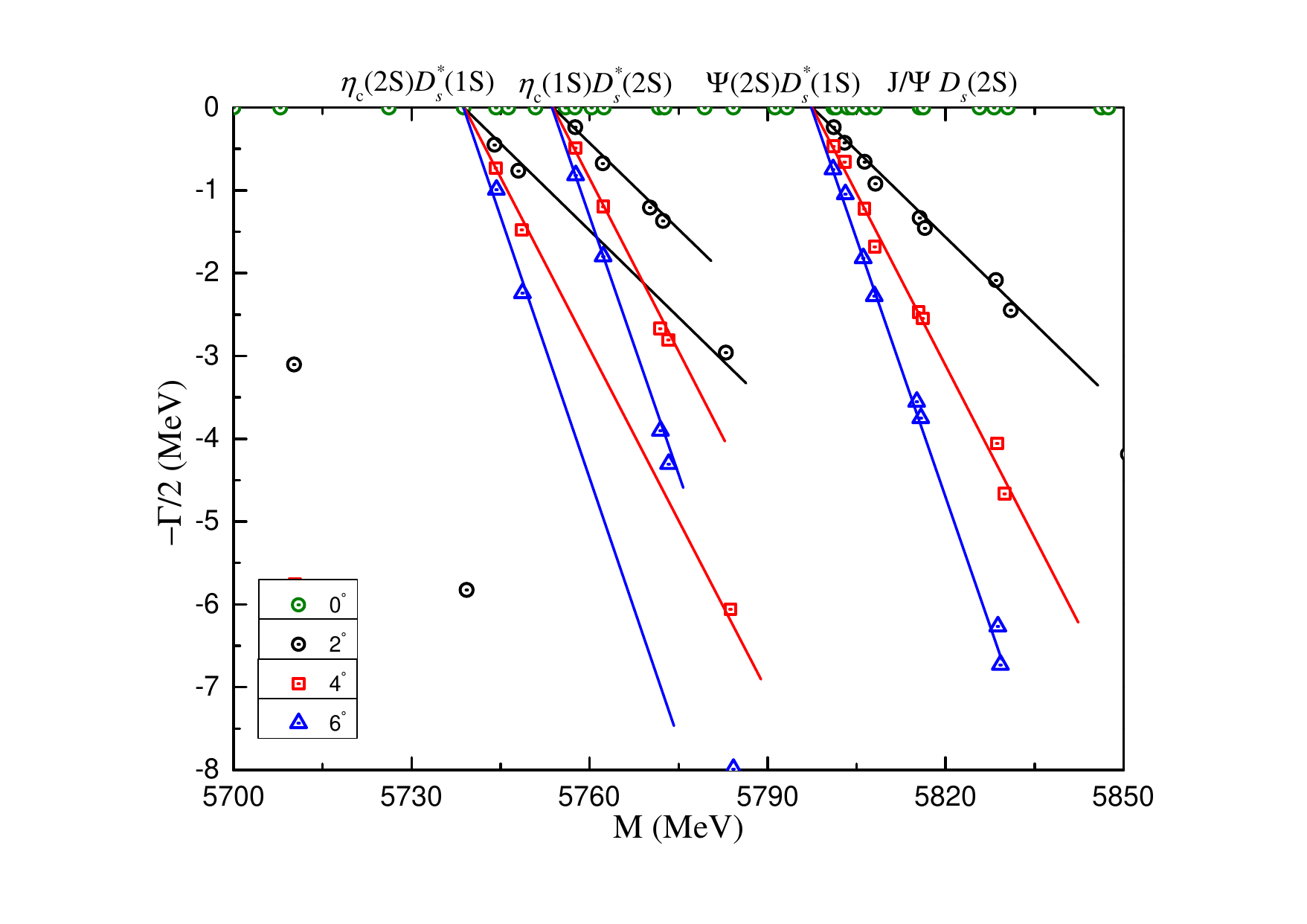}
\caption{\label{PP5} The complete coupled-channels calculation of $\bar{c}c\bar{s}c$ tetraquark system with $I(J^P)=0(1^+)$ quantum numbers. Particularly, the bottom panel is enlarged parts of dense energy region from $5.70\,\text{GeV}$ to $5.85\,\text{GeV}$.}
\end{figure}

\begin{table}[!t]
\caption{\label{GresultR5} Compositeness of exotic resonances obtained in a complete coupled-channel calculation in the $0(1^+)$ state of $\bar{c}c\bar{s}c$ tetraquark. Results are similarly organized as those in Table~\ref{GresultR1}.}
\begin{ruledtabular}
\begin{tabular}{rccc}
Resonance       & \multicolumn{3}{c}{Structure} \\[2ex]
$5863-i6.6$   & \multicolumn{3}{c}{$\mu=0.578$} \\
  & \multicolumn{3}{c}{$r_{c \bar{c}}:1.62$;\,\,\,\,\,$r_{\bar{c}\bar{s}}:2.26$;\,\,\,\,\,$r_{c\bar{s}}:1.68$;\,\,\,\,\,$r_{cc}:2.20$} \\
$Set$ I: & \multicolumn{3}{c}{$S$: 6.6\%;\, $H$: 14.6\%;\, $Di$: 1.1\%;\, $K$: 77.7\%}\\
$Set$ II: & \multicolumn{3}{c}{$S$: 3.0\%;\, $H$: 21.4\%;\, $Di$: 10.2\%;\, $K$: 65.4\%}\\[2ex]
$5941-i1.8$   & \multicolumn{3}{c}{$\mu=0.503$} \\
  & \multicolumn{3}{c}{$r_{c \bar{c}}:1.13$;\,\,\,\,\,$r_{\bar{c}\bar{s}}:1.52$;\,\,\,\,\,$r_{c\bar{s}}:1.33$;\,\,\,\,\,$r_{cc}:1.53$} \\
$Set$ I: & \multicolumn{3}{c}{$S$: 21.1\%;\, $H$: 7.8\%;\, $Di$: 1.3\%;\, $K$: 69.8\%}\\
$Set$ II: & \multicolumn{3}{c}{$S$: 16.0\%;\, $H$: 8.8\%;\, $Di$: 5.4\%;\, $K$: 69.8\%}\\[2ex]
$5942-i5.6$   & \multicolumn{3}{c}{$\mu=0.068$} \\
  & \multicolumn{3}{c}{$r_{c \bar{c}}:1.31$;\,\,\,\,\,$r_{\bar{c}\bar{s}}:1.95$;\,\,\,\,\,$r_{c\bar{s}}:1.60$;\,\,\,\,\,$r_{cc}:1.79$} \\
$Set$ I: & \multicolumn{3}{c}{$S$: 17.3\%;\, $H$: 5.7\%;\, $Di$: 0.5\%;\, $K$: 76.5\%}\\
$Set$ II: & \multicolumn{3}{c}{$S$: 17.8\%;\, $H$: 15.6\%;\, $Di$: 1.2\%;\, $K$: 65.5\%}\\
\end{tabular}
\end{ruledtabular}
\end{table}

{\bf The $\bm{I(J^P)=0(1^+)}$ sector:} 33 channels listed in Table~\ref{GresultCC5} are investigated in this case. Generally, the three meson-meson configurations in singlet-color channels are $\eta_c D^*_s$, $J/\psi D_s$ and $J/\psi D^*_s$. Their lowest-lying masses are $5.10$ GeV, $5.09$ GeV and $5.21$ GeV, respectively. Hence, bound states are still unavailable in this sector. Masses of the hidden-color, diquark-antidiquark, $K_3$ and $K_5$ channels are $\sim5.45$ GeV. This value is consistent with most channels in the $K_1$, $K_2$ and $K_4$ configurations, except for one $K_1$ channel at $5.38$ GeV, one $K_2$ channel at $5.34$ GeV, and one $K_4$ channel at $5.92$ GeV. After partially coupled-channel calculations are performed in each configuration, the lowest-lying mass, which is the non-interacting $J/\psi D_s$ theoretical threshold value, remains at $5.09$ GeV. Besides, the extremely weak coupling effect is persistent in a fully coupled-channels study.

In a fully coupled-channel investigation by the CSM, three resonance poles are obtained and they are circled in Fig.~\ref{PP5}. Particularly, in the top panel, whose energy ranges within the interval $5.05-6.00$ GeV, scattering states of the ground and first radial excitations of $\eta_c D^*_s$, $J/\psi D_s$ and $J/\psi D^*_s$ are well presented. Moreover, the energy region from $5.70$ GeV to $5.85$ GeV is enlarged in the bottom panel. Therein, the four radial excitations, $\eta_c(2S)D^*_s(1S)$, $\eta_c(1S)D^*_s(2S)$, $\psi(2S)D^*_s(1S)$ and $J/\psi D_s(2S)$, are clearly plotted. Apart from the scattering dots, three stable resonance poles are obtained at around $5.9$ GeV. The complex energies are $5863-i6.6$ MeV, $5941-i1.8$ MeV and $5942-i5.6$ MeV.

The nature of three resonances are listed in Table~\ref{GresultR5}. Considering the magnetic moment, the first two resonances have similar values, $0.578\mu_N$ and $0.503\mu_N$, respectively. However, the value $0.068\mu_N$ is obtained for the third resonance. The size of the resonance at $5.86$ GeV is $1.6-2.2$ fm, and the other two higher resonances present a size less than $1.95$ fm. Finally, the K-type channels takes up more than $65\%$ for all of these resonances. Meanwhile, there is also $\sim20\%$ di-meson configurations in color-singlet channels for resonances at $5.94$ GeV. Additionally, the hidden-color channels of the first resonance state have a similar proportion. Herein, the effect of off-diagonal elements is still weak, with less than $13\%$ modification.


\begin{table}[!t]
\caption{\label{GresultCC6} Lowest-lying $\bar{c}c\bar{s}c$ tetraquark states with $I(J^P)=0(2^+)$ calculated within the real range formulation of the chiral quark model. Results are similarly organized as those in Table~\ref{GresultCC1} (unit: MeV).}
\begin{ruledtabular}
\begin{tabular}{lcccc}
~~Channel   & Index & $\chi_J^{\sigma_i}$;~$\chi_j^c$ & $M$ & Mixed~~ \\
        &   &$[i; ~j]$ &  \\[2ex]
$(J/\psi D^*_s)^1 (5209)$  & 1  & [1;~1]   & $5212$ &  \\[2ex]
$(J/\psi D^*_s)^8$            & 2  & [1;~3]   & $5452$ &  \\[2ex]
$(cc)^*(\bar{c}\bar{d})^*$  & 3  & [1;~7]   & $5497$ & \\[2ex]
$K_1$  & 4  & [1;~8]   & $5459$ & \\
            & 5  & [1;~10]   & $5487$ & $5391$ \\[2ex]
$K_2$  & 6  & [1;~11]   & $5448$ & \\
             & 7  & [1;~12]   & $5465$ & $5437$ \\[2ex]
$K_3$  & 8  & [1;~14]   & $5496$ & \\[2ex]
$K_4$  & 9  & [1;~8]   & $5930$ & \\
             & 10  & [1;~10]   & $5498$ & $5483$ \\[2ex]
$K_5$  & 11  & [1;~11]   & $5506$ & \\[2ex]
\multicolumn{4}{c}{Complete coupled-channels:} & $5212$
\end{tabular}
\end{ruledtabular}
\end{table}

\begin{figure}[!t]
\includegraphics[clip, trim={3.0cm 1.9cm 3.0cm 1.0cm}, width=0.45\textwidth]{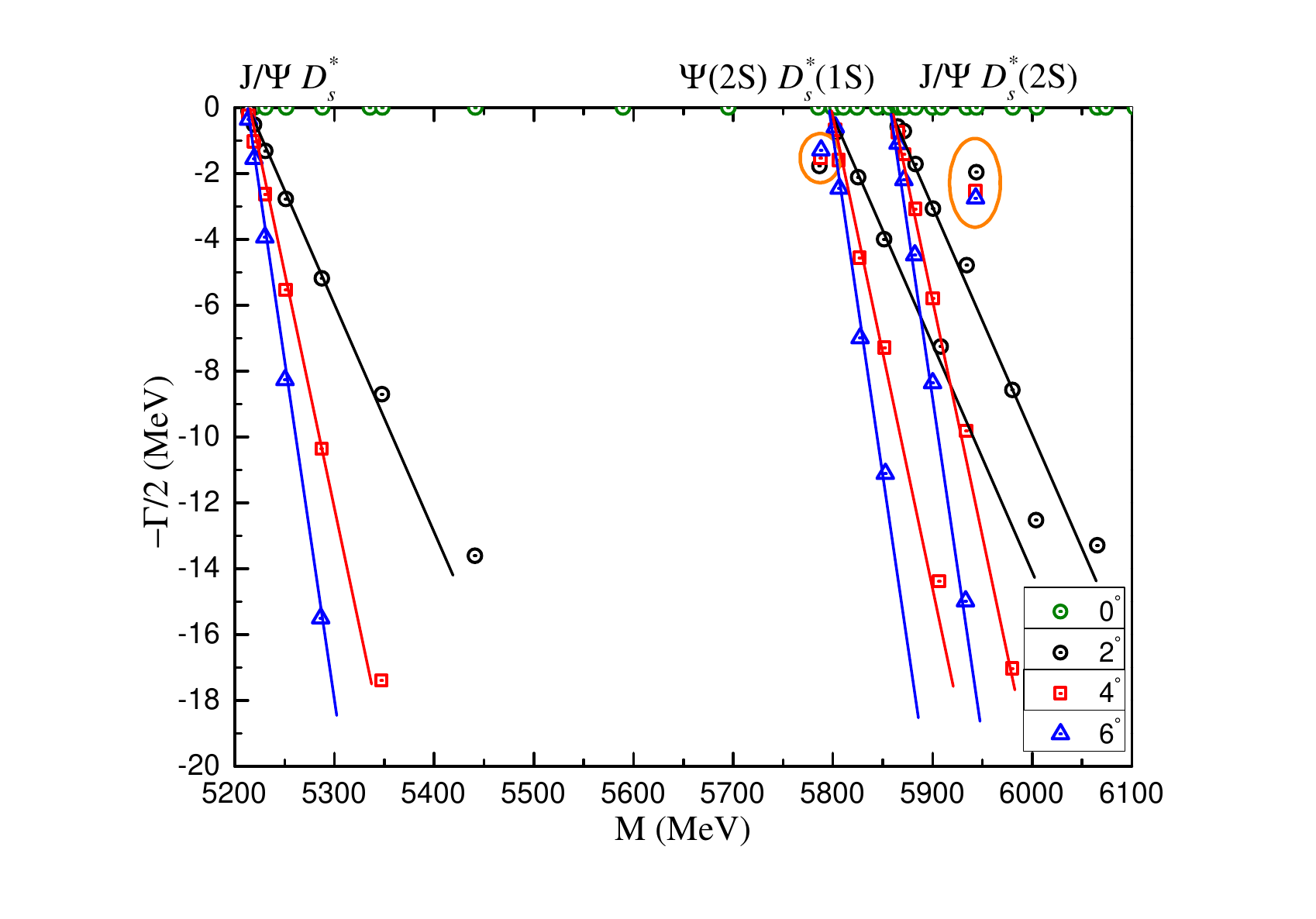}
\caption{\label{PP6} The complete coupled-channels calculation of $\bar{c}c\bar{s}c$ tetraquark system with $I(J^P)=0(2^+)$ quantum numbers.}
\end{figure}

\begin{table}[!t]
\caption{\label{GresultR6} Compositeness of exotic resonances obtained in a complete coupled-channel calculation in the $0(2^+)$ state of $\bar{c}c\bar{s}c$ tetraquark. Results are similarly organized as those in Table~\ref{GresultR1}.}
\begin{ruledtabular}
\begin{tabular}{rccc}
Resonance       & \multicolumn{3}{c}{Structure} \\[2ex]
$5788-i3.1$   & \multicolumn{3}{c}{$\mu=0.921$} \\
  & \multicolumn{3}{c}{$r_{c \bar{c}}:0.84$;\,\,\,\,\,$r_{\bar{c}\bar{s}}:1.08$;\,\,\,\,\,$r_{c\bar{s}}:1.02$;\,\,\,\,\,$r_{cc}:1.05$} \\
$Set$ I: & \multicolumn{3}{c}{$S$: 7.2\%;\, $H$: 35.2\%;\, $Di$: 2.6\%;\, $K$: 55.0\%}\\
$Set$ II: & \multicolumn{3}{c}{$S$: 8.5\%;\, $H$: 35.5\%;\, $Di$: 3.2\%;\, $K$: 52.8\%}\\[2ex]
$5943-i5.1$   & \multicolumn{3}{c}{$\mu=0.921$} \\
  & \multicolumn{3}{c}{$r_{c \bar{c}}:1.42$;\,\,\,\,\,$r_{\bar{c}\bar{s}}:2.03$;\,\,\,\,\,$r_{c\bar{s}}:1.65$;\,\,\,\,\,$r_{cc}:1.96$} \\
$Set$ I: & \multicolumn{3}{c}{$S$: 9.0\%;\, $H$: 2.5\%;\, $Di$: 2.0\%;\, $K$: 86.5\%}\\
$Set$ II: & \multicolumn{3}{c}{$S$: 18.4\%;\, $H$: 6.4\%;\, $Di$: 8.7\%;\, $K$: 66.5\%}\\
\end{tabular}
\end{ruledtabular}
\end{table}

{\bf The $\bm{I(J^P)=0(2^+)}$ sector:} There are 11 channels in the highest spin state and Table~\ref{GresultCC6} lists the calculated results. Several conclusions can be drawn. Firstly, no bound state is found in neither a single channel nor a coupled-channel computation. Secondly, the color-singlet channel of $J/\psi D^*_s$ is $5.21$ GeV, and the hidden-color channel is $5.45$ GeV. Besides, the other exotic color configurations also locate at this energy level, except for one $K_4$ channel whose mass is $5.93$ GeV. One can appreciate a notable mass shift in each K-type configuration when coupling, however, it is not strong enough to poduce a bound state.

Figure~\ref{PP6} shows the distribution of complex energies obtained in a complete coupled-channel computation by the CSM. Therein, three scattering states of $J/\psi D^*_s$, $\psi(2S)D^*_s(1S)$ and $J/\psi D^*_s(2S)$ are well presented. Besides, two resonance poles are obtained in the complex plane. Their mass and width are denoted as $5788-i3.1$ MeV and $5943-i5.1$ MeV, respectively.

The electromagnetic and geometric properties of the resonances can be found in Table~\ref{GresultR6}. In particular, the magnetic moment of these two resonance states is $0.92\mu_N$. A compact $\bar{c}c\bar{s}c$ tetraquark is obtained for the lower resonance with a size less than $1.10$ fm; however, the resonance at $5.94$ GeV is an extended object with a size of $1.42-2.03$ fm. The dominant wavefunction components of these two states are different, being the dominant ones for the first resonance the hidden-color ($35\%$) and K-type ($55\%$) channels while, for the second resonance, one has the color-singlet ($18\%$) and K-type ($67\%$) configurations as the dominant ones. There is almost no difference between sets I and II for the component assessment of the first resonance, and the difference is small for the second one.


\subsection{The $\mathbf{\bar{b}b\bar{d}b}$ tetraquarks}

\begin{table}[!t]
\caption{\label{GresultCC7} Lowest-lying $\bar{b}b\bar{d}b$ tetraquark states with $I(J^P)=\frac{1}{2}(0^+)$ calculated within the real range formulation of the chiral quark model. Results are similarly organized as those in Table~\ref{GresultCC1} (unit: MeV).}
\begin{ruledtabular}
\begin{tabular}{lcccc}
~~Channel   & Index & $\chi_J^{\sigma_i}$;~$\chi_j^c$ & $M$ & Mixed~~ \\
        &   &$[i; ~j]$ &  \\[2ex]
$(\eta_b B)^1 (14580)$          & 1  & [1;~1]  & $14732$ & \\
$(\Upsilon B^*)^1 (14785)$  & 2  & [2;~1]   & $14824$ & $14732$  \\[2ex]
$(\eta_b B)^8$          & 3  & [1;~2]  & $15050$ & \\
$(\Upsilon B^*)^8$       & 4  & [2;~2]   & $15049$ & $15020$  \\[2ex]
$(bb)(\bar{b}\bar{d})$      & 5     & [3;~4]  & $15042$ & \\
$(bb)^*(\bar{b}\bar{d})^*$  & 6  & [4;~3]   & $15061$ & $15034$ \\[2ex]
$K_1$  & 7  & [5;~5]   & $15021$ & \\
            & 8  & [5;~6]   & $15001$ & \\
            & 9  & [6;~5]   & $15028$ & \\
            & 10  & [6;~6]   & $14964$ & $14898$ \\[2ex]
$K_2$  & 11  & [7;~7]   & $14952$ & \\
             & 12  & [7;~8]   & $15027$ & \\
             & 13  & [8;~7]   & $14908$ & \\
             & 14  & [8;~8]   & $15035$ & $14894$ \\[2ex]
$K_3$  & 15  & [9;~10]   & $15053$ & \\
             & 16  & [10;~9]   & $15025$ & $15017$ \\[2ex]
$K_4$  & 17  & [11;~12]   & $15063$ & \\
             & 18  & [12;~12]   & $15457$ & \\
             & 19  & [11;~11]   & $15364$ & \\
             & 20  & [12;~11]   & $15036$ & $14959$ \\[2ex]
$K_5$  & 21  & [13;~14]   & $15063$ & \\
             & 22  & [14;~13]   & $15016$ & $15010$ \\[2ex]
\multicolumn{4}{c}{Complete coupled-channels:} & $14732$
\end{tabular}
\end{ruledtabular}
\end{table}

\begin{figure}[!t]
\includegraphics[clip, trim={3.0cm 1.9cm 3.0cm 1.0cm}, width=0.45\textwidth]{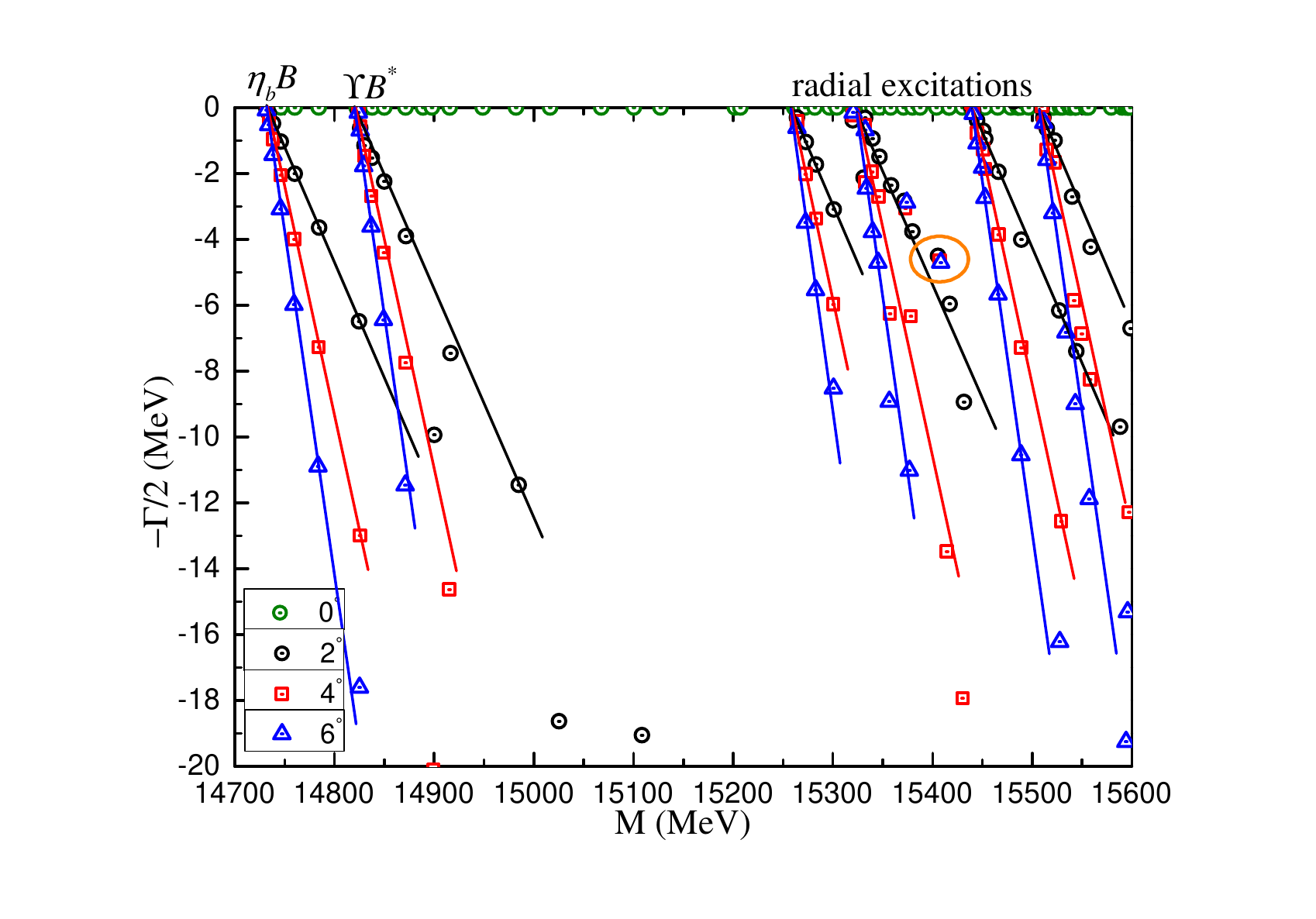} \\[1ex]
\includegraphics[clip, trim={3.0cm 1.9cm 3.0cm 1.0cm}, width=0.45\textwidth]{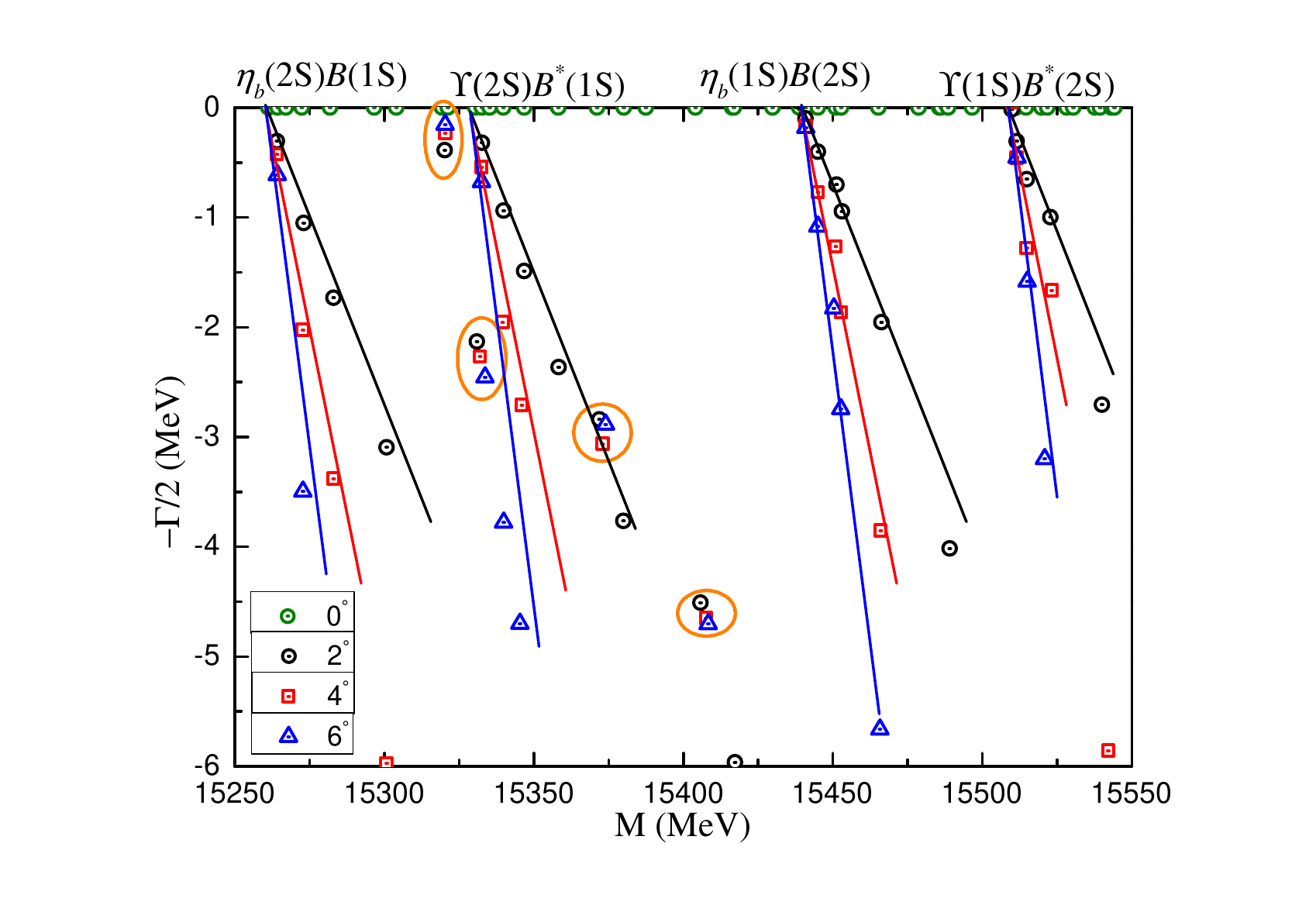}
\caption{\label{PP7} The complete coupled-channels calculation of $\bar{b}b\bar{d}b$ tetraquark system with $I(J^P)=\frac{1}{2}(0^+)$ quantum numbers. Particularly, the bottom panel is enlarged parts of dense energy region from $15.25\,\text{GeV}$ to $15.55\,\text{GeV}$.}
\end{figure}

\begin{table}[!t]
\caption{\label{GresultR7} Compositeness of exotic resonances obtained in a complete coupled-channel calculation in the $\frac{1}{2}(0^+)$ state of $\bar{b}b\bar{d}b$ tetraquark. Results are similarly organized as those in Table~\ref{GresultR1}.}
\begin{ruledtabular}
\begin{tabular}{rccc}
Resonance       & \multicolumn{3}{c}{Structure} \\[2ex]
$15320-i0.5$   & \multicolumn{3}{c}{$\mu=0$} \\
  & \multicolumn{3}{c}{$r_{b \bar{b}}:0.96$;\,\,\,\,\,$r_{\bar{b}\bar{d}}:1.52$;\,\,\,\,\,$r_{b\bar{d}}:1.26$;\,\,\,\,\,$r_{bb}:1.33$} \\
$Set$ I: & \multicolumn{3}{c}{$S$: 6.0\%;\, $H$: 0.6\%;\, $Di$: 21.2\%;\, $K$: 72.2\%}\\
$Set$ II: & \multicolumn{3}{c}{$S$: 8.2\%;\, $H$: 0.1\%;\, $Di$: 4.1\%;\, $K$: 87.6\%}\\[2ex]
$15331-i4.5$   & \multicolumn{3}{c}{$\mu=0$} \\
  & \multicolumn{3}{c}{$r_{b \bar{b}}:1.23$;\,\,\,\,\,$r_{\bar{b}\bar{d}}:1.77$;\,\,\,\,\,$r_{b\bar{d}}:1.32$;\,\,\,\,\,$r_{bb}:1.64$} \\
$Set$ I: & \multicolumn{3}{c}{$S$: 2.5\%;\, $H$: 1.6\%;\, $Di$: 25.1\%;\, $K$: 70.8\%}\\
$Set$ II: & \multicolumn{3}{c}{$S$: 4.5\%;\, $H$: 2.5\%;\, $Di$: 23.2\%;\, $K$: 69.8\%}\\[2ex]
$15372-i6.1$   & \multicolumn{3}{c}{$\mu=0$} \\
  & \multicolumn{3}{c}{$r_{b \bar{b}}:1.12$;\,\,\,\,\,$r_{\bar{b}\bar{d}}:1.70$;\,\,\,\,\,$r_{b\bar{d}}:1.37$;\,\,\,\,\,$r_{bb}:1.54$} \\
$Set$ I: & \multicolumn{3}{c}{$S$: 2.5\%;\, $H$: 0.7\%;\, $Di$: 20.9\%;\, $K$: 75.9\%}\\
$Set$ II: & \multicolumn{3}{c}{$S$: 6.2\%;\, $H$: 5.8\%;\, $Di$: 16.6\%;\, $K$: 71.4\%}\\[2ex]
$15407-i9.3$   & \multicolumn{3}{c}{$\mu=0$} \\
  & \multicolumn{3}{c}{$r_{b \bar{b}}:1.05$;\,\,\,\,\,$r_{\bar{b}\bar{d}}:1.59$;\,\,\,\,\,$r_{b\bar{d}}:1.31$;\,\,\,\,\,$r_{bb}:1.45$} \\
$Set$ I: & \multicolumn{3}{c}{$S$: 1.2\%;\, $H$: 1.0\%;\, $Di$: 23.7\%;\, $K$: 74.1\%}\\
$Set$ II: & \multicolumn{3}{c}{$S$: 4.5\%;\, $H$: 3.1\%;\, $Di$: 22.7\%;\, $K$: 69.7\%}\\
\end{tabular}
\end{ruledtabular}
\end{table}

{\bf The $\bm{I(J^P)=\frac{1}{2}(0^+)}$ sector:} 22 channels are studied in this quantum state; particularly, there are two channels in the color-singlet, hidden-color and diquark-antidiquark congigurations, respectively. The other 16 channels are from the K-type structures. As shown in Table~\ref{GresultCC7}, within real-range method, the lowest-lying channel is an $\eta_b B$ scattering state with mass at $14.73$ GeV, and the mass of $\Upsilon B^*$ state is $14.82$ GeV. The other channels are at $\sim15.05$ GeV, except for two $K_4$ channels with masses $15.36$ GeV and $15.46$ GeV, respectively. The effect of channel coupling is weak. The lowest-lying mass remains at $14.73$ GeV, $K_1$ and $K_2$ configurations are almost degenerate in mass, $14.89$ GeV, and the other coupled-mass of exotic structures are at around $15.0$ GeV.

The computed complex energies obtained from a fully-coupled channels CSM calculation are plotted in Fig.~\ref{PP7}. Firstly, $\eta_b B$ and $\Upsilon B^*$ in both ground and radial excited states are clearly shown in the top panel (energy region $14.7-15.6$ GeV). Meanwhile, four radial excitations, $\eta_b(2S) B(1S)$, $\Upsilon(2S)B^*(1S)$, $\eta_b(1S) B(2S)$ and $\Upsilon(1S)B^*(2S)$ are presented in the bottom panel, which is an enlarged part of the energy region from $15.25$ to $15.55$ GeV. Apart from these scattering states, four resonance states are observed with complex energies $15320-i0.5$ MeV, $15331-i4.5$ MeV, $15372-i6.1$ MeV and $15407-i9.3$ MeV. In particular, the two resonances at $15.33$ GeV and $15.41$ GeV are consistent with those obtained in Ref.~\cite{Zhu:2023lbx}.

The nature of these resonances can be guessed from Table~\ref{GresultR7}. Firstly, their magnetic moment are all 0. Secondly, they are relatively extended structures, with sizes of $1.0-1.7$ fm. Thirdly, their dominant wavefunction components are K-type channels but there are also considerable diquark-antidiquark components. Note that the differences between set I and II is less than $17\%$ after comparison. 


\begin{table}[!t]
\caption{\label{GresultCC8} Lowest-lying $\bar{b}b\bar{d}b$ tetraquark states with $I(J^P)=\frac{1}{2}(1^+)$ calculated within the real range formulation of the chiral quark model. Results are similarly organized as those in Table~\ref{GresultCC1} (unit: MeV).}
\begin{ruledtabular}
\begin{tabular}{lcccc}
~~Channel   & Index & $\chi_J^{\sigma_i}$;~$\chi_j^c$ & $M$ & Mixed~~ \\
        &   &$[i; ~j]$ &  \\[2ex]
$(\eta_b B^*)^1 (14625)$   & 1  & [1;~1]  & $14773$ & \\
$(\Upsilon B)^1 (14740)$  & 2  & [2;~1]   & $14783$ &  \\
$(\Upsilon B^*)^1 (14785)$  & 3  & [3;~1]   & $14824$ & $14773$  \\[2ex]
$(\eta_b B^*)^8$       & 4  & [1;~2]   & $15043$ &  \\
$(\Upsilon B)^8$      & 5  & [2;~2]  & $15046$ & \\
$(\Upsilon B^*)^8$  & 6  & [3;~2]   & $15041$ & $15022$ \\[2ex]
$(bb)^*(\bar{b}\bar{d})^*$  & 7  & [6;~3]  & $15065$ & \\
$(bb)^*(\bar{b}\bar{d})$    & 8  & [5;~4]   & $15036$ &  \\
$(bb)(\bar{b}\bar{d})^*$  & 9  & [4;~3]   & $15053$ & $15031$  \\[2ex]
$K_1$      & 10  & [7;~5]   & $15040$ &  \\
      & 11 & [8;~5]  & $14998$ & \\
      & 12  & [9;~5]   & $15020$ & \\
      & 13   & [7;~6]  & $14991$ & \\
      & 14   & [8;~6]  & $15027$ & \\
      & 15   & [9;~6]  & $14976$ & $14913$ \\[2ex]
$K_2$      & 16  & [10;~7]   & $14972$ & \\
                 & 17  & [11;~7]   & $14940$ &  \\
                 & 18  & [12;~7]   & $14923$ & \\
                 & 19  & [10;~8]   & $15010$ & \\
                 & 20  & [11;~8]   & $15042$ & \\
                 & 21  & [12;~8]   & $15027$ & $14904$ \\[2ex]
$K_3$      & 22  & [13;~10]   & $15048$ & \\
                 & 23  & [14;~10]   & $15058$ & \\
                 & 24  & [15;~9]    & $15018$ & $15014$ \\[2ex]
$K_4$      & 25  & [16;~11]   & $15067$ & \\
                 & 26  & [17;~11]   & $15081$ & \\
                 & 27  & [18;~11]   & $15387$ & \\
                 & 28  & [16;~12]   & $15070$ & \\
                 & 29  & [17;~12]   & $15069$ & \\
                 & 30  & [18;~12]   & $15055$ & $14963$ \\[2ex]
$K_5$      & 31  & [19;~14]   & $15066$ & \\
                 & 32  & [20;~14]   & $15062$ & \\
                 & 33  & [21;~13]   & $15009$ & $15006$ \\[2ex]  
\multicolumn{4}{c}{Complete coupled-channels:} & $14773$
\end{tabular}
\end{ruledtabular}
\end{table}

\begin{figure}[!t]
\includegraphics[clip, trim={3.0cm 1.9cm 3.0cm 1.0cm}, width=0.45\textwidth]{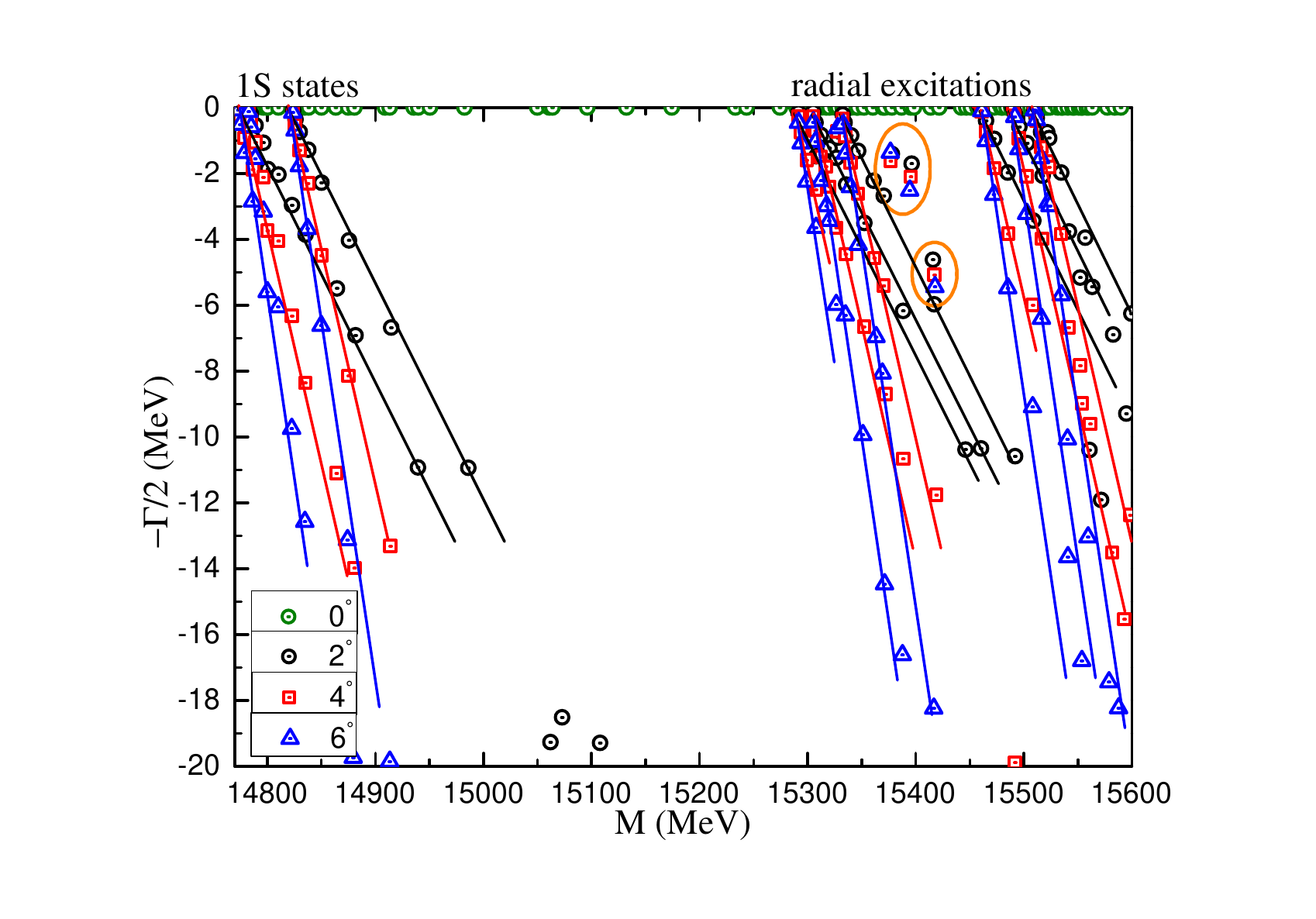} \\[1ex]
\includegraphics[clip, trim={3.0cm 1.9cm 3.0cm 1.0cm}, width=0.45\textwidth]{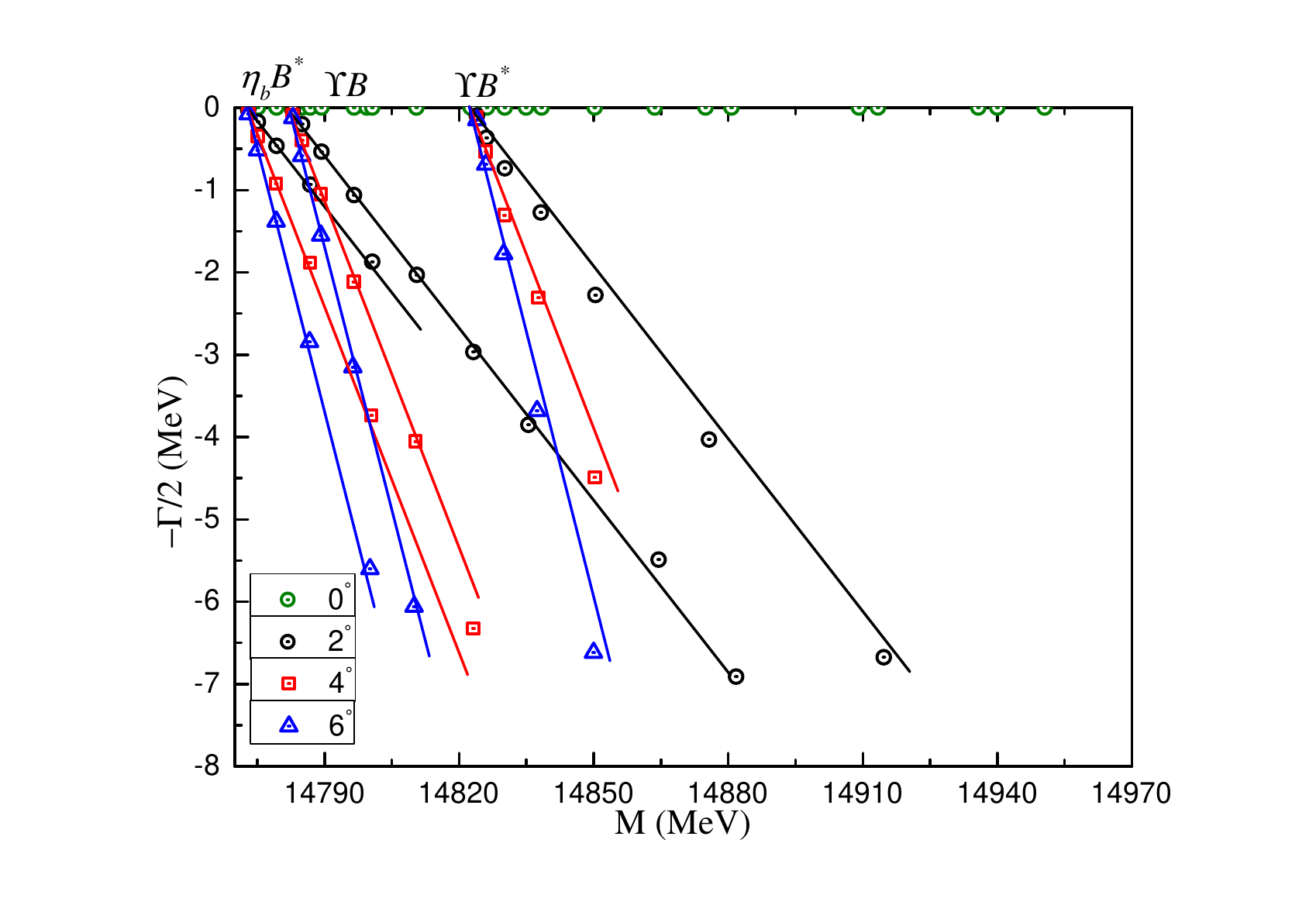}\\[1ex]
\includegraphics[clip, trim={3.0cm 1.9cm 3.0cm 1.0cm}, width=0.45\textwidth]{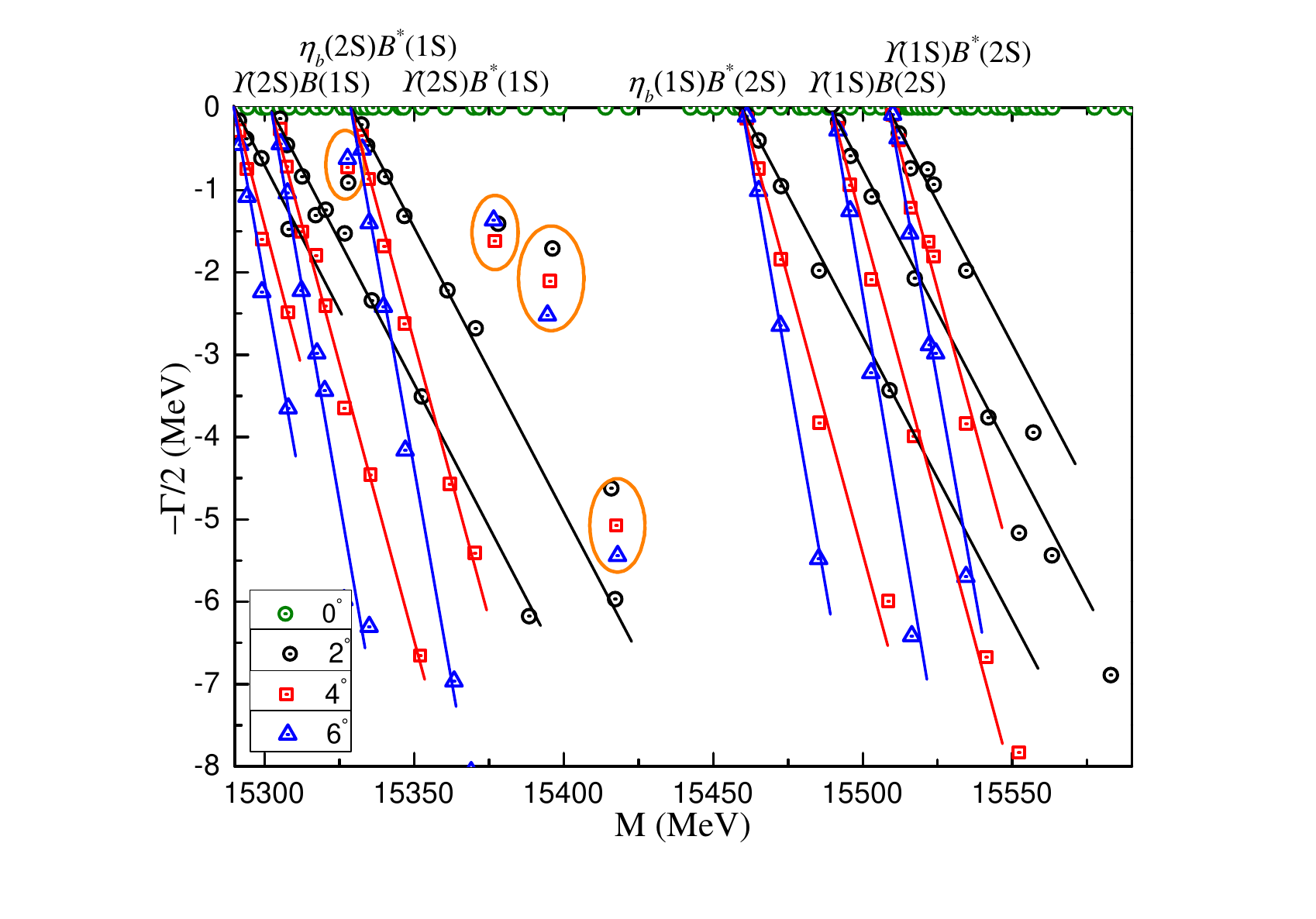}
\caption{\label{PP8} The complete coupled-channels calculation of $\bar{b}b\bar{u}b$ tetraquark system with $I(J^P)=\frac{1}{2}(1^+)$ quantum numbers. Particularly, the middle panel is enlarged parts of dense energy region from $14.77\,\text{GeV}$ to $14.97\,\text{GeV}$, and the bottom one is enlarged parts of dense energy region from $15.28\,\text{GeV}$ to $15.58\,\text{GeV}$.}
\end{figure}

\begin{table}[!t]
\caption{\label{GresultR8} Compositeness of exotic resonances obtained in a complete coupled-channel calculation in the $\frac{1}{2}(1^+)$ state of $\bar{b}b\bar{d}b$ tetraquark. Results are similarly organized as those in Table~\ref{GresultR1}.}
\begin{ruledtabular}
\begin{tabular}{rccc}
Resonance       & \multicolumn{3}{c}{Structure} \\[2ex]
$15327-i1.5$   & \multicolumn{3}{c}{$\mu=-1.453$} \\
  & \multicolumn{3}{c}{$r_{b \bar{b}}:1.08$;\,\,\,\,\,$r_{\bar{b}\bar{d}}:1.58$;\,\,\,\,\,$r_{b\bar{d}}:1.22$;\,\,\,\,\,$r_{bb}:1.44$} \\
$Set$ I: & \multicolumn{3}{c}{$S$: 1.9\%;\, $H$: 1.5\%;\, $Di$: 26.3\%;\, $K$: 70.3\%}\\
$Set$ II: & \multicolumn{3}{c}{$S$: 1.4\%;\, $H$: 4.0\%;\, $Di$: 8.4\%;\, $K$: 86.2\%}\\[2ex]
$15376-i3.2$   & \multicolumn{3}{c}{$\mu=-1.817$} \\
  & \multicolumn{3}{c}{$r_{b \bar{b}}:1.04$;\,\,\,\,\,$r_{\bar{b}\bar{d}}:1.63$;\,\,\,\,\,$r_{b\bar{d}}:1.33$;\,\,\,\,\,$r_{bb}:1.43$} \\
$Set$ I: & \multicolumn{3}{c}{$S$: 1.1\%;\, $H$: 1.1\%;\, $Di$: 24.9\%;\, $K$: 72.9\%}\\
$Set$ II: & \multicolumn{3}{c}{$S$: 3.9\%;\, $H$: 6.2\%;\, $Di$: 13.4\%;\, $K$: 76.5\%}\\[2ex]
$15395-i4.2$   & \multicolumn{3}{c}{$\mu=-1.456$} \\
  & \multicolumn{3}{c}{$r_{b \bar{b}}:0.95$;\,\,\,\,\,$r_{\bar{b}\bar{d}}:1.48$;\,\,\,\,\,$r_{b\bar{d}}:1.26$;\,\,\,\,\,$r_{bb}:1.31$} \\
$Set$ I: & \multicolumn{3}{c}{$S$: 1.6\%;\, $H$: 0.7\%;\, $Di$: 14.8\%;\, $K$: 82.9\%}\\
$Set$ II: & \multicolumn{3}{c}{$S$: 5.4\%;\, $H$: 3.1\%;\, $Di$: 12.3\%;\, $K$: 79.2\%}\\[2ex]
$15417-i10.2$   & \multicolumn{3}{c}{$\mu=-1.771$} \\
  & \multicolumn{3}{c}{$r_{b \bar{b}}:0.99$;\,\,\,\,\,$r_{\bar{b}\bar{d}}:1.48$;\,\,\,\,\,$r_{b\bar{d}}:1.20$;\,\,\,\,\,$r_{bb}:1.36$} \\
$Set$ I: & \multicolumn{3}{c}{$S$: 2.7\%;\, $H$: 0.5\%;\, $Di$: 17.5\%;\, $K$: 79.3\%}\\
$Set$ II: & \multicolumn{3}{c}{$S$: 5.4\%;\, $H$: 1.5\%;\, $Di$: 2.9\%;\, $K$: 90.2\%}\\
\end{tabular}
\end{ruledtabular}
\end{table}

{\bf The $\bm{I(J^P)=\frac{1}{2}(1^+)}$ sector:} 33 channels have to be studied for this quantum state and they are shown in Table~\ref{GresultCC8}. The three meson-meson configurations of $\eta_b B^*$, $\Upsilon B$ and $\Upsilon B^*$ in color-singlet channels are all unbound. Their lowest-lying masses just equal to the corresponding theoretical threshold values, \emph{i.e.} $14.77$ GeV, $14.78$ GeV and $14.82$ GeV, respectively. The hidden-color channels are almost degenerate in mass at around $15.04$ GeV. Meanwhile, the three diquark-antidiquark channels locate at $\sim15.05$ GeV. Masses of the 24 K-type channels are generally within an energy region from $14.92$ GeV to $15.39$ GeV. Coupled-channels effect helps a little in pushing the lowest mass down and it is still weak to obtain a bound state in each kind of calculation.

In a further step towards a complex-range study of complete coupled-channels calculation, four resonance states are obtained and they are circled in Fig.~\ref{PP8}. Particularly, three panels are presented. The energy point distributions of $1S$ and $2S$ states of the system are plotted in the top panel with an energy region $14.75-15.60$ GeV. Since there are dense distributions of energy dots in $14.77-14.97$ GeV and $15.28-15.58$ GeV, enlarged parts of these two energy regions are plotted in the middle and bottom panels, respectively. Firstly, three scattering states of $\eta_b B^*$, $\Upsilon B$ and $\Upsilon B^*$ are well presented in the middle panel of Fig.~\ref{PP8}. Besides, six radial excitation states, which include $\Upsilon(2S)B(1S)$, $\eta_b(2S)B^*(1S)$, $\Upsilon(2S)B^*(1S)$, $\eta_b(1S)B^*(2S)$, $\Upsilon(1S)B(2S)$ and $\Upsilon(1S)B^*(2S)$, are also shown in the bottom one. Therein, four stable poles are obtained, the resonance states are $15327-i1.5$ MeV, $15376-i3.2$ MeV, $15395-i4.2$ MeV and $15417-i10.2$ MeV, respectively. Moreover, the later three resonances are also compatible with results in Ref.~\cite{Zhu:2023lbx}.

Table~\ref{GresultR8} summarizes properties of the four resonance states. First of all, magnetic moments of them are negative: $-1.453\mu_N$, $-1.817\mu_N$, $-1.456\mu_N$ and $-1.771\mu_N$, respectively. Their sizes are within the range $1.0-1.6$ fm. The dominant components ($>70\%$) are K-type channels, and there are also considerable diquark-antidiquark components. The influence of off-diagonal elements continues to be small in this sector.


\begin{table}[!t]
\caption{\label{GresultCC9} Lowest-lying $\bar{b}b\bar{d}b$ tetraquark states with $I(J^P)=\frac{1}{2}(2^+)$ calculated within the real range formulation of the chiral quark model. Results are similarly organized as those in Table~\ref{GresultCC1} (unit: MeV).}
\begin{ruledtabular}
\begin{tabular}{lcccc}
~~Channel   & Index & $\chi_J^{\sigma_i}$;~$\chi_j^c$ & $M$ & Mixed~~ \\
        &   &$[i; ~j]$ &  \\[2ex]
$(\Upsilon B^*)^1 (14785)$  & 1  & [1;~1]   & $14824$ &  \\[2ex]
$(\Upsilon B^*)^8$            & 2  & [1;~3]   & $15040$ &  \\[2ex]
$(bb)^*(\bar{b}\bar{d})^*$  & 3  & [1;~7]   & $15073$ & \\[2ex]
$K_1$  & 4  & [1;~8]   & $15020$ & \\
            & 5  & [1;~10]   & $15038$ & $14964$ \\[2ex]
$K_2$  & 6  & [1;~11]   & $14986$ & \\
             & 7  & [1;~12]   & $15027$ & $14977$ \\[2ex]
$K_3$  & 8  & [1;~14]   & $15067$ & \\[2ex]
$K_4$  & 9  & [1;~8]   & $15391$ & \\
             & 10  & [1;~10]   & $15075$ & $15021$ \\[2ex]
$K_5$  & 11  & [1;~11]   & $15077$ & \\[2ex]
\multicolumn{4}{c}{Complete coupled-channels:} & $14824$
\end{tabular}
\end{ruledtabular}
\end{table}

\begin{table}[!t]
\caption{\label{GresultR9} Compositeness of exotic resonances obtained in a complete coupled-channel calculation in the $\frac{1}{2}(2^+)$ state of $\bar{b}b\bar{d}b$ tetraquark. Results are similarly organized as those in Table~\ref{GresultR1}.}
\begin{ruledtabular}
\begin{tabular}{rccc}
Resonance       & \multicolumn{3}{c}{Structure} \\[2ex]
$15308-i9.3$   & \multicolumn{3}{c}{$\mu=-2.061$} \\
  & \multicolumn{3}{c}{$r_{b \bar{b}}:0.64$;\,\,\,\,\,$r_{\bar{b}\bar{d}}:0.92$;\,\,\,\,\,$r_{b\bar{d}}:0.81$;\,\,\,\,\,$r_{bb}:0.62$} \\
$Set$ I: & \multicolumn{3}{c}{$S$: 4.8\%;\, $H$: 4.0\%;\, $Di$: 19.0\%;\, $K$: 72.2\%}\\
$Set$ II: & \multicolumn{3}{c}{$S$: 0.6\%;\, $H$: 8.8\%;\, $Di$: 25.6\%;\, $K$: 65.0\%}\\[2ex]
$15449-i1.4$   & \multicolumn{3}{c}{$\mu=-2.061$} \\
  & \multicolumn{3}{c}{$r_{b \bar{b}}:0.79$;\,\,\,\,\,$r_{\bar{b}\bar{d}}:1.29$;\,\,\,\,\,$r_{b\bar{d}}:1.19$;\,\,\,\,\,$r_{bb}:1.07$} \\
$Set$ I: & \multicolumn{3}{c}{$S$: 3.7\%;\, $H$: 2.4\%;\, $Di$: 24.2\%;\, $K$: 69.7\%}\\
$Set$ II: & \multicolumn{3}{c}{$S$: 3.2\%;\, $H$: 5.1\%;\, $Di$: 12.8\%;\, $K$: 78.9\%}\\[2ex]
$15558-i5.4$   & \multicolumn{3}{c}{$\mu=-2.061$} \\
  & \multicolumn{3}{c}{$r_{b \bar{b}}:1.00$;\,\,\,\,\,$r_{\bar{b}\bar{d}}:1.54$;\,\,\,\,\,$r_{b\bar{d}}:1.30$;\,\,\,\,\,$r_{bb}:1.22$} \\
$Set$ I: & \multicolumn{3}{c}{$S$: 2.2\%;\, $H$: 3.4\%;\, $Di$: 31.6\%;\, $K$: 62.8\%}\\
$Set$ II: & \multicolumn{3}{c}{$S$: 1.2\%;\, $H$: 2.8\%;\, $Di$: 26.7\%;\, $K$: 69.3\%}\\
\end{tabular}
\end{ruledtabular}
\end{table}

\begin{figure}[!t]
\includegraphics[clip, trim={3.0cm 1.9cm 3.0cm 1.0cm}, width=0.45\textwidth]{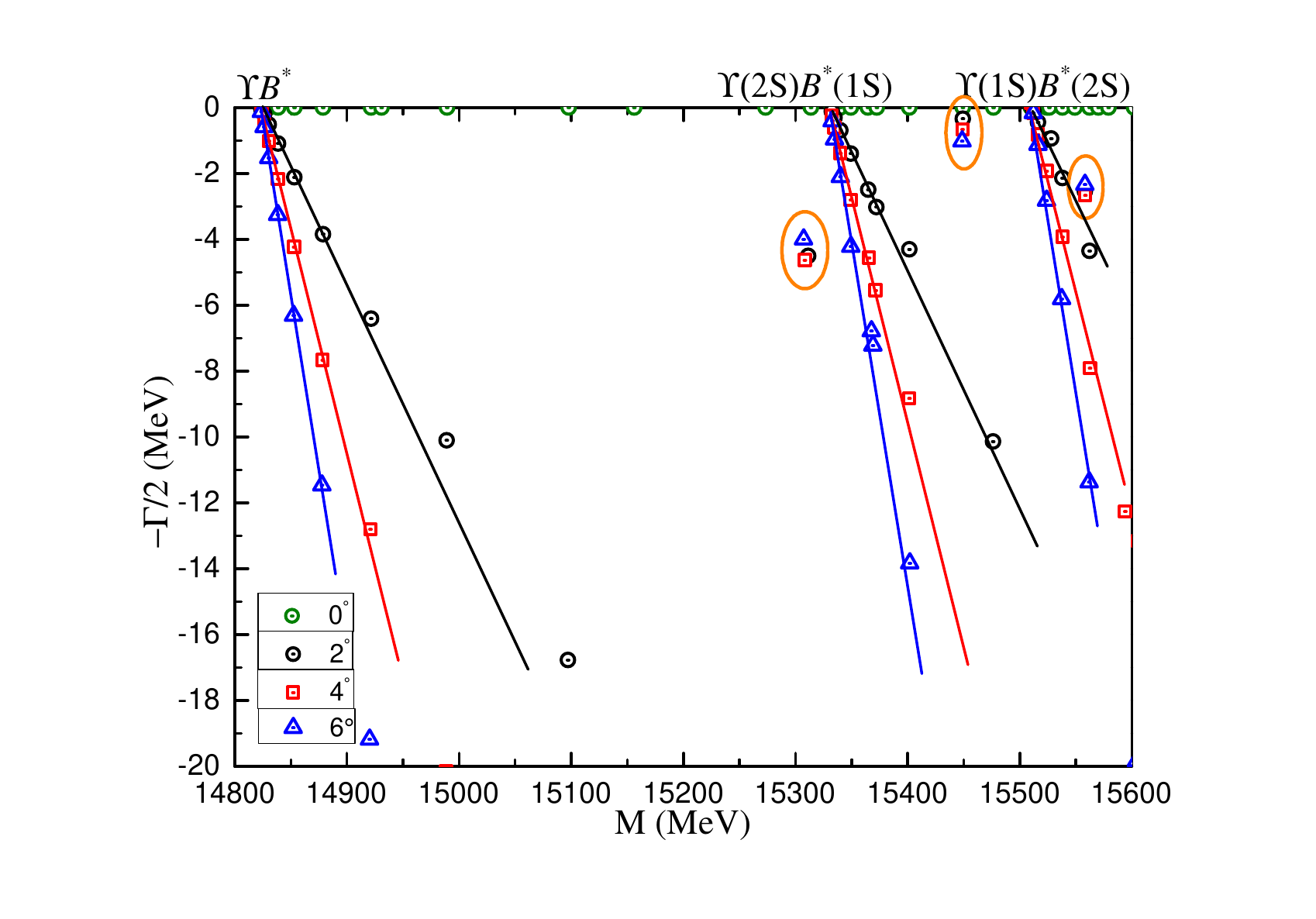}
\caption{\label{PP9} The complete coupled-channels calculation of $\bar{b}b\bar{u}b$ tetraquark system with $I(J^P)=\frac{1}{2}(2^+)$ quantum numbers.}
\end{figure}

{\bf The $\bm{I(J^P)=\frac{1}{2}(2^+)}$ sector:} The $\Upsilon B^*$ in both color-singlet and hidden-color channels, a $(bb)^*(\bar{b}\bar{d})^*$ channel, along with 8 K-type channels are comprehensively studied in Table~\ref{GresultCC9}. A bound state is still excluded in this sector, and the lowest-lying mass is $14.82$ GeV, which is the $\Upsilon B^*$ theoretical threshold value. The other channels locate within $14.99-15.39$ GeV energy interval. In coupled-channel calculations, the lowest-lying mass of each K-type configuration is $\sim15.0$ GeV.

Three resonances are obtained in a complete coupled-channel by the CSM. Figure~\ref{PP9} shows the distribution of complex energies within $14.8-15.6$ GeV. Apart from three scattering states of $\Upsilon B^*$, $\Upsilon(2S)B^*(1S)$ and $\Upsilon(1S)B^*(2S)$, three stable poles are circled. The resonances are $15308-i9.3$ MeV, $15449-i1.4$ MeV and $15558-i5.4$ MeV.

In Table~\ref{GresultR9} one can find that the magnetic moment of the three resonance states is approximately $-2.1\mu_N$. Besides, a compact $\bar{b}b\bar{d}b$ tetraquark structure is found for the first resonance whereas the size of the other two states is larger. Furthermore, diquark-antidiquark and K-type channels are the dominant components ($>90\%$) of these three resonances.


\begin{figure}[!t]
\includegraphics[clip, trim={3.0cm 1.9cm 3.0cm 1.0cm}, width=0.45\textwidth]{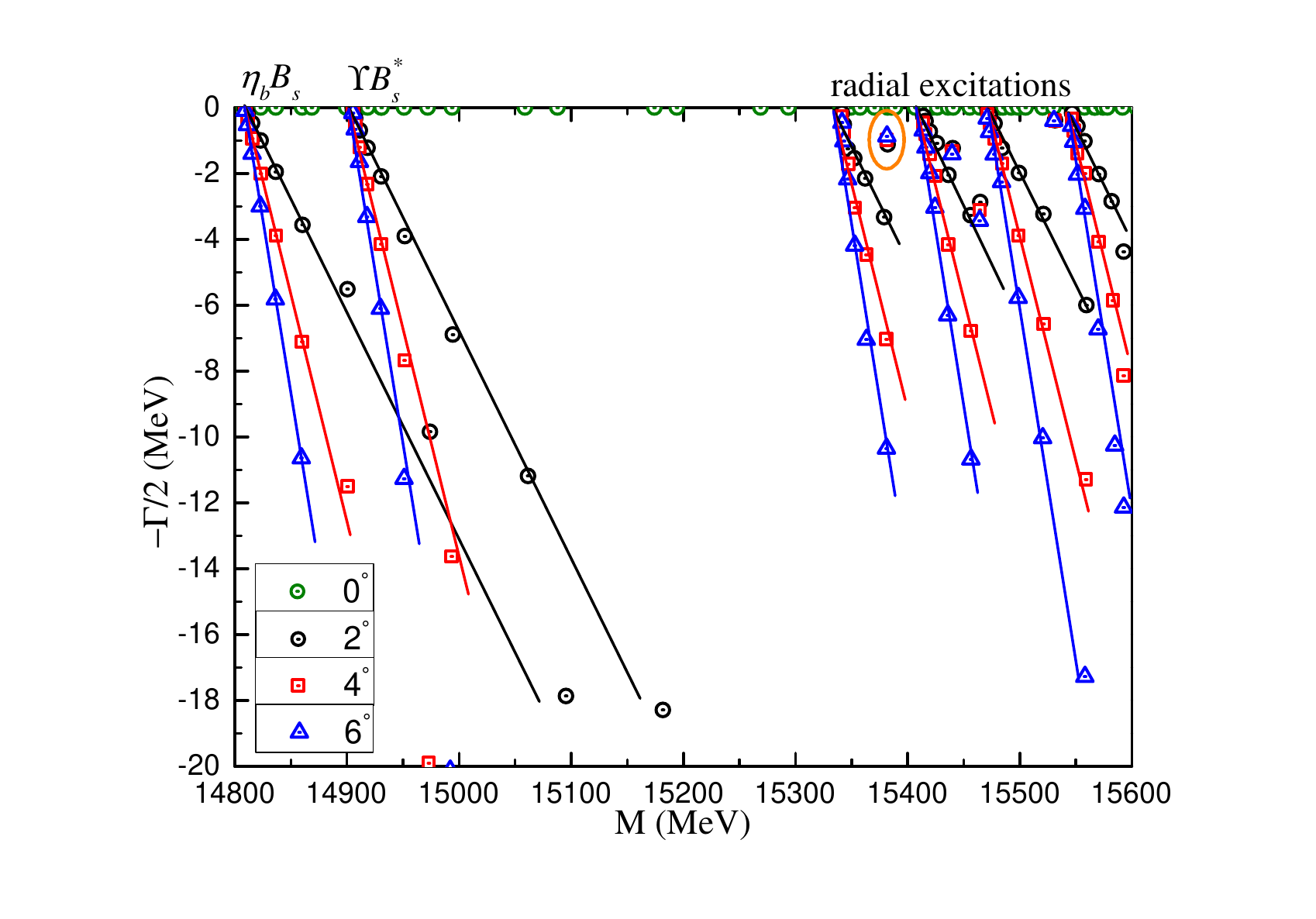} \\[1ex]
\includegraphics[clip, trim={3.0cm 1.9cm 3.0cm 1.0cm}, width=0.45\textwidth]{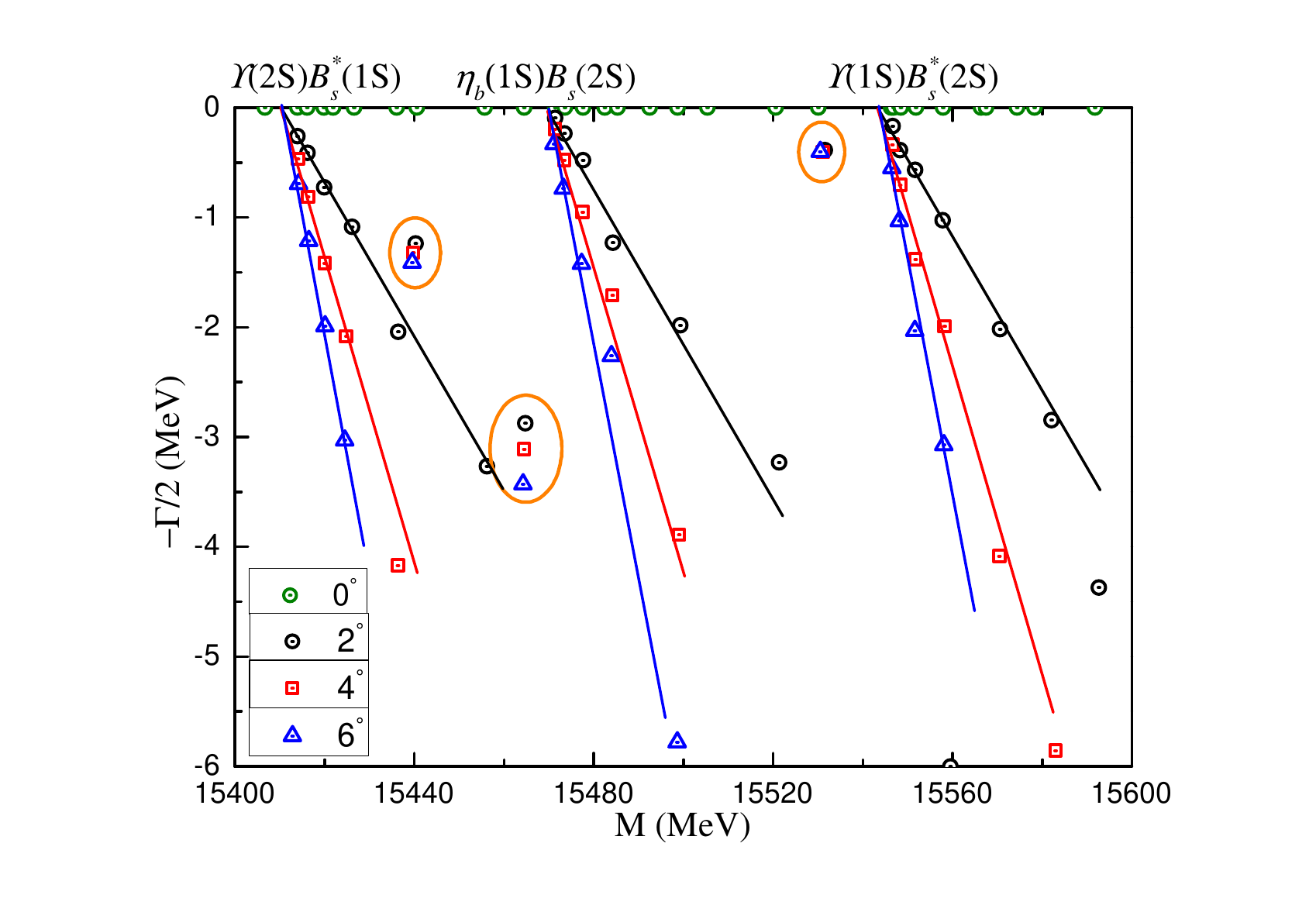}
\caption{\label{PP10} The complete coupled-channels calculation of $\bar{b}b\bar{s}b$ tetraquark system with $I(J^P)=0(0^+)$ quantum numbers. Particularly, the bottom panel is enlarged parts of dense energy region from $15.4\,\text{GeV}$ to $15.6\,\text{GeV}$.}
\end{figure}

\begin{table}[!t]
\caption{\label{GresultCC10} Lowest-lying $\bar{b}b\bar{s}b$ tetraquark states with $I(J^P)=0(0^+)$ calculated within the real range formulation of the chiral quark model. Results are similarly organized as those in Table~\ref{GresultCC1} (unit: MeV).}
\begin{ruledtabular}
\begin{tabular}{lcccc}
~~Channel   & Index & $\chi_J^{\sigma_i}$;~$\chi_j^c$ & $M$ & Mixed~~ \\
        &   &$[i; ~j]$ &  \\[2ex]
$(\eta_b B_s)^1 (14667)$          & 1  & [1;~1]  & $14809$ & \\
$(\Upsilon B^*_s)^1 (14875)$  & 2  & [2;~1]   & $14905$ & $14809$  \\[2ex]
$(\eta_b B_s)^8$          & 3  & [1;~2]  & $15138$ & \\
$(\Upsilon B^*_s)^8$       & 4  & [2;~2]   & $15137$ & $15109$  \\[2ex]
$(bb)(\bar{b}\bar{s})$      & 5     & [3;~4]  & $15129$ & \\
$(bb)^*(\bar{b}\bar{s})^*$  & 6  & [4;~3]   & $15149$ & $15122$ \\[2ex]
$K_1$  & 7  & [5;~5]   & $15111$ & \\
            & 8  & [5;~6]   & $15103$ & \\
            & 9  & [6;~5]   & $15119$ & \\
            & 10  & [6;~6]   & $15067$ & $14993$ \\[2ex]
$K_2$  & 11  & [7;~7]   & $15052$ & \\
             & 12  & [7;~8]   & $15118$ & \\
             & 13  & [8;~7]   & $15011$ & \\
             & 14  & [8;~8]   & $15127$ & $14997$ \\[2ex]
$K_3$  & 15  & [9;~10]   & $15142$ & \\
             & 16  & [10;~9]   & $15117$ & $15109$ \\[2ex]
$K_4$  & 17  & [11;~12]   & $15150$ & \\
             & 18  & [12;~12]   & $15555$ & \\
             & 19  & [11;~11]   & $15477$ & \\
             & 20  & [12;~11]   & $15124$ & $15065$ \\[2ex]
$K_5$  & 21  & [13;~14]   & $15152$ & \\
             & 22  & [14;~13]   & $15107$ & $15101$ \\[2ex]
\multicolumn{4}{c}{Complete coupled-channels:} & $14809$
\end{tabular}
\end{ruledtabular}
\end{table}

\begin{table}[!t]
\caption{\label{GresultR10} Compositeness of exotic resonances obtained in a complete coupled-channel calculation in the $0(0^+)$ state of $\bar{b}b\bar{s}b$ tetraquark. Results are similarly organized as those in Table~\ref{GresultR1}.}
\begin{ruledtabular}
\begin{tabular}{rccc}
Resonance       & \multicolumn{3}{c}{Structure} \\[2ex]
$15381-i2.0$   & \multicolumn{3}{c}{$\mu=0$} \\
  & \multicolumn{3}{c}{$r_{b \bar{b}}:0.69$;\,\,\,\,\,$r_{\bar{b}\bar{s}}:1.10$;\,\,\,\,\,$r_{b\bar{s}}:1.01$;\,\,\,\,\,$r_{bb}:0.92$} \\
$Set$ I: & \multicolumn{3}{c}{$S$: 17.2\%;\, $H$: 5.3\%;\, $Di$: 8.2\%;\, $K$: 69.3\%}\\
$Set$ II: & \multicolumn{3}{c}{$S$: 11.2\%;\, $H$: 6.4\%;\, $Di$: 9.0\%;\, $K$: 73.4\%}\\[2ex]
$15439-i2.6$   & \multicolumn{3}{c}{$\mu=0$} \\
  & \multicolumn{3}{c}{$r_{b \bar{b}}:1.52$;\,\,\,\,\,$r_{\bar{b}\bar{s}}:2.17$;\,\,\,\,\,$r_{b\bar{s}}:1.62$;\,\,\,\,\,$r_{bb}:2.11$} \\
$Set$ I: & \multicolumn{3}{c}{$S$: 1.6\%;\, $H$: 0.8\%;\, $Di$: 1.9\%;\, $K$: 95.7\%}\\
$Set$ II: & \multicolumn{3}{c}{$S$: 6.9\%;\, $H$: 4.4\%;\, $Di$: 1.2\%;\, $K$: 87.5\%}\\[2ex]
$15464-i6.2$   & \multicolumn{3}{c}{$\mu=0$} \\
  & \multicolumn{3}{c}{$r_{b \bar{b}}:1.31$;\,\,\,\,\,$r_{\bar{b}\bar{s}}:1.89$;\,\,\,\,\,$r_{b\bar{s}}:1.43$;\,\,\,\,\,$r_{bb}:1.81$} \\
$Set$ I: & \multicolumn{3}{c}{$S$: 2.5\%;\, $H$: 1.9\%;\, $Di$: 2.7\%;\, $K$: 92.9\%}\\
$Set$ II: & \multicolumn{3}{c}{$S$: 6.8\%;\, $H$: 6.1\%;\, $Di$: 5.6\%;\, $K$: 81.5\%}\\[2ex]
$15531-i0.8$   & \multicolumn{3}{c}{$\mu=0$} \\
  & \multicolumn{3}{c}{$r_{b \bar{b}}:0.88$;\,\,\,\,\,$r_{\bar{b}\bar{s}}:1.31$;\,\,\,\,\,$r_{b\bar{s}}:1.14$;\,\,\,\,\,$r_{bb}:1.22$} \\
$Set$ I: & \multicolumn{3}{c}{$S$: 1.4\%;\, $H$: 1.2\%;\, $Di$: 3.1\%;\, $K$: 94.3\%}\\
$Set$ II: & \multicolumn{3}{c}{$S$: 1.2\%;\, $H$: 3.5\%;\, $Di$: 7.8\%;\, $K$: 87.5\%}\\
\end{tabular}
\end{ruledtabular}
\end{table}


\subsection{The $\mathbf{\bar{b}b\bar{s}b}$ tetraquarks}

{\bf The $\bm{I(J^P)=0(0^+)}$ sector:} Firstly, two meson-meson configurations, $\eta_b B_s$ and $\Upsilon B^*_s$, are studied in both color-singlet and hidden-color channels. No bound state is found, the theoretical thresholds of these two cases are $14.81$ GeV and $14.91$ GeV, respectively. Besides, the two hidden-color channels are almost degenerate at $15.14$ GeV. As for the rest channels, diquark-antidiquark and K-type, their masses are $\sim15.10$ GeV, except for two $K_4$ channels which are $15.48$ GeV and $15.56$ GeV, respectively. We do not find bound states in partially and fully coupled-channel real-range calculations, the lowest-lying mass remains at $14.81$ GeV. Meanwhile, the coupled-mass is $14.99$ GeV for both $K_1$ and $K_2$ states, and the other exotic structures are around $15.1$ GeV.

Four resonance states are available in a CSM fully coupled-channel investigation. Scattering energy dots of $\eta_b B_s$ and $\Upsilon B^*_s$ in both ground and radial excitations are well presented in Fig.~\ref{PP10}. One narrow resonance is also found in the top panel of Fig.~\ref{PP10}, and three more resonances are circled in the bottom panel. Complex energies of these resonances are $15381-i2.0$ MeV, $15439-i2.6$ MeV, $15464-i6.2$ MeV and $15531-i0.8$ MeV, respectively. Furthermore, the second resonance at $15.44$ GeV is quite compatible with the result in Ref.~\cite{Zhu:2023lbx}.

Table~\ref{GresultR10} list electromagnetic, geometric and wave-function component properties of the found resonances. The magnetic moment $\mu$ is 0 for all of the four resonances. Compact structure is concluded for resonances at $15.38$ GeV and $15.53$ GeV, whereas the other two seems to be more extended objects. The dominant wave-function components are of K-type for all these four and they are larger than $70\%$ of the total.


\begin{figure}[!t]
\includegraphics[clip, trim={3.0cm 1.9cm 3.0cm 1.0cm}, width=0.45\textwidth]{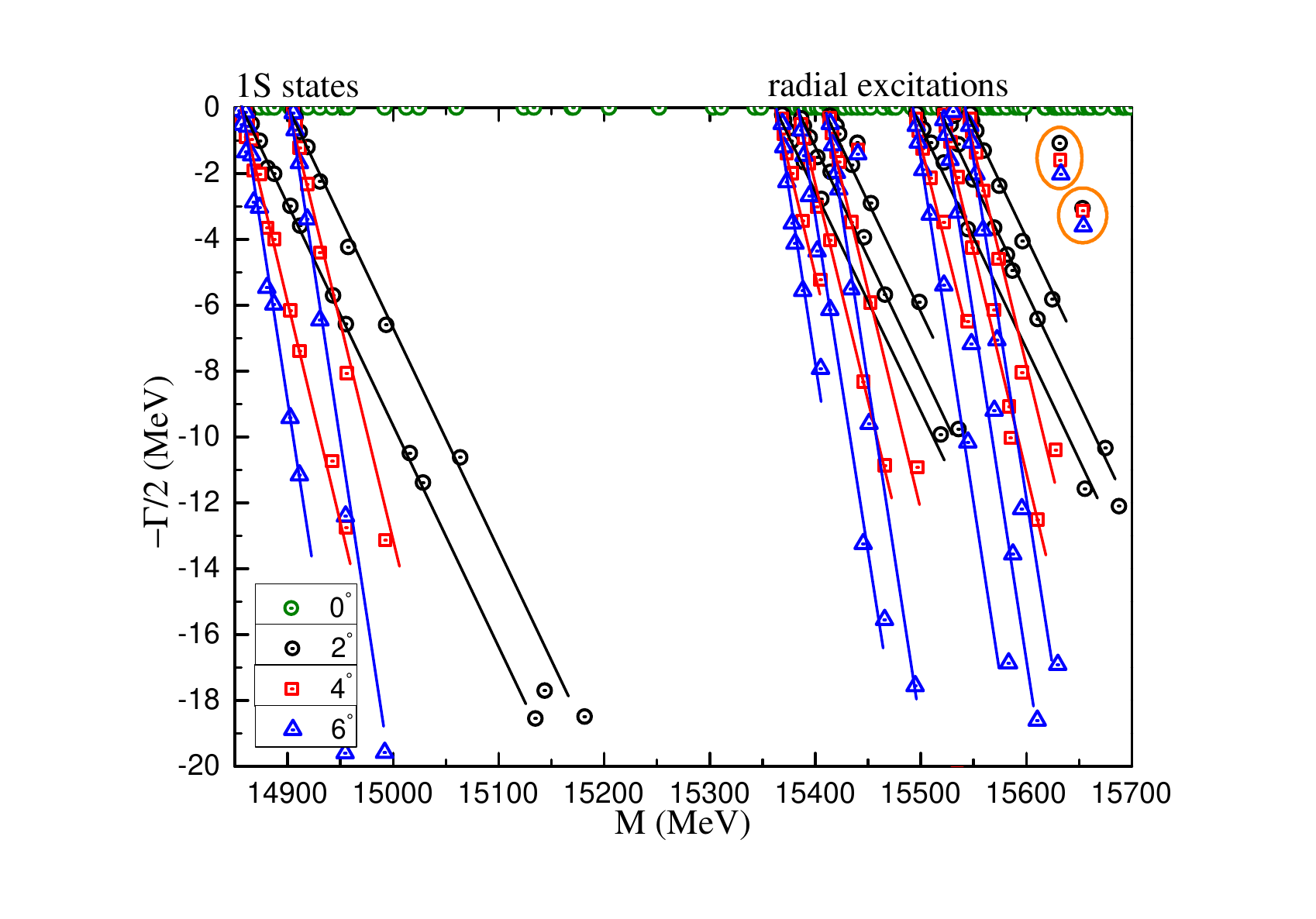} \\[1ex]
\includegraphics[clip, trim={3.0cm 1.9cm 3.0cm 1.0cm}, width=0.45\textwidth]{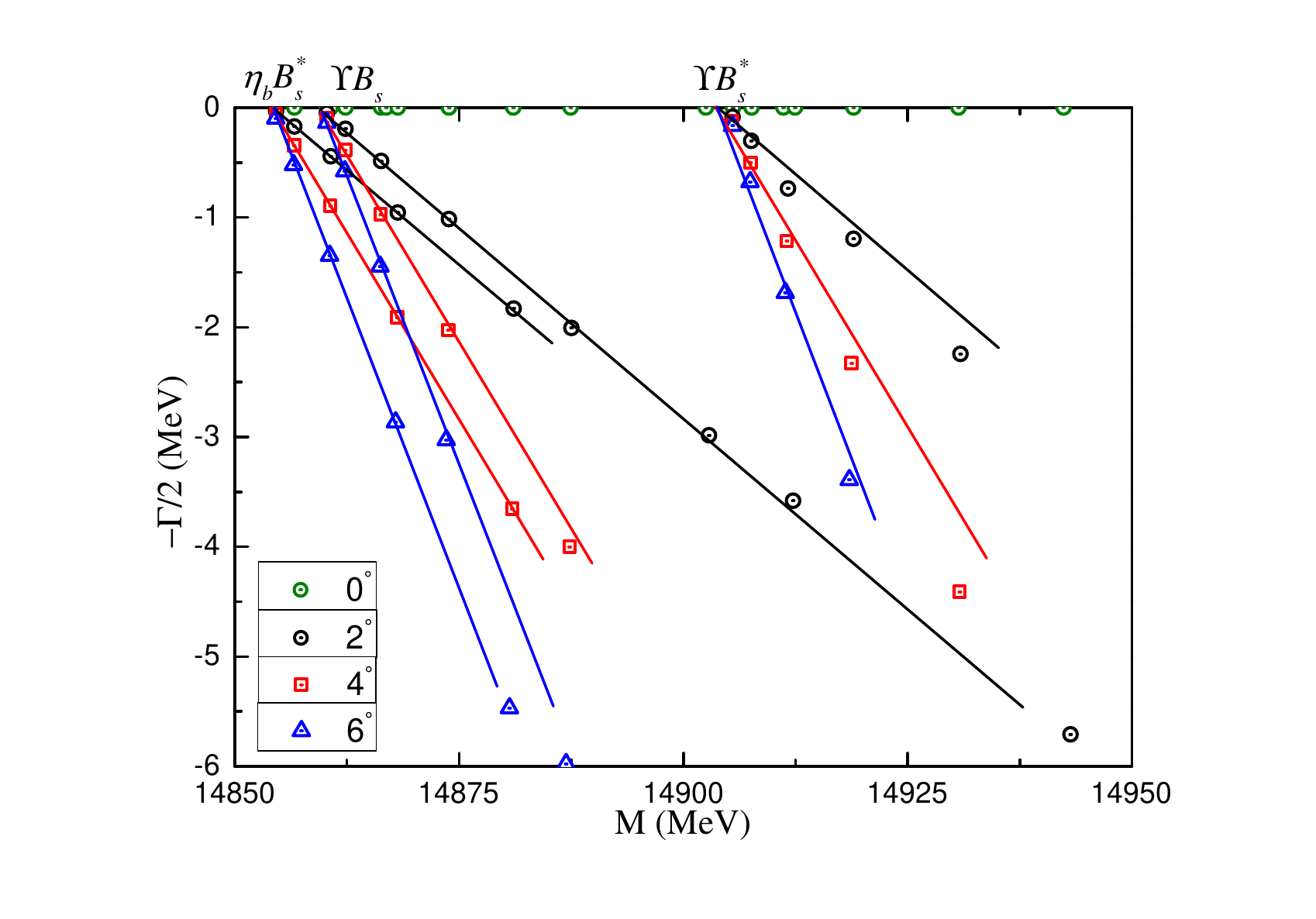}\\[1ex]
\includegraphics[clip, trim={3.0cm 1.9cm 3.0cm 1.0cm}, width=0.45\textwidth]{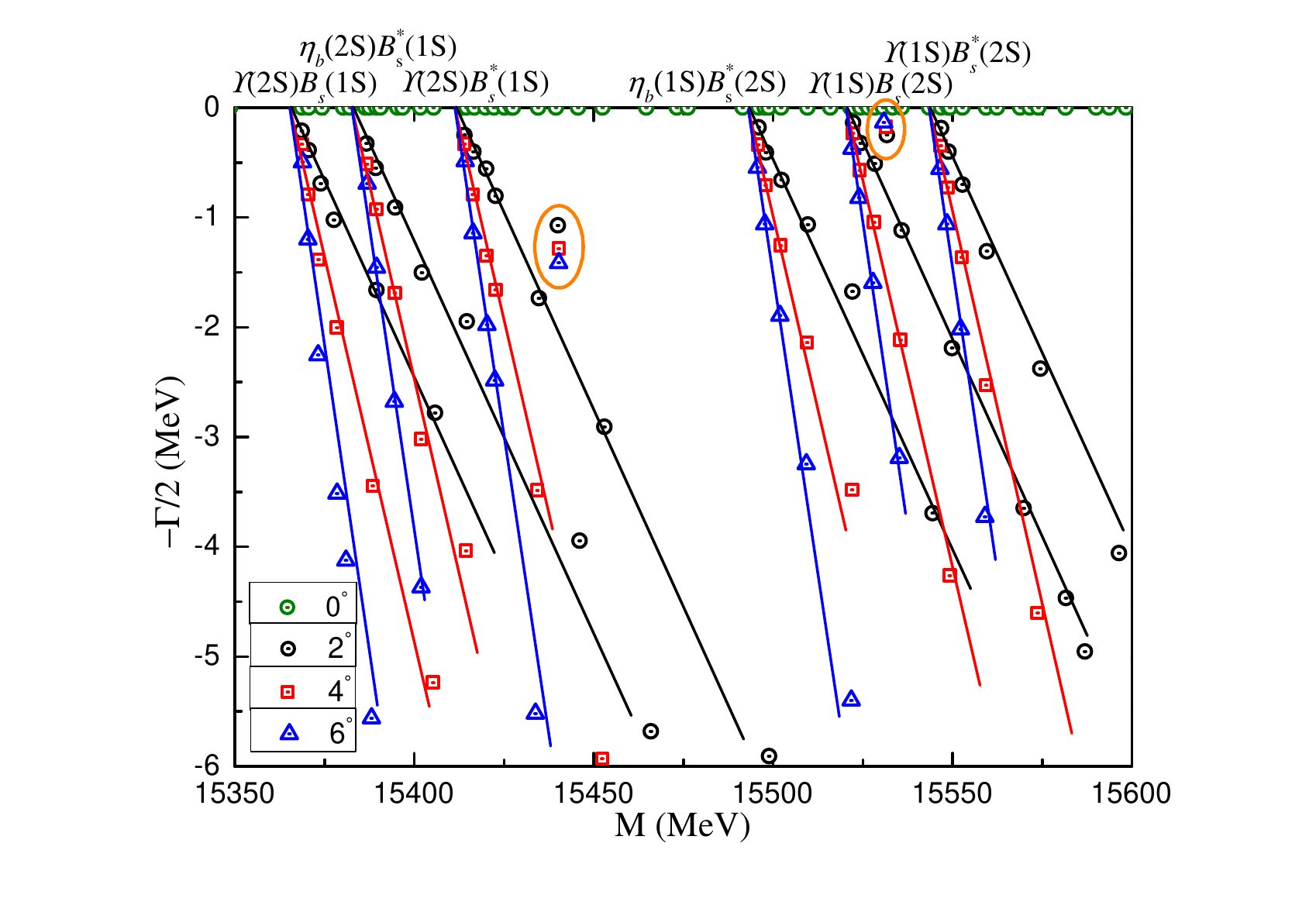}
\caption{\label{PP11} The complete coupled-channels calculation of $\bar{b}b\bar{s}b$ tetraquark system with $I(J^P)=0(1^+)$ quantum numbers. Particularly, the middle panel is enlarged parts of dense energy region from $14.85\,\text{GeV}$ to $14.95\,\text{GeV}$, and the bottom one is enlarged parts of dense energy region from $15.35\,\text{GeV}$ to $15.60\,\text{GeV}$.}
\end{figure}

\begin{table}[!t]
\caption{\label{GresultCC11} Lowest-lying $\bar{b}b\bar{s}b$ tetraquark states with $I(J^P)=0(1^+)$ calculated within the real range formulation of the chiral quark model. Results are similarly organized as those in Table~\ref{GresultCC1} (unit: MeV).}
\begin{ruledtabular}
\begin{tabular}{lcccc}
~~Channel   & Index & $\chi_J^{\sigma_i}$;~$\chi_j^c$ & $M$ & Mixed~~ \\
        &   &$[i; ~j]$ &  \\[2ex]
$(\eta_b B^*_s)^1 (14715)$   & 1  & [1;~1]  & $14854$ & \\
$(\Upsilon B_s)^1 (14827)$  & 2  & [2;~1]   & $14860$ &  \\
$(\Upsilon B^*_s)^1 (14875)$  & 3  & [3;~1]   & $14905$ & $14854$  \\[2ex]
$(\eta_b B^*_s)^8$       & 4  & [1;~2]   & $15132$ &  \\
$(\Upsilon B_s)^8$      & 5  & [2;~2]  & $15134$ & \\
$(\Upsilon B^*_s)^8$  & 6  & [3;~2]   & $15130$ & $15111$ \\[2ex]
$(bb)^*(\bar{b}\bar{s})^*$  & 7  & [6;~3]  & $15153$ & \\
$(bb)^*(\bar{b}\bar{s})$    & 8  & [5;~4]   & $15124$ &  \\
$(bb)(\bar{b}\bar{s})^*$  & 9  & [4;~3]   & $15141$ & $15120$  \\[2ex]
$K_1$      & 10  & [7;~5]   & $15129$ &  \\
      & 11 & [8;~5]  & $15091$ & \\
      & 12  & [9;~5]   & $15111$ & \\
      & 13   & [7;~6]  & $15094$ & \\
      & 14   & [8;~6]  & $15127$ & \\
      & 15   & [9;~6]  & $15077$ & $15008$ \\[2ex]
$K_2$      & 16  & [10;~7]   & $15073$ & \\
                 & 17  & [11;~7]   & $15042$ &  \\
                 & 18  & [12;~7]   & $15024$ & \\
                 & 19  & [10;~8]   & $15103$ & \\
                 & 20  & [11;~8]   & $15131$ & \\
                 & 21  & [12;~8]   & $15119$ & $15005$ \\[2ex]
$K_3$      & 22  & [13;~10]   & $15138$ & \\
                 & 23  & [14;~10]   & $15147$ & \\
                 & 24  & [15;~9]    & $15111$ & $15106$ \\[2ex]
$K_4$      & 25  & [16;~11]   & $15156$ & \\
                 & 26  & [17;~11]   & $15169$ & \\
                 & 27  & [18;~11]   & $15497$ & \\
                 & 28  & [16;~12]   & $15157$ & \\
                 & 29  & [17;~12]   & $15156$ & \\
                 & 30  & [18;~12]   & $15142$ & $15068$ \\[2ex]
$K_5$      & 31  & [19;~14]   & $15155$ & \\
                 & 32  & [20;~14]   & $15152$ & \\
                 & 33  & [21;~13]   & $15101$ & $15098$ \\[2ex]  
\multicolumn{4}{c}{Complete coupled-channels:} & $14854$
\end{tabular}
\end{ruledtabular}
\end{table}

\begin{table}[!t]
\caption{\label{GresultR11} Compositeness of exotic resonances obtained in a complete coupled-channel calculation in the $0(1^+)$ state of $\bar{b}b\bar{s}b$ tetraquark. Results are similarly organized as those in Table~\ref{GresultR1}.}
\begin{ruledtabular}
\begin{tabular}{rccc}
Resonance       & \multicolumn{3}{c}{Structure} \\[2ex]
$15440-i2.6$   & \multicolumn{3}{c}{$\mu=0.382$} \\
  & \multicolumn{3}{c}{$r_{b \bar{b}}:1.01$;\,\,\,\,\,$r_{\bar{b}\bar{s}}:1.49$;\,\,\,\,\,$r_{b\bar{s}}:1.19$;\,\,\,\,\,$r_{bb}:1.39$} \\
$Set$ I: & \multicolumn{3}{c}{$S$: 2.3\%;\, $H$: 2.5\%;\, $Di$: 2.2\%;\, $K$: 93.0\%}\\
$Set$ II: & \multicolumn{3}{c}{$S$: 10.1\%;\, $H$: 3.8\%;\, $Di$: 5.0\%;\, $K$: 81.1\%}\\[2ex]
$15531-i0.4$   & \multicolumn{3}{c}{$\mu=0.496$} \\
  & \multicolumn{3}{c}{$r_{b \bar{b}}:1.01$;\,\,\,\,\,$r_{\bar{b}\bar{s}}:1.48$;\,\,\,\,\,$r_{b\bar{s}}:1.20$;\,\,\,\,\,$r_{bb}:1.37$} \\
$Set$ I: & \multicolumn{3}{c}{$S$: 0.8\%;\, $H$: 2.3\%;\, $Di$: 1.7\%;\, $K$: 95.2\%}\\
$Set$ II: & \multicolumn{3}{c}{$S$: 2.8\%;\, $H$: 2.9\%;\, $Di$: 3.0\%;\, $K$: 91.3\%}\\[2ex]
$15632-i3.2$   & \multicolumn{3}{c}{$\mu=0.355$} \\
  & \multicolumn{3}{c}{$r_{b \bar{b}}:0.83$;\,\,\,\,\,$r_{\bar{b}\bar{s}}:1.22$;\,\,\,\,\,$r_{b\bar{s}}:1.02$;\,\,\,\,\,$r_{bb}:1.07$} \\
$Set$ I: & \multicolumn{3}{c}{$S$: 1.5\%;\, $H$: 1.8\%;\, $Di$: 3.0\%;\, $K$: 93.7\%}\\
$Set$ II: & \multicolumn{3}{c}{$S$: 8.9\%;\, $H$: 7.1\%;\, $Di$: 5.3\%;\, $K$: 78.7\%}\\[2ex]
$15653-i6.3$   & \multicolumn{3}{c}{$\mu=0.226$} \\
  & \multicolumn{3}{c}{$r_{b \bar{b}}:0.89$;\,\,\,\,\,$r_{\bar{b}\bar{s}}:1.28$;\,\,\,\,\,$r_{b\bar{s}}:1.09$;\,\,\,\,\,$r_{bb}:1.03$} \\
$Set$ I: & \multicolumn{3}{c}{$S$: 1.7\%;\, $H$: 1.1\%;\, $Di$: 3.3\%;\, $K$: 93.9\%}\\
$Set$ II: & \multicolumn{3}{c}{$S$: 8.9\%;\, $H$: 2.7\%;\, $Di$: 6.2\%;\, $K$: 82.2\%}\\
\end{tabular}
\end{ruledtabular}
\end{table}

{\bf The $\bm{I(J^P)=0(1^+)}$ sector:} Table~\ref{GresultCC11} lists the 33 channels under investigation for this quantum state. Firstly, masses of the color-singlet channels $\eta_b B^*_s$, $\Upsilon B_s$ and $\Upsilon B^*_s$ are $14.85$ GeV, $14.86$ GeV and $14.91$ GeV, respectively. The three hidden-color channels are almost degenerate with mass $\sim15.13$ GeV. Furthermore, diquark-antidiquark and K-type channels are also locate at around $15.1$ GeV, except for a $K_4$ channel with a mass of $15.50$ GeV. Bound states are not found and coupled-channel calculations do not change this fact. In particular, the lowest mass remains at $14.85$ GeV in the color-singlet and fully coupled-channel computations, the other configurations masses are in an energy region of $15.0-15.1$ GeV.

Four resonances are obtained in a complex-range assessment of the fully-coupled channels case. Figure~\ref{PP11} shows the distribution of complex energy dots within the range $14.85-15.70$ GeV. The ground and radial excitations show scattering nature and they are presented in the top panel. An enlarged part of $14.85-14.95$ GeV energy region is shown in the middle panel. Therein, the $\eta_b B^*_s$, $\Upsilon B_s$ and $\Upsilon B^*_s$ scattering states are well identified. In the bottom panel of Fig.~\ref{PP11}, whose energy interval is $15.35-15.60$ GeV, six $2S$ states of $\bar{b}b\bar{s}b$ tetraquarks are well presented: $\Upsilon(2S)B_s(1S)$, $\eta_b(2S)B^*_s(1S)$, $\Upsilon(2S)B^*_s(1S)$, $\eta_b(1S)B^*_s(2S)$, $\Upsilon(1S)B_s(2S)$ and $\Upsilon(1S)B^*_s(2S)$. Among most unstable dots, four resonance poles are circled, the complex energies read as $15440-i2.6$ MeV, $15531-i0.4$ MeV, $15632-i3.2$ MeV and $15653-i6.3$ MeV.

Some insights about the nature of these resonances can be found in Table~\ref{GresultR11}. Considering the electromagnetic structure the magnetic moment is $0.382\mu_N$, $0.496\mu_N$, $0.355\mu_N$ and $0.226\mu_N$ for each state. Meanwhile, the first two resonances have similar properties related to their size and components. Particularly, their inner quark distances are both within $1.0-1.49$ fm and K-type channels dominate. There are also common features for the other two resonance states. Their sizes are $0.8-1.2$ fm and more than $80\%$ is of K-type component. 


\begin{figure}[!t]
\includegraphics[clip, trim={3.0cm 1.9cm 3.0cm 1.0cm}, width=0.45\textwidth]{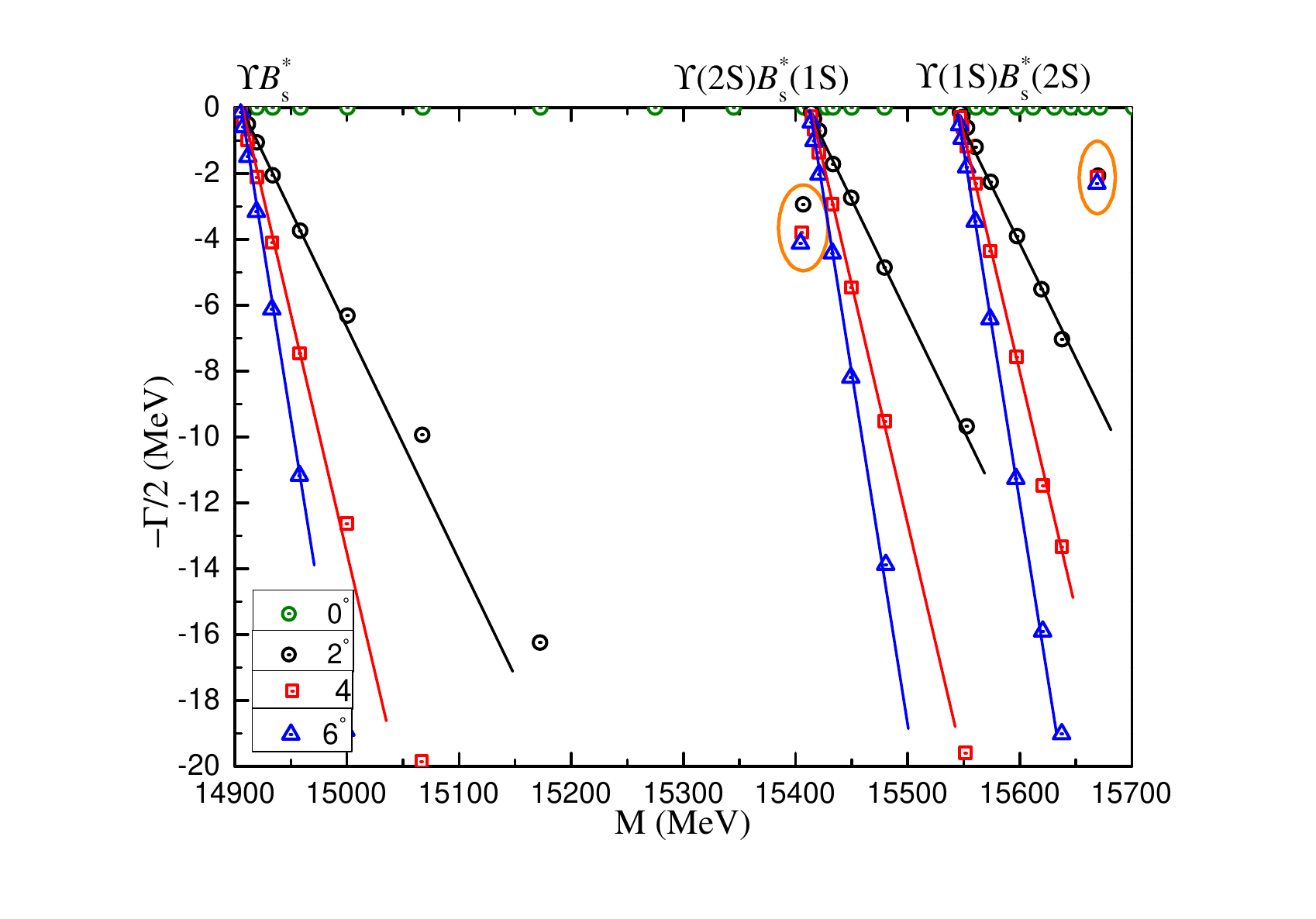}
\caption{\label{PP12} The complete coupled-channels calculation of $\bar{b}b\bar{s}b$ tetraquark system with $I(J^P)=0(2^+)$ quantum numbers.}
\end{figure}

\begin{table}[!t]
\caption{\label{GresultCC12} Lowest-lying $\bar{b}b\bar{s}b$ tetraquark states with $I(J^P)=0(2^+)$ calculated within the real range formulation of the chiral quark model. Results are similarly organized as those in Table~\ref{GresultCC1} (unit: MeV).}
\begin{ruledtabular}
\begin{tabular}{lcccc}
~~Channel   & Index & $\chi_J^{\sigma_i}$;~$\chi_j^c$ & $M$ & Mixed~~ \\
        &   &$[i; ~j]$ &  \\[2ex]
$(\Upsilon B^*_s)^1 (14875)$  & 1  & [1;~1]   & $14905$ &  \\[2ex]
$(\Upsilon B^*_s)^8$            & 2  & [1;~3]   & $15127$ &  \\[2ex]
$(bb)^*(\bar{b}\bar{s})^*$  & 3  & [1;~7]   & $15161$ & \\[2ex]
$K_1$  & 4  & [1;~8]   & $15112$ & \\
            & 5  & [1;~10]   & $15137$ & $15055$ \\[2ex]
$K_2$  & 6  & [1;~11]   & $15086$ & \\
             & 7  & [1;~12]   & $15119$ & $15076$ \\[2ex]
$K_3$  & 8  & [1;~14]   & $15155$ & \\[2ex]
$K_4$  & 9  & [1;~8]   & $15500$ & \\
             & 10  & [1;~10]   & $15162$ & $15121$ \\[2ex]
$K_5$  & 11  & [1;~11]   & $15166$ & \\[2ex]
\multicolumn{4}{c}{Complete coupled-channels:} & $14905$
\end{tabular}
\end{ruledtabular}
\end{table}

\begin{table}[!t]
\caption{\label{GresultR12} Compositeness of exotic resonances obtained in a complete coupled-channel calculation in the $0(2^+)$ state of $\bar{b}b\bar{s}b$ tetraquark. Results are similarly organized as those in Table~\ref{GresultR1}.}
\begin{ruledtabular}
\begin{tabular}{rccc}
Resonance       & \multicolumn{3}{c}{Structure} \\[2ex]
$15405-i7.6$   & \multicolumn{3}{c}{$\mu=0.503$} \\
  & \multicolumn{3}{c}{$r_{b \bar{b}}:0.68$;\,\,\,\,\,$r_{\bar{b}\bar{s}}:0.89$;\,\,\,\,\,$r_{b\bar{s}}:0.74$;\,\,\,\,\,$r_{bb}:0.71$} \\
$Set$ I: & \multicolumn{3}{c}{$S$: 0.6\%;\, $H$: 0.7\%;\, $Di$: 2.3\%;\, $K$: 96.4\%}\\
$Set$ II: & \multicolumn{3}{c}{$S$: 4.5\%;\, $H$: 6.4\%;\, $Di$: 6.5\%;\, $K$: 82.6\%}\\[2ex]
$15669-i4.2$   & \multicolumn{3}{c}{$\mu=0.503$} \\
  & \multicolumn{3}{c}{$r_{b \bar{b}}:0.99$;\,\,\,\,\,$r_{\bar{b}\bar{s}}:1.56$;\,\,\,\,\,$r_{b\bar{s}}:1.31$;\,\,\,\,\,$r_{bb}:1.35$} \\
$Set$ I: & \multicolumn{3}{c}{$S$: 0.6\%;\, $H$: 1.1\%;\, $Di$: 1.9\%;\, $K$: 96.4\%}\\
$Set$ II: & \multicolumn{3}{c}{$S$: 3.8\%;\, $H$: 3.7\%;\, $Di$: 4.5\%;\, $K$: 88.0\%}\\
\end{tabular}
\end{ruledtabular}
\end{table}

{\bf The $\bm{I(J^P)=0(2^+)}$ sector:} Table~\ref{GresultCC12} lists the calculated results on the highest spin state of $\bar{b}b\bar{s}b$ tetrquark. Bound states remain unavailable. The masses of the color-singlet and hidden-color channels of $\Upsilon B^*_s$ are $14.91$ GeV and $15.13$ GeV, respectively. The $(bb)^*(\bar{b}\bar{s})^*$ channel is slightly higher in mass, at $15.16$ GeV. Eight K-type channels are generally located at an energy interval from $15.09$ GeV to $15.50$ GeV. The lowest-lying mass is around $15.10$ GeV when partially coupled-channel calculations are performed in the $K_1$, $K_2$ and $K_4$ configurations.

When performing a fully-coupled channels investigation within the CSM method, three scattering states of $\Upsilon B^*_s$, $\Upsilon(2S)B^*_s(1S)$ and $\Upsilon(1S)B^*_s(2S)$ are clearly presented in the $14.9-15.7$ GeV energy region of Fig.~\ref{PP12}. Two narrow resonances are obtained in the complex plane, $15405-i7.6$ MeV and $15669-i4.2$ MeV. In particular, the higher resonance is consistent with the result in Ref.~\cite{Mutuk:2023yev}.

Some properties of these two resonance states are summarized in Table~\ref{GresultR12}. Particularly, their magnetic moments are both $0.503\mu_N$. Besides, the first resonance has a size of $\sim0.7$ fm and the other state is a little larger with size of $1.0-1.5$ fm. The dominant component, which is more than $80\%$, comes from K-type channels for these two resonances.


\begin{table}[!t]
\caption{\label{GresultCCT} Summary of resonance structures found in the $\bar{Q}Q\bar{q}Q$ $(q=u,\,d,\,s;\,Q=c,\,b)$ tetraquark systems. The first column shows the isospin, total spin and parity of each singularity. The second column refers to the theoretical resonance with notation: $E=M-i\Gamma$ (unit: MeV). Size ($r$, unit: fm) and magnetic moment ($\mu$, unit: $\mu_N$) of resonance are presented in the last column.}
\begin{ruledtabular}
\begin{tabular}{lcc}
~ $I(J^P)$ & Theoretical resonance   & Structure~~ \\
                   & $E=M-i\Gamma$   & $r,\,\mu$ \\
\hline
\multicolumn{3}{c}{$\bar{c}c\bar{q}c$ tetraquarks}\\
~~$\frac{1}{2}(0^+)$  & $5592-i2.5$  & $0.95\sim1.45$, $0$~~  \\
                                    & $5714-i1.8$   & $1.30\sim1.90$, $0$~~  \\
                                    & $5828-i2.6$   & $1.30\sim1.85$, $0$~~  \\[2ex]
~~$\frac{1}{2}(1^+)$   & $5693-i3.0$   & $1.25\sim1.75$, $-1.06$~~ \\[2ex]
~~$\frac{1}{2}(2^+)$   & $5691-i5.9$   & $0.90\sim1.20$, $-1.64$~~ \\
                                     & $5827-i2.3$   & $1.25\sim1.80$, $-1.64$~~ \\[2ex]
~~$0(0^+)$    & $5682-i4.6$   & $1.50\sim2.10$, $0$~~ \\
                       & $5855-i2.8$   & $1.05\sim1.40$, $0$~~ \\[2ex]
~~$0(1^+)$     & $5863-i6.6$  & $1.60\sim2.30$, $0.578$~~  \\
                       & $5941-i1.8$   & $1.15\sim1.55$, $0.503$~~  \\
                       & $5942-i5.6$   & $1.30\sim1.95$, $0.068$~~  \\[2ex]
~~$0(2^+)$    & $5788-i3.1$   & $0.80\sim1.10$, $0.921$~~ \\
                       & $5943-i5.1$   & $1.40\sim2.05$, $0.921$~~ \\
\hline
\multicolumn{3}{c}{$\bar{b}b\bar{q}b$ tetraquarks}\\
~~$\frac{1}{2}(0^+)$  & $15320-i0.5$  & $0.95\sim1.55$, $0$~~  \\
                                    & $15331-i4.5$   & $1.20\sim1.80$, $0$~~  \\
                                    & $15372-i6.1$   & $1.10\sim1.70$, $0$~~  \\
                                    & $15407-i9.3$   & $1.05\sim1.60$, $0$~~  \\[2ex]
~~$\frac{1}{2}(1^+)$   & $15327-i1.5$  & $1.05\sim1.60$, $-1.453$~~  \\
                                    & $15376-i3.2$   & $1.00\sim1.65$, $-1.817$~~  \\
                                    & $15395-i4.2$   & $0.95\sim1.50$, $-1.456$~~  \\
                                    & $15417-i10.2$   & $0.95\sim1.50$, $-1.771$~~  \\[2ex]
~~$\frac{1}{2}(2^+)$   & $15308-i9.3$   & $0.60\sim0.95$, $-2.061$~~ \\
                                     & $15449-i1.4$   & $0.75\sim1.30$, $-2.061$~~ \\
                                     & $15558-i5.4$   & $1.00\sim1.55$, $-2.061$~~ \\[2ex]
~~$0(0^+)$    & $15381-i2.0$  & $0.65\sim1.10$, $0$~~  \\
                       & $15439-i2.6$   & $1.50\sim2.20$, $0$~~  \\
                       & $15464-i6.2$   & $1.30\sim1.90$, $0$~~  \\
                       & $15531-i0.8$   & $0.85\sim1.35$, $0$~~  \\[2ex]
~~$0(1^+)$     & $15440-i2.6$  & $1.00\sim1.50$, $0.382$~~  \\
                       & $15531-i0.4$   & $1.00\sim1.50$, $0.496$~~  \\
                       & $15632-i3.2$   & $0.80\sim1.25$, $0.355$~~  \\
                       & $15653-i6.3$   & $0.85\sim1.30$, $0.226$~~  \\[2ex]
~~$0(2^+)$    & $15405-i7.6$   & $0.65\sim0.90$, $0.503$~~ \\
                       & $15669-i4.2$   & $0.95\sim1.60$, $0.503$~~
\end{tabular}
\end{ruledtabular}
\end{table}

\section{Summary}
\label{sec:summary}

Triply heavy tetraquarks, $\bar{Q}Q\bar{q}Q$ $(q=u,\,d,\,s;\,Q=c,\,b)$, with spin-parity $J^P=0^+$, $1^+$ and $2^+$, and isospin $0$ and $\frac{1}{2}$, have been systematically investigated within a constituent quark model framework which has been widely used in heavy quark sectors. Moreover, meson-meson, diquark-antidiquark and K-type arrangements are comprehensively considered, taking into account all corresponding color, spin and flavor configurations. Furthermore, a highly accurate and efficient numerical method, dealing simultaneously with bound, resonant and scattering states, is employed to solve the 4-body Schr\"odinger-like equation. For each $I(J^P)$ case, single channel calculation is performed firstly, whereas partially and fully coupled-channel computations are developed later. In the last case, narrow resonances are obtained in each $I(J^P)$ case.

Our findings for $\bar{c}c\bar{q}c$ and $\bar{b}b\bar{q}b$ tetraquarks are summarized in Table~\ref{GresultCCT}. In particular, complex mass, average size and magnetic moment of resonances are collected according to the quantum numbers $I(J^P)$, which are shown in the first column; the second column is about the resonance's pole position, and the third one shows average size and magnetic moment. 

Some conclusions can be highlighted. Firstly, several narrow resonances with widths less than $~11$ MeV are found within an energy region $5.6-5.9$ GeV and $15.3-15.7$ GeV for triply charm and bottom tetraquarks, respectively. Secondly, the magnetic moment of $0^+$ states is zero for both charm and bottom cases; meanwhile, it is negative valued for $1^+$ and $2^+$ states with $I=\frac{1}{2}$, but positive for the isoscalar cases. Thirdly, the average size of those resonances obtained in the $\bar{b}b\bar{q}b$ tetraquark sector is less than $1.8$ fm, except for the $I(J^P)=0(0^+)$ $\bar{b}b\bar{d}b$ tetraquark with a size of $1.5-2.2$ fm. In the case of $\bar{c}c\bar{q}c$ tetraquarks, one can find compact and loosely bound molecules. Finally, the dominant component of all these resonances is usually of K-type configuration.


\begin{acknowledgments}
Work partially financed by National Natural Science Foundation of China under Grant Nos. 12305093, 11535005 and 11775118; Zhejiang Provincial Natural Science Foundation under Grant No. LQ22A050004; Ministerio Espa\~nol de Ciencia e Innovaci\'on under grant No. PID2022-140440NB-C22; Junta de Andaluc\'\i a under contract Nos. PAIDI FQM-370 and PCI+D+i under the title: ”Tecnolog\'\i as avanzadas para la exploraci\'on del universo y sus componentes” (Code AST22-0001).
\end{acknowledgments}


\bibliography{THT}

\end{document}